\documentclass[longauth]{aa}

\usepackage{graphicx}
\usepackage{txfonts}

\usepackage[T1]{fontenc}
\usepackage{ae,aecompl}
\usepackage{natbib}
\bibpunct{(}{)}{;}{a}{}{,} 
\usepackage{amsmath}	
\usepackage{amssymb}	
\usepackage{tikz,graphics,color,float,epsf} 
\usepackage{rotating} 
\usepackage{lscape} 
\usepackage{rotating}
\usepackage{xcolor}
\usepackage{calc}
\usepackage[english]{babel}
\usepackage{lipsum}
\usepackage{hyperref}
\usepackage{textcomp}
\usepackage{longtable}
\usepackage{booktabs}
\usepackage{caption}
\usepackage{geometry}
\usepackage{amsmath}
\usepackage{placeins}

\newcommand{\lsun}{\mbox{L}_\odot}

\newcommand{\msun}{\mbox{M}_\odot}
\newcommand{\pab}{\mbox{Pa}\beta}
\newcommand{\brg}{\mbox{Br}\gamma}
\newcommand{\lacc}{L_{\rm acc}}
\newcommand{\macc}{\dot{M}_{\rm acc}}
\newcommand{\lstar}{L_\star}
\newcommand{\mstar}{M_\star}

\newcommand{\teff}{T_{\rm eff}}

\newcommand{\lline}{L_{\rm line}}

\begin{document}

   \title{PENELLOPE VII: Revisiting empirical relations to measure accretion luminosity \thanks{Based on observations collected at the European Southern Observatory under ESO programme 106.20Z8.002, 106.20Z8.004, 106.20Z8.006, and 106.20Z8.008.}} 

\titlerunning{PENELLOPE VII.}   

   \author{E. Fiorellino
          \inst{1,2,3}
          \and J. M. Alcal\'a\inst{1} 
          \and C. F. Manara\inst{4}
          \and C. Pittman\inst{5,6}  \and 
          P. \'Abrah\'am\inst{7,8,9} \and L. Venuti\inst{10} \and S. Cabrit\inst{11,12} \and R. Claes\inst{4} \and M. Fang\inst{13}  \and \'A. K\'osp\'al\inst{7,8,14} \and {G. Lodato}\inst{15} \and K. Mauco\inst{4} \and \L{}. Tychoniec\inst{16}
          }

   \institute{INAF-Osservatorio Astronomico di Capodimonte, via Moiariello 16, 80131 Napoli, Italy\\
\email{eleonora.fiorellino@inaf.it}
%2
\and Instituto de Astrofísica de Canarias, IAC, Vía Láctea s/n, 38205 La Laguna (S.C.Tenerife), Spain
\and
Departamento de Astrofísica, Universidad de La Laguna, 38206 La Laguna (S.C.Tenerife), Spain
\and
European Southern Observatory, Karl-Schwarzschild-Strasse 2, 85748, Garching bei München, Germany
\and 
Department of Astronomy, Boston University, 725 Commonwealth Avenue, Boston, MA 02215, USA
\and
Institute for Astrophysical Research, Boston University, 725 Commonwealth Avenue, Boston, MA 02215, USA
\and
Konkoly Observatory, HUN-REN Research Centre for Astronomy and Earth Sciences, MTA Centre of Excellence, Konkoly-Thege Mikl\'os \'ut 15-17, 1121 Budapest, Hungary
\and
Institute of Physics and Astronomy, ELTE E\"otv\"os Lor\'and University, P\'azm\'any P\'eter s\'et\'any 1/A, 1117 Budapest, Hungary
\and
Institute for Astronomy (IfA), University of Vienna,T\"urkenschanzstrasse 17, A-1180 Vienna, Austria
\and
SETI Institute, 339 Bernardo Ave., Suite 200, Mountain View, CA 94043, USA
\and
Observatoire de Paris - PSL University, Sorbonne Université, LERMA, CNRS, Paris, France 
\and Univ. grenoble Alpes, CNRS, IPAG, Grenoble, France
\and 
Purple Mountain Observatory, Chinese Academy of Sciences, 10 Yuanhua Road, Nanjing 210023, People's Republic of China
\and Max-Planck-Insitut f\"ur Astronomie, K\"onigstuhl 17, 69117 Heidelberg, Germany
\and 
Dipartimento di Fisica, Università degli Studi di Milano, Via Celoria 16, 20133 Milano, Italy
\and
Leiden Observatory, Leiden University, PO Box 9513, 2300RA, Leiden, The Netherlands
             }

   \date{}

% \abstract{}{}{}{}{} 
% 5 {} token are mandatory
 
  \abstract
   {The accretion luminosity ($\lacc$) in young, low-mass stars is crucial for understanding stellar formation, but direct measurements are often hindered by limited spectral coverage and challenges in UV-excess modeling. Empirical relations linking $\lacc$ to various accretion tracers are widely used to overcome these limitations.}
   {This work revisits these empirical relations using the PENELLOPE dataset, evaluating their applicability across different star-forming regions and to accreting young objects other than Classical T\,Tauri Stars (CTTSs; Class\,II sources).}
   {We analyzed the PENELLOPE VLT/X-Shooter dataset of 64 CTTSs, measuring fluxes of several accretion tracers and adopting the stellar and accretion parameters derived from PENELLOPE works. For 61 sources, we supplemented our analysis with the ODYSSEUS HST data set, which covers a wider spectral range in NUV bands.}
   {We compared the $\lacc$ values obtained in the PENELLOPE and ODYSSEUS surveys, which employed a single hydrogen slab model (XS-fit) and a multi-column accretion shock model (HST-fit), respectively, and found statistically consistent results. Our analysis confirms that existing empirical relations, previously derived for the Lupus sample, provide reliable $\lacc$ estimates for CTTSs in several other star-forming regions. We revisit empirical relations for accretion tracers in our dataset, based on HST-fit, with coefficients which are consistent within 1$\sigma$ with XS-fit results for most lines. We also propose a method to estimate extinction using these relations and investigate the empirical relations for Brackett lines (Br8 to Br21).}
   {The $\lacc - \lline$ empirical relations can be successfully used for statistical studies of accretion on young forming objects in different star-forming regions. These relations also offer a promising approach to independently estimate extinction in CTTSs, provided a sufficient number of flux-calibrated tracers is available across a broad spectral range. We confirm that near-infrared lines (Pa$\beta$ and Br$\gamma$) reliably trace $\lacc$ in high accretors, making them valuable tools for probing accretion properties of high accreting young stars not accessible in the UVB.}


\maketitle

%-------------------------------------------------------------------

\section{Introduction}

\begin{figure*}
    \centering
    \includegraphics[width=\textwidth]{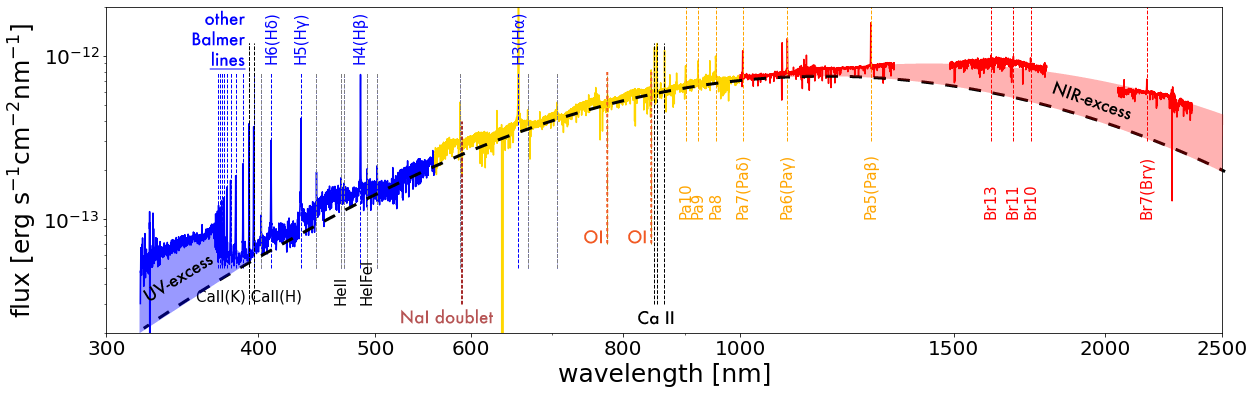}
        \caption{X-Shooter spectrum of the CTTS VW\,Cha. The UVB, VIS and NIR spectra are plotted in blue, yellow, and red, respectively. The UV-excess (blue region), tracing the accretion activity, and the NIR-excess (red region), tracing the presence of an active disk, are highlighted. The black curved dashed line corresponds to a stylized stellar photosphere\,$+$\,continuum veiling. All the accretion tracers for which \citet{alc17} provided empirical relations are shown in different colors, but not all the lines are labelled in the figure because of limited space. Also Br13, Br11, and Br10 lines are shown. Note that the spectrum starting wavelength was artificially set at $\sim$320\,nm for plotting purposes, while the actual UV-cut of X-Shooter is at $\sim$300\,nm. }
    \label{fig:XS-spectrum}
\end{figure*}

The stellar mass in low-mass ($M_\star < 2\,M_\odot$) stars is set by post-collapse accretion \citep{lyn74, bertout1988, har98}, with young stellar objects (YSOs) gaining mass by accreting material from the envelope and circumstellar disk. 
Depending on their Spectral Energy Distributions (SEDs), YSOs can be classified in Classes from the most embedded protostars (Class\,0, I, and Flat\,Spectrum) to the optically visible pre-main-sequence stars, i.e., Class\,II and III \citep[see e.g.,][]{gre94, and95}. Although, in the current star formation scenario, the most embedded sources should also be the youngest and the most accreting YSOs, the classification into Classes might not be directly linked to the corresponding evolutionary stage \citep{eno09}.

The accretion luminosity ($\lacc$) is the parameter that quantifies the energy emitted from the accretion process and play an important role in constraining the evolutionary path of protoplanetary discs. 
In Class\,II pre-main sequence stars or Classical T\,Tauri Stars (CTTSs) the accretion flow from the disk onto the forming star follows the stellar magnetic field lines, as described by the magnetospheric accretion paradigm \citep{bouvier1986, har16}. 

The accretion luminosity has been constrained in CTTSs \citep[][and references therein]{val93, Gullbring1998, her08, alc14, alc17, pittman2022,Manara2023}, where it can be directly measured by modeling the ultraviolet continuum excess emission (UV-excess), generated by the accretion shock, over the stellar photosphere \citep{calvet1998, Schneider2020} and by using empirical correlations between $\lacc$ and U-band excess luminosity \citep{Gullbring1998, Sicilia-Aguilar2010}. 
Alternatively, $\lacc$ can be derived using accretion-tracing emission lines (see Fig.\,\ref{fig:XS-spectrum}) and empirical relations between $\lacc$ and the luminosity of the lines \citep{muz98,her08,alc14,alc17}.

Commonly used $\lacc - L_{\rm line}$ relations are provided in \citet{alc17}.
In that work, $\lacc$ was measured adopting the methodologies described in \citet{man13b} namely, by fitting the continuum UV-excess.  
Among other stellar and accretion parameters (see Section\,\ref{sect:XS-fit}), the procedure yields an estimate of the interstellar extinction, $A_V$.
These $A_V$ values were used to deredden the line fluxes of several accretion tracers (see Fig.\,\ref{fig:XS-spectrum}). 
The line fluxes were then converted in luminosity, $\lline$, by adopting a distance. 
This provided a set of $\lacc$ and $\lline$ values for every accretion tracer, from which a relation of the form $\log (\lacc) = a \, \log (\lline) + b$ was derived. 
The sample of CTTSs in \citet{alc17} is composed by 89 objects in the Lupus star-forming region (hereafter the \href{#label}{A17} sample) with spectral types ranging from M8.5 to K0, $\mstar \sim 0.20 - 2.15\,\msun$, $ \lacc \sim 10^{-5.4} - 10^{-0.25}\, \lsun$, and $\lstar \sim 0.003 - 5.420\, \lsun$.

Empirical relations have been used to constrain the accretion luminosity not only on CTTS \citep[e.g.,][]{Rugel2018, fio22dqtau, fiorellino2022wx, Pouilly2024}, but also on younger stars classified as protostars \citep[e.g.,][]{fio21, Fiorellino2023, Tychoniec2024}, on accreting brown dwarfs \citep[e.g.,][]{Whelan2018, Almendros-Abad2024}, on YSOs experiencing episodic accretion, as EXors-like burst \citep[e.g.,][]{Singh2024, giannini2024}, and even on forming-planets \citep[e.g.,][]{Haffert2019, Plunkett2025, Bowler2025}. 

While accurate $\lacc$ measurements must be drawn from data acquired in the widest possible spectral range in the UV, it is important to note that $\lacc$ determinations based on X-Shooter data may have the caveat of the spectral range cut at 300\,nm (see Fig.\,\ref{fig:XS-spectrum}).
The ULLYSES\footnote{\url{https://ullyses.stsci.edu/}} \citep[UV Legacy Library of Young Stars as Essential Standards,][]{Roman-Duval2020} survey aims at solving such limitations. 
ULLYSES utilizes the Hubble Space Telescope (HST) to provide an unprecedented UV spectroscopic library of about 70 CTTSs. 
The ODYSSEUS\footnote{\url{https://sites.bu.edu/odysseus/}} \citep[Outflows and Disks around Young Stars: Synergies for the Exploration of Ullyses Spectra,][]{Espaillat2022} collaboration builds on the ULLYSES survey, utilizing multi-wavelength observations from X-ray to submillimeter in addition to COS and STIS observations, using over 500 HST orbits. 
The ESO Large Program PENELLOPE\footnote{\url{https://doi.eso.org/10.18727/archive/88}} \citep[][]{manara2021} complements ULLYSES-ODYSSEUS dataset with high-resolution (UVES/ESPRESSO) and flux-calibrated medium-resolution (X-Shooter) optical to near-infrared spectra from the Very Large Telescope (VLT), contemporaneous to the ULLYSES observations. 
The ULLYSES-ODYSSEUS-PENELLOPE synergy enables a comprehensive analysis of key accretion and stellar parameters, extinction, and gas kinematics. 

The goal of this work is twofold: first, to test whether empirical relation by \href{#label}{A17} can be successfully applied to the overall PENELLOPE sample, composed by CTTSs belonging to several star-forming regions, and, second, compare the empirical relations with those derived from the HST ULLYSES data. We then discuss the applicability of the empirical relations on young accreting objects other than CTTSs.

The paper is structured as follows. 
In Sect.\,\ref{sect:sample} the samples are described; Sect.\,\ref{sect:analysis} summarizes the methodologies for deriving $\lacc$ for the PENELLOPE and ULLYSES/ODYSSEUS samples, and describe the analysis of the X-Shooter data. We revisit the empirical relations between $\lacc$ and $\lline$ using the ULLYSSES-HST data set, we test the reliability of $\lacc$ estimates obtained through empirical relations by comparing them with values derived from direct UV-excess modeling. We also provide an independent method to constrain the $A_V$ of CTTSs using empirical relations, and investigate the possibility for new empirical relations of the Brackett series. Finally, we discuss the results in Sect.\,\ref{sect:discussion}, drawing our conclusions in Sect.\,\ref{sect:conclusions}.

\section{Data Sample}
\label{sect:sample}
This work analyses the X-Shooter spectra from the PENELLOPE project. For a detailed description of the sample and the primary goals of the PENELLOPE observing program, we refer the reader to \citet{manara2021}. We excluded variability monitoring targets, for which dedicated work has already been published \citep{armeni2023, armeni2024, Wendeborn2024I, Wendeborn2024III}.
This results in a sample of 68 YSOs belonging to several star forming regions: Orion\,OB1 (8), $\sigma$Ori (3), Chamaeleon\,I (13), $\epsilon$\,Cha (1), $\eta$\,Cha (7), Corona Australis (2), Taurus (8), and Lupus (30), see Tab.\ref{tab:sample1}. Among these objects, RECX\,5, Lk\,Ca4, and RXJ0438.6$+$1546 are Class\,III YSOs (or Weak-line T\,Tauri Stars, WTTSs); AA\,Tau is a dipper \citep{Bouvier1999} which lost its regular dipper appearance in 2011 \citep{Bouvier2013}, and its Kepler (K2) light curve from 2017 is dominated by stochastic variability \citep{Cody2022}; the remaining 64 CTTS constitute the sample we analyzed in this work.

\begin{figure}
    \centering
    \includegraphics[width=0.7\columnwidth]{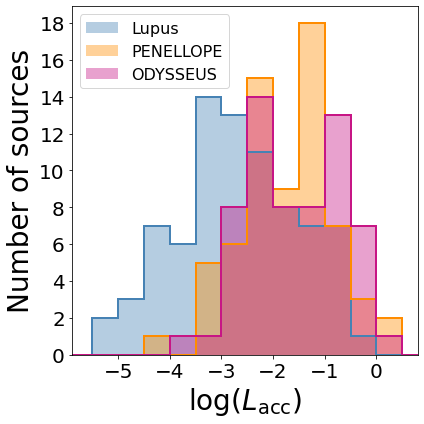}
\includegraphics[width=0.7\columnwidth]{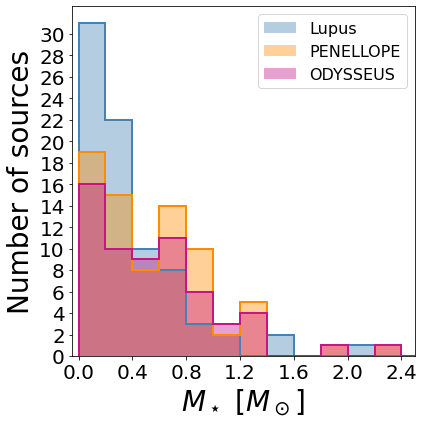}
    \caption{Histograms of accretion luminosity (top) and stellar mass (bottom) for the Lupus sample from \citep[][blue]{alc17}, the PENELLOPE sample (orange), and the ODYSSEUS sample (pink). All histograms are constructed using the same bin width: 0.5\,dex for $\log \lacc$ and 0.2\,$M_\odot$ for stellar mass.}
    \label{fig:HistoLacc}
\end{figure}
The details on data reduction of the PENELLOPE sample are described in \citet{manara2021}. In a nutshell, the X-Shooter data were reduced using a dedicated ESO pipeline \citep{modigliani2010}, which follows the standard steps including flat, bias, and dark correction, wavelength calibration, spectral rectification, extraction of 1D spectra, and flux calibration using a standard star obtained during the same night.
In cases where the SNR of the UVB arm was low, or for resolved binaries, the 1D spectrum was extracted with IRAF\footnote{iraf is distributed by the National Optical Astronomy Observatories, which are operated by the Association of Universities for Research in Astronomy, Inc., under the cooperative agreement with the National Science Foundation. NOAO stopped supporting IRAF, see \url{https: //iraf- community.github.io/}.}.
Telluric correction for the VIS and NIR arms was performed using {\it molecfit} \citep{sme15, kausch2015}. 
Each target is observed first using a set of wide slit (5.0$\arcsec$) in the three arms, leading to a low resolution
observation with no slit losses. Then, using 1.0$\arcsec$/0.4$\arcsec$/0.4$\arcsec$–wide slits
for the UVB, VIS, and NIR arms, respectively.
The flux calibration has been performed by scaling the narrow slits spectra to the wide-slit ones to correct for slit losses. A comparison of the flux calibrated X-Shooter and HST spectra, in the overlapping spectral regions, is shown in \citet{manara2021}. This comparison shows that the flux calibration of the X-Shooter spectra is comparable to that of HST. This is important because the precision on the $\lacc - L_{\rm line}$ relations relies on the precision on $L_{\rm line}$ determinations, which in turn depends on the goodness of the flux calibration.
All the reduced spectra are publicly available on Zenodo \footnote{\url{https://zenodo.org/communities/odysseus/records?q=&l=list&p=1&s=10&sort=newest}} and on the ESO Phase\,3\footnote{https://doi.eso.org/10.18727/archive/88}.

The HST data reduction was completed using a custom pipeline at STScI, which was created by the ULLYSES team \citep{Roman-Duval2025}.
Details on the data reduction are described in \citet{Roman-Duval2020, Espaillat2022}. The reader is referred to \citet{pittman2025} for a detail description of the HST sample analysed in this work. 
For technical reasons, the VLT/X-Shooter and HST samples slightly differ in number, resulting in a common sample of 61 sources (see Tab.\,\ref{tab:Lacc-Av-varie1}). 

Fig.\,\ref{fig:HistoLacc} shows the comparison between the $\log \lacc$ and $\mstar$ distributions of the \href{#label}{A17} Lupus, PENELLOPE, and ODYSSEUS samples. The $\lacc$ values were calculated as explained in Sect.\,\ref{sect:analysis}.
The \href{#label}{A17} Lupus sample is the only one covering the low-accretion regime down to $\log \lacc = -5.4$, while the ODYSSEUS and PENELLOPE samples are skewed toward higher accretion luminosities, with $\log \lacc > -4$ and $\log \lacc > -4.5$ respectively. The $\mstar$ histograms are very alike. We note that the PENELLOPE sample has more sources between 0.6 and 1.0\,$M_\odot$, compared to Lupus and ODYSSEUS samples. 
We highlight that the ODYSSEUS sample analysed here is a subset of the PENELLOPE catalogue.

\section{Analysis}
\label{sect:analysis}
A major goal of our analysis is to compare the $\lacc$ computed using the hydrogen X-Shooter (XS) slab modeling \citep{man13b} with the multicolumn accretion shock HST model \citep{pittman2022}, and with those from empirical relations. 
To do so we need to: 
\begin{itemize}
    \item have a set of $\lacc$ values from both the XS slab modeling (XS-fit) and from the HST multi-columns modeling (HST-fit); 
    \item derive $\lline$ for every accretion tracer of the PENELLOPE sample; 
    \item develop empirical relations using the HST-fit results and compare these with those from the \href{#label}{A17} sample; 
    \item compare $\lacc$ drawn from the $\lacc - \lline$ relations with $\lacc$ computed from XS- and HST-fit;  
    \item  finally investigate the $\lacc$ validity range of the empirical relations.
\end{itemize}

Since $A_V$ is a key parameter impacting the $\lacc$ measurement, a further goal is to discuss the possibility of computing $A_V$ independently, using empirical relations alone.
Lastly, since in our sample several Br series line are detected, we also investigate the possibility of developing new empirical relations in the NIR. 
The different modelling approaches to determine $\lacc$ and stellar parameters, are broadly explained in the PENELLOPE \citep[][Manara et al. in prep.]{manara2021} and ODYSSEUS \citep{pittman2022, pittman2025} papers.

\subsection{The X-Shooter slab-model}
\label{sect:XS-fit}

Stellar and accretion parameters were determined using the method originally described by \citet{man13a} and used in the general PENELLOPE papers by \citet[][]{manara2021}. This method has been recently further developed  by \citet{Claes2024}. 
Briefly, this approach involves dereddening the observed spectrum across a range of extinction values assuming the reddening law of \citet{car89} with $R_V = 3.1$, and fitting the data with a combination of a photospheric template spectrum and a hydrogen slab model. 
The slab model, with uniform gas density and temperature, is used to replicate the continuum excess emission observed in the spectrum due to accretion from the disk onto the forming-star. 

Photospheric templates, from \citet{man13b, man17b}, cover spectral types (SpT) from G- to late M-type. A more complete grid of templates is provided in \citet{Claes2024}.
The integrated flux of the best-fit slab model is used to estimate $\lacc$, i.e. the excess luminosity due to accretion. 
The SpT gives the $\teff$ based on the relation by \citet{luh03,kenyon-hartmann1995}. 
The stellar mass is then inferred by interpolating evolutionary tracks from \citet{bar15} or \citet{sie00}, depending on the mass. 

In this study we will focus extensively on two of the aformentioned parameters: $A_V$ and $\lacc$, as listed in Tab.\ref{tab:sample1}, adopted from Manara et al. (in prep.). The typical error on $A_V$ and $\lacc$ is 0.5\,mag and 0.25\,dex, respectively, as discussed in \href{#label}{A17}.

\subsection{The HST shock model}
\label{sect:HSTmodel}

\citet{ing13} introduced an alternative approach to measure the accretion luminosity of CTTSs using the accretion shock model of \citet{calvet1998}.
They used multiple accretion columns with varying energy fluxes to account for continuum excesses in the NUV through optical.  
Recent studies by the ODYSSEUS collaboration \citep{Espaillat2022, pittman2022} applied this method to the first HST observations of the ULLYSES program in Orion OB1b, and it has now been extended to all eight star forming regions in the ULLYSES-PENELLOPE sample by \citep{pittman2025}.
Building on the XS-fit parameters as first guess input parameters, \citet{pittman2022, pittman2025} derived the extinction and accretion parameters by minimizing the fit to the full 0.2-1.0\,$\mu$m continuum. 
We refer the reader to Pittman's works for a more detailed description of this method.

\begin{figure*}
    \centering
    \includegraphics[width=0.77\columnwidth]{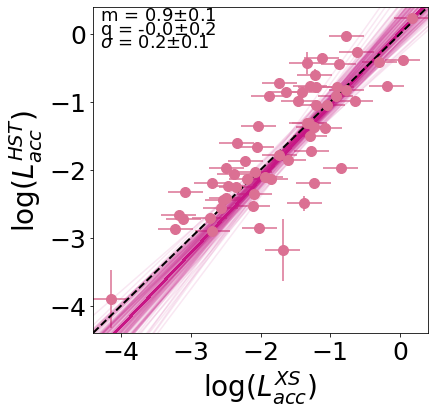}
    \includegraphics[width=0.8\columnwidth]{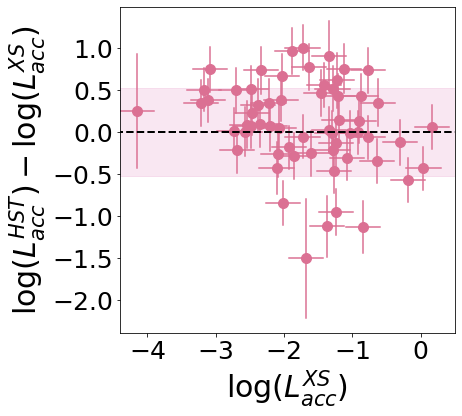}
    \caption{Comparison of $\lacc$ computed using the HST modeling with that from the XS-fit. {\it Left}: $\log \lacc^{HST}\,{\rm vs.}\, \log \lacc^{XS}$, the pink line show the best fit, which is linear within the error with its dispersion. {\it Right}: differential analysis, the pink region corresponds to the standard deviation of $\log \lacc^{HST}-\log \lacc^{XS}$, computed as the quadrature sum of the individual errors.}
    \label{fig:LaccMIADMslab}
\end{figure*}

\citet{pittman2025} results can be used to compare the XS- and the HST- fit results, as the PENELLOPE X-Shooter observations followed-up the ULLYSES-HST  observations quasi-simultaneously. More in detail, the median time difference between HST and XS data is 1.15\,days, while $\lacc$ varies on timescales of days, due to stellar rotation and intrinsic variability. 
Fig.\,\ref{fig:LaccMIADMslab} shows the comparison between the accretion luminosity computed using the HST- and the XS-fit. 
The left panel shows $\log \lacc^{HST}$ as a function of the $\log \lacc^{XS}$, where $\log \lacc^{HST}$ is measured from the accretion shock model \citep[see][]{pittman2025}. We performed a linear fit  using the hierarchical Bayesian method from \citet{kel07}, which accounts for errors on both axes, obtaining:

\begin{equation}
    \log \lacc^{HST} = (0.9 \pm 0.1) \log \lacc^{XS} + (0.0 \pm 0.2)
\end{equation}
with a standard deviation $\sigma = 0.2 \pm 0.1$ and a correlation factor of 0.9. 
Thus, the best-fit relation between $\log \lacc^{HST}$ and $\log \lacc^{XS}$ is consistent with a one-to-one correlation, indicating that the two estimates are {\it statistically comparable} on average. Indeed, for 41\% of this sample (26/61) the accretion luminosity computed with the HST- and the XS-fit is in agreement within the error. 
However, considering individual cases large residuals show-up. The HST-fit method is higher (lower) than the XS-fit method in 24(12) sources, i.e. 39\%(20\%) of the sample. 
The right panel in Fig.\,\ref{fig:LaccMIADMslab} shows the difference of the two accretion luminosity population as a function of the $\log \lacc^{XS}$. The standard deviation of $\log \lacc^{HST} - \log \lacc^{XS}$ is $\sigma=0.5$. While only 9 sources (15\%) fall completely outside the 1$\sigma$ (0.5) region (pink area), eleven additional sources have central values beyond the 1$\sigma$ range, but still fall within it when uncertainties are taken into account. 
The observed scatter ($\sigma \sim 0.5$\,dex) exceeds that expected from formal fitting uncertainties alone ($\sim 0.2-0.3$\,dex), suggesting either underestimated errors or the presence of intrinsic differences between the two methods. 
We check how the difference in time between HST and XS observations affected the agreement between the two $\lacc$ measurement (see Appendix\,\ref{app:time-variability}), without finding any significant correlation. Similarly, there is no trend between accretion or stellar parameters and the level of agreement between the two models. Individual discrepancies and a detailed analysis of the large spread in the two distributions will be the focus of a forthcoming paper.

\begin{figure}
    \centering
    \includegraphics[width=0.49\columnwidth]{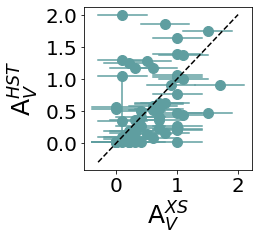}
    \includegraphics[width=0.49\columnwidth]{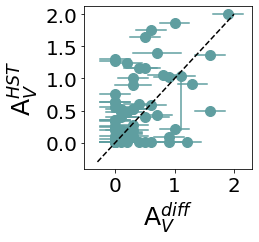}
    
    \caption{{\it Left}: comparison of $A_V$ computed using the HST- and XS- modelling. {\it Right}: comparison of $A_V$ using the HST-fit and the difference method of Sect.\,\ref{sect:Av}). The dashed line shows the one-to-one relation. The $\sigma$ in both cases is 0.4.}
    \label{fig:Av-miaCae}
\end{figure}

The left panel of Fig.\,\ref{fig:Av-miaCae} shows the comparison between the A$_V$ values obtained using the HST- and XS-fit. 
The $\sigma$ of the distribution is 0.4. 
The spread is the same for the plot in the right panel, showing the comparison between $A_V^{HST}$ and the extinction computed using empirical relations (see Sect.\,\ref{sect:Av}).

To investigate whether the scatter in $\lacc$ is driven by differences in $A_V$, we compared the $\log \lacc$ residuals and the $A_V$ residuals. We found a correlation (correlation factor 0.8) between the $\log \lacc$ and $A_V$ residuals. This correlation highlights the known degeneracy between $\lacc$ and $A_V$. We note that a more comprehensive analysis, where the XS spectra are fit fixing $A_V^{HST}$, is something that must be investigated and that we will perform in a forthcoming paper.

Overall, our analysis shows that $\lacc$ derived from HST- and XS-fit are statistically equivalent within the sample. Consequently, we expect empirical relations based on HST data to be statistically equivalent to those from X-shooter.

\subsection{Revisiting the $\lacc - \lline$ relations with HST data}
\label{sect:revised-rel}

Since HST and XS observations were quasi-simultaneous, we use the lines luminosity from the X-Shooter PENELLOPE data and the $\lacc$ values drawn from the HST modelling to revisit the $\lacc-\lline$ relationships and compare with the previous \href{#label}{A17} X-Shooter results. Note, however, that the HST data cover a smaller range in $\lacc$ with respect to the \href{#label}{A17} sample.

For the 61 objects observed by both XS and HST, we dereddened the observed PENELLOPE line fluxes (see Appendix\,\ref{sect:Lacc}) using extinction values derived from the HST-fit, denoted as $A_V^{HST}$, and we computed their luminosities: $L_i =4\pi d^2 F_i$, where $d$ is the distance to each source, computed from inverting the parallaxes
from Gaia EDR3 \citep{GaiaCollaboration2021}. 
For each line, i, a best linear fit $\log \lacc^{i} = a_i \log \lacc^{HST} + b_i$ was computed as explained in Sect\,\ref{sect:HSTmodel}. See Appendix\,\ref{app:MIADM} for the details.

Fig.\,\ref{fig:new-rel-check} shows that the slope and the intercept obtained using the PENELLOPE sample and the HST-fit are consistent within $<3\sigma$ of those in \href{#label}{A17}. 
The color code represents the wavelengths of a certain accretion tracer, as detailed in the colorbar.
Dark and light gray regions in the figure represent levels of agreement: dark for 1$\sigma$, light for 2$\sigma$. 
We plot accretion tracers whose slope or intercept is not in agreement within 1$\sigma$ as squares, and within 2$\sigma$ as triangles. 
We note that the slope and the intercept for the He\,I line at 471.31\,nm and for the O\,I at 844.64\,nm do not agree within 2$\sigma$ and 1$\sigma$, respectively, with the previous version. Indeed, these lines are not suggested for calculating $\lacc$ due to the large spread in the distribution and high number of upper limits (i.e., non-detection of these line, see Appendix\,E of \href{#label}{A17} and Figs.\,\ref{fig:new-rel-hydrogen} and \ref{fig:new-rel-other}). 
The Pa10 slope is in agreement with \href{#label}{A17} results, but its intercept slitghly differ. Also this line is not suggested to derive accretion luminosity by \href{#label}{A17}.
For the other accretion tracers, the $\lacc-\lline$ relations drawn from the HST-fit are well consistent with those previously derived from the Lupus sample using the XS-fit by \href{#label}{A17}. However, we note that both a and b are systematically higher, although consistent, when computed using the HST-sample than in \href{#label}{A17}. We speculate that this could be related to the fact that the ODYSSEUS sample does not include as many low-mass stars as the Lupus sample (see Fig.,\ref{fig:HistoLacc}). 
To check this, we fitted again \href{#label}{A17} $\log \lacc - \log L^{\rm line}$, limiting the $\lacc$ range to the one of our sample ($-3.90 \leq \log \lacc \leq 0.23$). Using all H\,I emission lines, we find that limiting the fit results in both the slope and intercept being lower than the original \href{#label}{A17} parameters. More specifically, the mean values of the slope and intercept are lower by 0.1 and 0.6, respectively. 
We also note that the trend for which higher values of the slope and intercept have larger differences is confirmed even with the limited \href{#label}{A17} sample. This suggests that the systematic trend is not simply a product of the sample population.

\begin{figure}
    \centering
    \includegraphics[width=0.65\columnwidth]{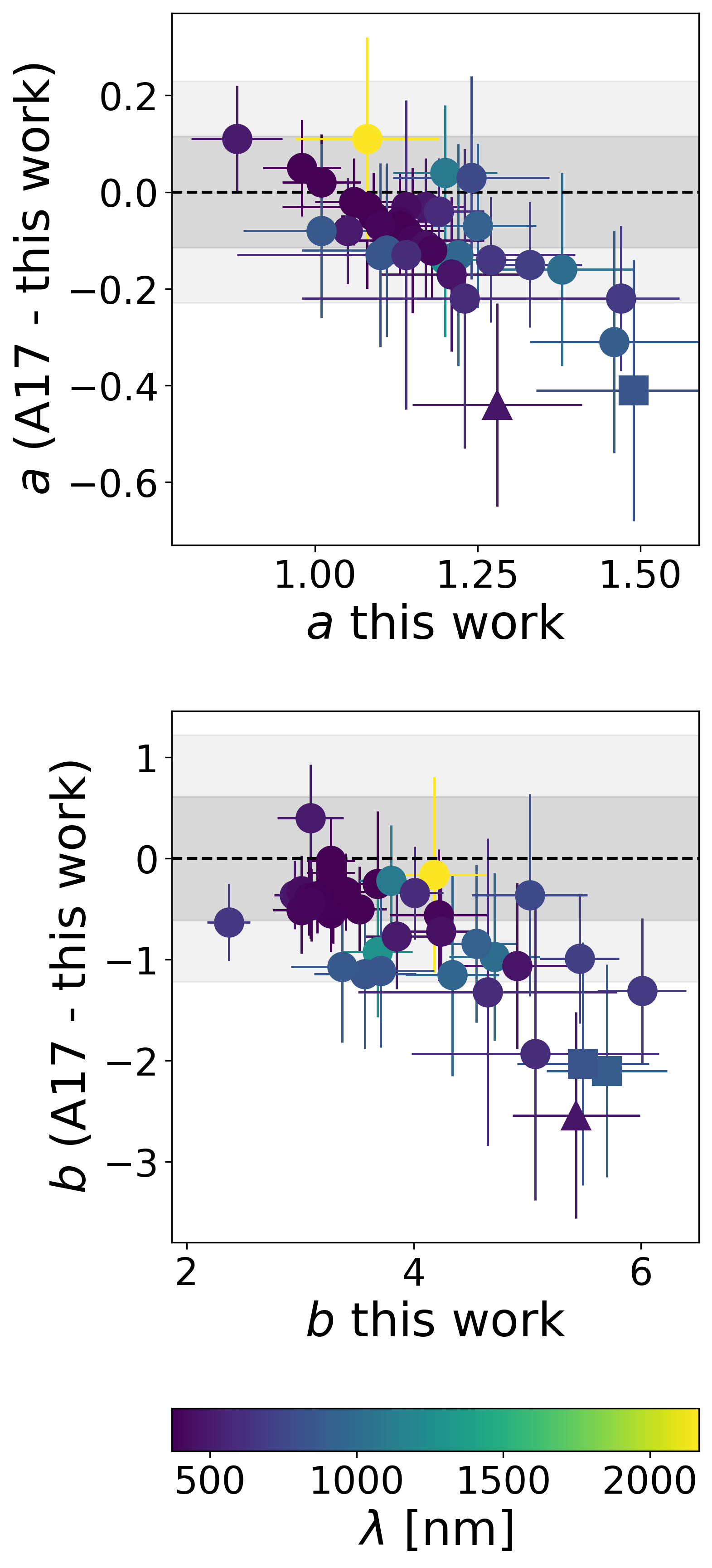} 
    \caption{Comparison of the slopes (top) and intercepts (bottom) of the $\log \lacc - \log L_{\rm lines}$ empirical relations derived in this work with those from \citet{alc17}. The colorbar indicates the corresponding wavelength of each accretion tracer, ranging from UVB (blue) to NIR (yellow). Dark and light gray regions represent agreement within 1$\sigma$ and 2$\sigma$, respectively.
    Accretion tracers not in agreement within 1$\sigma$ (2$\sigma$) within the error are plotted as squares (triangles).
    }
    \label{fig:new-rel-check}
\end{figure}

\subsection{$\lacc$ from lines vs. $\lacc$ from modeling} \label{sect:Llines-XS}

We compared $\lacc$ values derived from direct UV-excess modeling (XS- and HST-fit) with those computed indirectly using empirical relations from \href{#label}{A17} and this work, respectively (see Sect.\,\ref{sect:revised-rel}). 
To ensure consistency, observed fluxes of each accretion tracer were respectively de-reddened using $A_V$ from both XS-fit and HST-fit (Tab.\,\ref{tab:sample1}, applying the \citet{car89} extinction law with $R_V = 3.1$. Line luminosities were then calculated using distances in Tab.\,\ref{tab:sample1}.

$\lacc^{XS}$ and $\lacc^{HST}$ for our sample were computed using empirical relations from \href{#label}{A17} and our newly derived relations, respectively (Tab.\,\ref{tab:sample1} for XS; Table.\,\ref{tab:Lacc-Av-varie1} for HST). The tabulated $\log \lacc^{\rm lines}$ values are the average of $\log \lacc^{i}$ computed from the several tracers, with errors as the standard deviation.

\begin{figure*}
    \centering
\includegraphics[width=0.8\columnwidth]{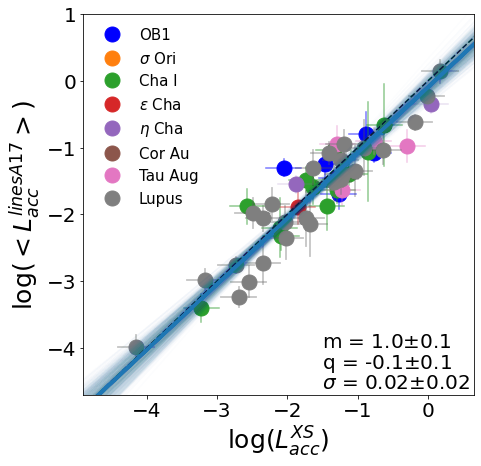}
\includegraphics[width=0.8\columnwidth]{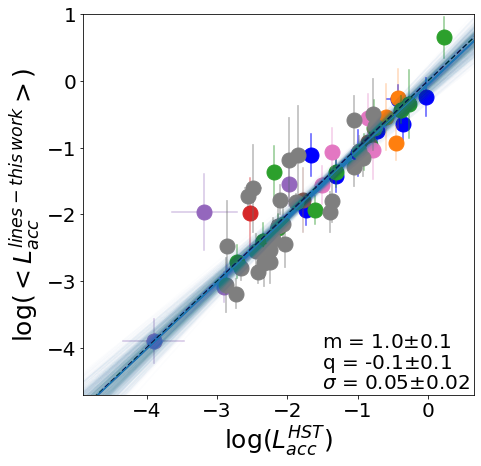}

    \caption{$\log(<\lacc^{lines}$>) vs. $\log \lacc^{\rm modeling}$. Colors of the points represent different star-forming regions as described in the legend. 
    The blue line corresponds to our best fit. The dashed black line shows the linear trend as a reference. 
    {\it Left}: $\log <\lacc>^{lines}$ is computed as the average from the $\lacc$'s from  many lines and using empirical relations by  \citet{alc17} and $\log \lacc^{XS}$ is obtained using the XS-fit method. {\it Right}: $\log <\lacc>^{lines}$ is computed as the average from the $\lacc$'s from  many lines and using empirical relations obtained in this work and $\log \lacc^{HST}$ is obtained using the HST-fit method.}
    \label{fig:logLaccComparison}
\end{figure*}

Fig.\,\ref{fig:logLaccComparison} illustrates the $\log \lacc$ vs. $\log \lacc^{lines}$ comparison for XS-fit and \href{#label}{A17} relations (left panel), and for HST-fit and our relations (right panel). Colors denote star-forming regions. Linear regression fits for both cases, performed as described in Sect.\,\ref{sect:HSTmodel}, show general agreement between UV-excess modeled and empirically derived accretion luminosities.

Specifically, the fit for the left panel is:
\begin{equation}
\log \lacc^{lines} = (1.0 \pm 0.1) \log \lacc^{XS} + (-0.1 \pm 0.1)
\label{eq:fit-lacc-no-low-acc}
\end{equation} 
with a standard deviation of $0.02 \pm 0.02$ and a correlation parameter of 1.0. 
We find $\lacc$ agreement within errors for 85\% (56/64) of the sources. Of the remainder, 11\% (7/64) show higher $\lacc$ up to 0.19\,dex via XS-fit, and 4\% (3/64) show lower up to 0.37\,dex. The $\log \lacc$ range differs slightly: $-4.49$ to 0.17 using the XS-fit, vs. $-4.20$ to $-0.39$ using \href{#label}{A17} empirical relations.

Similarly, the fit of the right panel is 
\begin{equation}
\log \lacc^{lines} = (1.0 \pm 0.1) \log \lacc^{XS} + (-0.1 \pm 0.1)
\label{eq:fit-lacc-no-low-acc-HST}
\end{equation} 
with a standard deviation of $0.05\pm 0.02$. 66\% (40/61) of the sample shows $\lacc$ agreement within errors. $\lacc$ is higher up to 0.26\,dex for 18\% (11/61) with HST-fit and lower up to 0.65\,dex for 16\% (10/61). The $\log \lacc$ range is $-4.15$ to 0.23 when computed using the HST-fit model , vs. $-4.15$ to $0.65$ when computed with corresponding empirical relations.

Using a number of tracers distributed in a wide spectral range improves $\lacc$ estimates \citep[e.g.,][]{rig12, alc19, fio22dqtau}. Thus, we investigated the impact of the $\brg$ line, our longest-wavelength tracer, which is not always detected (Tab.\,\ref{tab:sample1}). The $\brg$ line is also known to trace highly accreting CTTS. 
Repeating the analysis for the subset of sources with detected $\brg$ only, hereafter referred to as the $\brg$-sample, we found no statistical difference in $\lacc$ results, with linear regression fits yielding the same best-fit parameters as Eqs.\,\ref{eq:fit-lacc-no-low-acc}\,and\,\ref{eq:fit-lacc-no-low-acc-HST}. 
Thus, we conclude that in the the case of our sample, the $\lacc$ estimate does not depends on the $\brg$ detection.

\subsection{Continuum excess versus emission in lines}
\begin{figure}
    \centering
\includegraphics[width=0.48\columnwidth]{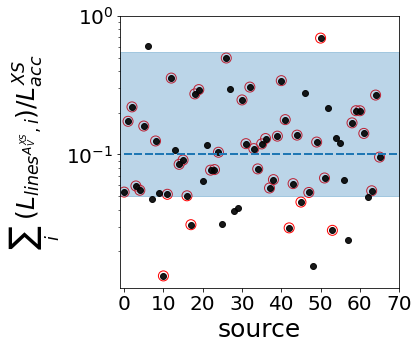}
\includegraphics[width=0.45\columnwidth]{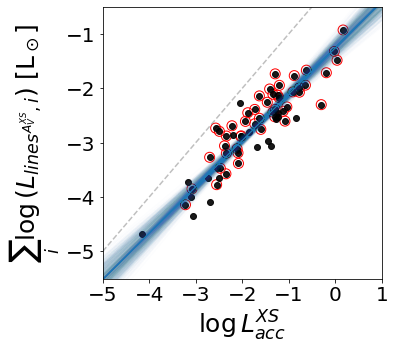}

\includegraphics[width=0.48\columnwidth]{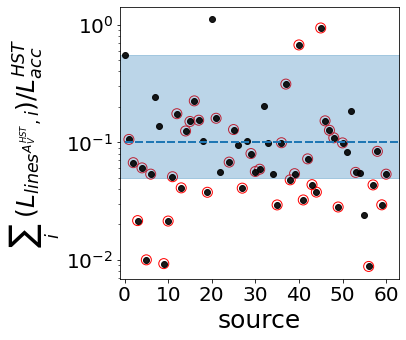}
\includegraphics[width=0.45\columnwidth]{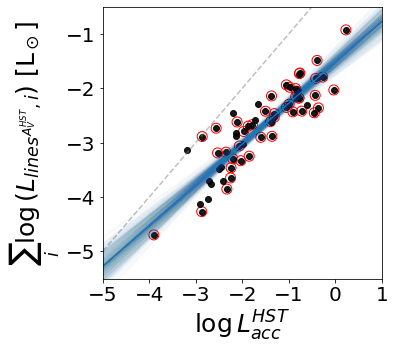}
    
    \caption{
    {\it Left panels}: Fraction of the total luminosity in lines relative to the accretion luminosity, i.e. the continuum excess luminosity. The light blue region corresponds to the range of values found in \citet{alc14}.
    {\it Right panels}: Total luminosity from the lines as a function of the $\log \lacc$ as indicated in the labels. The blue line in each panel corresponds to our best fit, while the light-blue lines give an indication of the spread of the fit (see Sect.\,\ref{sect:Llines-XS}). The dashed gray lines show the one-to-one relation.
    Red circled black dots correspond to sources where the $\brg$ line is detected.  
    }
    \label{fig:ratioLlinesLacc}
\end{figure}

It is interesting to  quantify how much energy per unit time is emitted in the continuum relative to that in lines, $\Sigma_i L_{\rm lines, i}$, and whether the $\Sigma_i L_{\rm lines, i}$ to $\lacc$ ratio depends on the stellar mass.

We computed the fraction of the total luminosity from the lines, $\Sigma_i L_{\rm lines, i}$, with respect to the accretion luminosity from the continuum excess emission $\lacc$. 
For every CTTS, we computed $\Sigma_i L_{\rm lines, i}$ as the sum of the luminosity of all the emission lines considered in this work. 
Fig.\,\ref{fig:ratioLlinesLacc} shows the results of our analysis. The top panels refer to $\log \lacc^{XS}$ and $\lline$ dereddened using $A_V^{XS}$; while the bottom panels refer to $\log \lacc^{HST}$ and $\lline$ dereddened using $A_V^{HST}$. 
The blue region on the left panels corresponds to the range of values found for the \href{#label}{A17} Lupus sample, i.e. $0.05 < \Sigma_i L_{\rm lines, i} / \lacc < 0.55$. The range of ratios in our sample is broader than in \href{#label}{A17}, ranging from 0.01 to 0.69 for $\Sigma_i L_{\rm lines^{A17}, i}/\lacc^{XS}$, and from 0.01 to 1.11 for $\Sigma_i L_{\rm lines^{TW}, i}/\lacc^{HST}$. 
The panels on the right show $\Sigma_i \, L_{\rm lines, i}$ as a function of accretion luminosity in a log scale. 
A linear regression ﬁt in this parameter space yields  
\begin{equation}
    \log (\Sigma_i \, L_{\rm lines,i} / \lsun) = (0.9 \pm 0.1) \,\times \, \log (\, \lacc / \lsun) - (1.3 \pm 0.1)
\label{eq:energy-lossXS}
\end{equation} 
with a correlation factor of 1.0 and $\sigma = 0.14 \pm 0.10$ for the XS-fit (top right); and 
\begin{equation}
    \log (\, \Sigma_i \, L_{\rm lines,i} / \lsun) = (0.8 \pm 0.1) \,\times \, \log (\, \lacc / \lsun) - (1.5 \pm 0.1)
\label{eq:energy-lossHST}
\end{equation} with a correlation factor of 0.9 and $\sigma = 0.36 \pm 0.16$ for the HST-fit (bottom right).
This result confirms the correlation between these two quantities, and is consistent with the best fit presented in \href{#label}{A17} sample.
The fit performed for the Br$\gamma$-sample (black dots surrounded by red circles) provides the same results. 
The dashed lines on the right panels of Fig.\,\ref{fig:ratioLlinesLacc} show the one-to-one relation. It is interesting to note that the total energy emitted in lines is more similar to that of the continuum at low $\lacc$ values, i.e. tendentially for objects with low masses. However, by plotting $\log (\, \Sigma_i \, L_{\rm lines,i} / \lsun)$ as a function of the stellar mass, there is no evident correlation.

\subsection{Extinction estimates with empirical relations}
\label{sect:Av}

Previous sections show that empirical relations are a practical tool to compute $\lacc$. However, these relations can only be applied to the extinction-corrected fluxes. Thus, computing $A_V$ is a key step in measuring $\lacc$ with empirical relations. 
Our dataset allows us to estimate $A_V$ with a direct modeling of the UV-excess, either the XS- or the HST-fit. 
However, this is not possible for very noisy UVB data or when the bluest part of the UVB spectrum ($\lambda \lesssim 370$\, nm) is not available. 
In these cases, the empirical relations can be used to estimate the extinction, provided that a good number of emission line fluxes in the widest possible wavelength range are available. 

We adopt three methodologies, independent of the UV-excess modelling results, to estimate $A_V$. 
These methodologies  are based on the assumption that the accretion luminosity, $L_{{\rm acc},\,i}$, as derived from different accretion tracers, must be the same for all the lines. 
Thus, in a plot of  $\log L_{{\rm acc},\,i}$ as a function of $\log \lambda$, the $\log L_{{\rm acc},\,i}$ differences can be minimized with the three adopted methods, yielding a flat $\log \lacc {\rm \,vs.\,} \log \lambda$ distribution and the $A_V$ that minimizes the $\log L_{{\rm acc},\,i}$ differences.
We used the XS sample only, because we demonstrated that results between XS-fit and HST-fit and their corresponding empirical relations are in agreement and the XS sample contains more sources that allows a better statistical analysis.

We proceeded as follows: we artificially corrected the observed line fluxes for extinction using a grid of $A_V$ values ranging from 0 to 3 mag, in steps of 0.05\,mag. This produces a corresponding grid of line luminosities, which we then used to derive a grid of accretion luminosities ($\lacc$) through the empirical relations.

\begin{figure*}
    \centering
    \includegraphics[width=.25\textwidth]{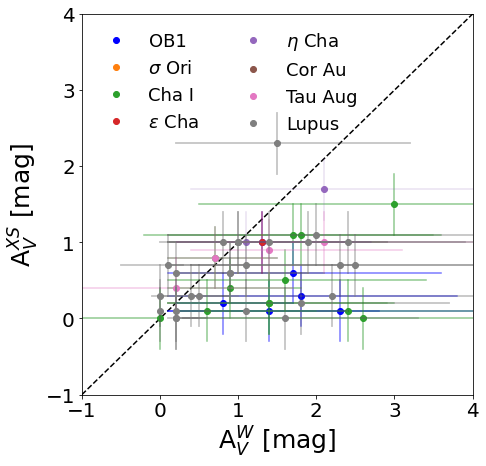}
    \includegraphics[width=.25\textwidth]{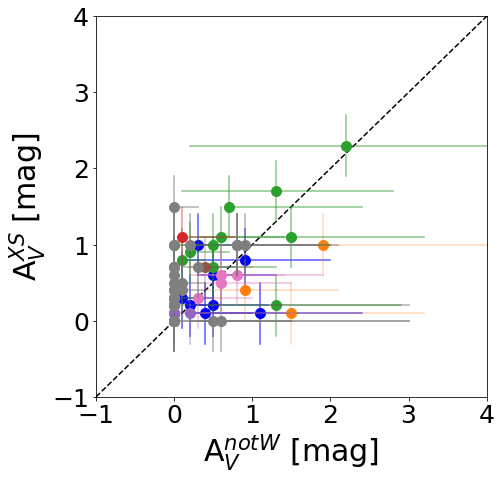}
    \includegraphics[width=.25\textwidth]{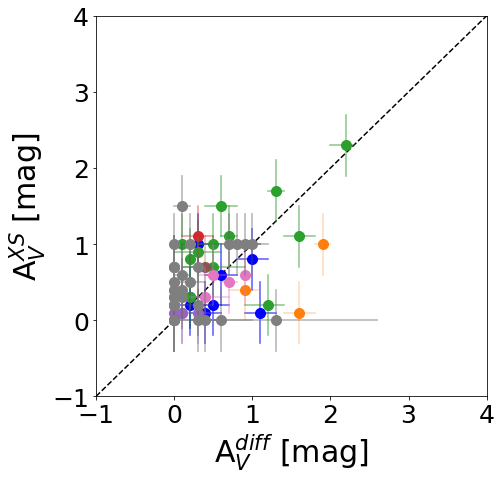}
    
    \includegraphics[width=0.25\textwidth]{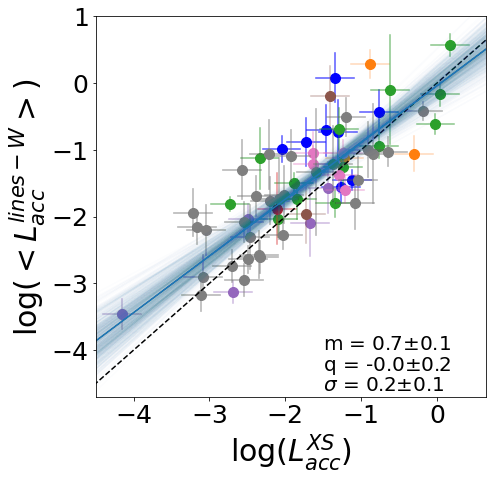}
    \includegraphics[width=0.25\textwidth]{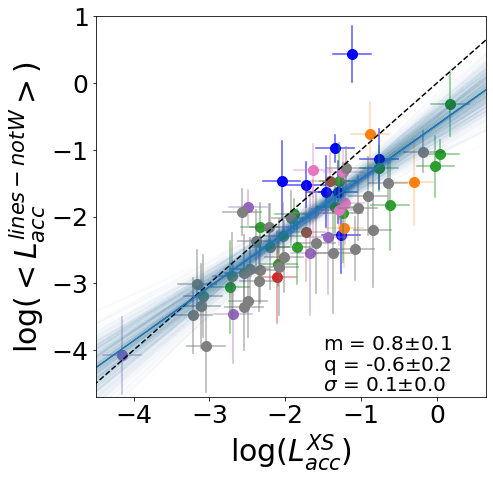}
    \includegraphics[width=0.25\textwidth]{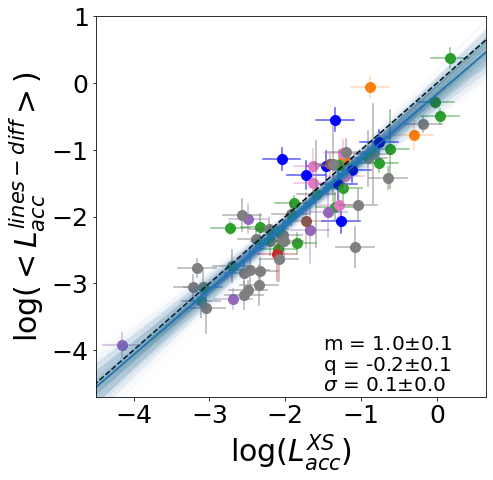}

    \caption{Top panels: Comparison between $A_V^{XS}$ and the extinction drawn from the coefficient method without weights (left), the method weighting for the error (central), and the difference method. Lower panels: Comparison between average $\lacc$ computed with the three methods and $\lacc^{XS}$. The blue line corresponds to the best ﬁt.}
    \label{fig:AvComparison}
\end{figure*}

The three adopted methods to minimize the $\lacc^{\rm lines}$ differences as a function of wavelength are:
\begin{itemize}
    \item Weighted coefficient method, $A_V^W$: selecting the $A_V$ that minimizes the slope of the $\lacc$ vs. $\log \lambda$ relation, using $1/\Delta \log \lacc$ as the weight of each point.
    \item Unweighted coefficient method, $A_V^{\rm notW}$: same as the previous method but without weighting for uncertainties.
    \item Difference method, $A_V^{\rm diff}$: selecting the $A_V$ that minimizes the dispersion among $\log \lacc$ values (including their error) derived from different accretion tracers. 
\end{itemize}
The results of these methods are graphically shown in the top panels of Fig.\,\ref{fig:AvComparison}: the left and central panels display the coefficient methods with and without weighting, respectively, while the right panels present the difference minimization method. The extinction values derived with the three approaches differ (see Tab.\,\ref{tab:Lacc-Av-varie1}).
To evaluate the accuracy of these methods, we compare the derived $A_V$ values with $A_V^{\rm XS}$, using it as a benchmark. The distribution of $A_V^{\rm XS}$ vs. $A_V^{\rm lines}$ is broad in all three cases and does not show a clear linear trend.
More in detail, the root mean square (rms) deviation between $A_V^{\rm XS}$ and $A_V$ is 0.8, 0.5 and 0.4 for the weighted, unweighted and difference methods, respectively.
As anticipated in Sect.\,\ref{sect:HSTmodel}, the result is the same between $A_V^{\rm HST}$ and $A_V^{\rm diff}$ (see the right panel of Fig.\,\ref{fig:Av-miaCae}). 

We computed the line luminosity and accretion luminosity as described in Sect.\,\ref{sect:Llines-XS} using the extinction values provided by the above three minimizing methods. 
The bottom panels of Fig.\,\ref{fig:AvComparison} show the linear regression performed as described in Sect.\,\ref{sect:HSTmodel} between the accretion luminosity obtained from the XS-fit and the accretion luminosity obtained from accretion tracers luminosities calculated with the tree different $A_V$ values. 
The relations found for the three methods are:
\begin{equation}
\log \lacc^{lines-W} = (0.8 \pm 0.1) \log \lacc^{XS} + (-0.0 \pm 0.2) \\
\label{eq:Lacc_linesAv-coeffW}\\
\end{equation}
\begin{equation}
\log \lacc^{lines-notW} = (0.8 \pm 0.1) \log \lacc^{XS} + (-0.6 \pm 0.2) \\
\label{eq:Lacc_linesAv-coeffnotW}\\
\end{equation}
\begin{equation}
\log \lacc^{lines-diff} = (1.0 \pm 0.1) \log \lacc^{XS} + (-0.2 \pm 0.1) 
\label{eq:Lacc_linesAv}\\
\end{equation}
with a standard deviation of $0.2$, $0.1$, and $0.1$, respectively, and a correlation factor of 0.9 in all three cases.  
Only the best fit obtained with the difference-method (Eq.\,\ref{eq:Lacc_linesAv}, low right panel in Fig.\,\ref{fig:AvComparison}) is consistent with the one-to-one line within the error. 
We conclude that the difference-method yields an $A_V$ value which best reproduces the accretion luminosity distribution from the XS modeling. 

\begin{figure}
    \centering
    \includegraphics[width=0.76\columnwidth]{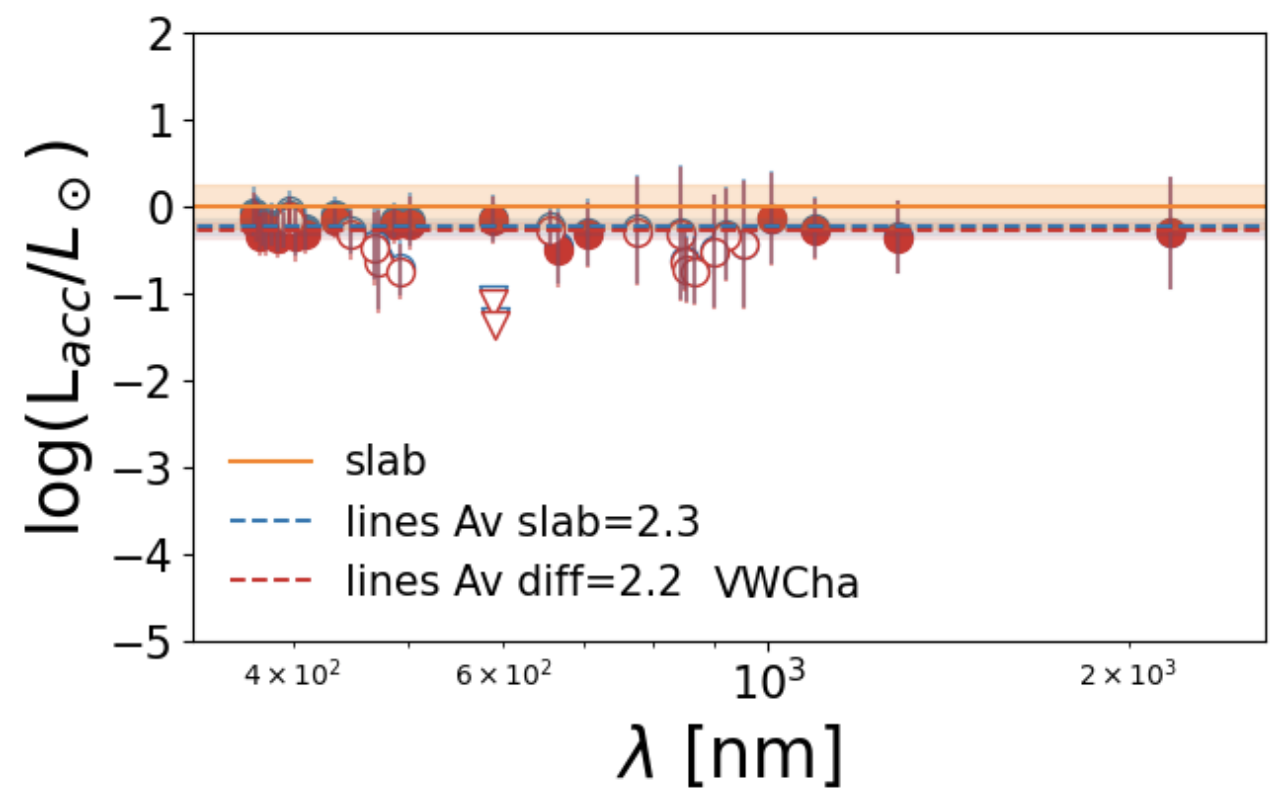}
    
\includegraphics[width=0.75\columnwidth]{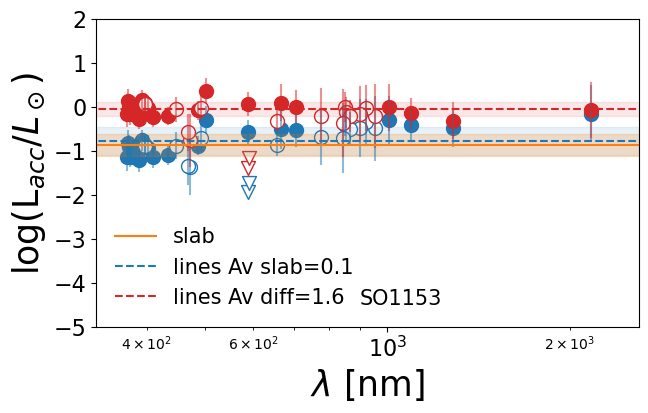}    
\caption{$\log \lacc$ as a function of the wavelength for two PENELLOPE CTTSs. Filled (empty) circles are $\log \lacc$ values computed using empirical relations, corresponding to suggested (non suggested) lines by \citet{alc14}, while triangles represent upper limits. 
In red and blue results obtained by dereddening the flux with $A_V^{diff}$ and $A_V^{XS}$ are shown, respectively. 
Horizontal red and blue dashed lines and regions correspond to mean value and standard deviation of $\lacc$ using $A_V^{diff}$ and $A_V^{XS}$, respectively. 
The orange solid line shows $\lacc^{XS}$ with its error bar (orange horizontal region). 
{\it Top}: Plot for the multiple system VW\,Cha, where the extinction and accretion luminosity estimated with the difference method and the XS-fit are in agreement and the $\log \lacc$ distributions are flat. 
{\it Bottom}: SO1152 young star, where $A_V^{diff}$ and $A_V^{XS}$ differ by 1.5\,mag and the distribution of $\log \lacc$ obtained by dereddening the fluxes with $A_V^{diff}$ is flatter than the one obtained using $A_{V}^{XS}$. The latter increases with the wavelength.
}
    \label{fig:RMplot-ex}
\end{figure}

Appendix\,\ref{app:logLacclogLambda} shows the $\log \lacc \,{\rm vs.}\, \log \lambda$ plots for the PENELLOPE sample. Those plots show the comparison of the $\log \lacc$  when using the $A_V^{XS}$  and $A_V^{diff}$ values to correct the line fluxes for extinction.
Fig.\,\ref{fig:RMplot-ex} shows two examples of these plots. The top panel shows the case of the triples system VW\,Cha, while the bottom panel shows the case of SO1153 young star. For VW\,Cha, the difference method and the XS-fit provide consistent results of $A_V$ and $\lacc$. Thus, the distribution of $\log \lacc \,{\rm vs.}\, \lambda$ is flat in both the cases, as in the 80\% of the sources in our sample (see Appendix\,\ref{app:logLacclogLambda}). 
Differently, applying the difference method to SO1153 YSO, we find $A_V^{diff} \neq A_V^{XS}$, this happens in 20\% of the sources in our sample. In this case, the  $\log \lacc \,{\rm vs.}\, \lambda$ distribution appears flatter when using one of the two methods (XS-fit or difference method), compared to the other. We discuss this further in Sect.\,\ref{sect:discussion}.

We note that in some sources the $\brg$ line provides systematically lower (0.5\,dex on average) $\lacc$ values with respect to the average drawn from the many other lines.
This could be attributed to photospheric line contamination, especially for sources with weak $\brg$ emission. This aligns with the observation that consistent results are obtained for objects displaying strong $\brg$ emission in our sample. Furthermore, uncertain K-band flux calibration is a non-negligible potential source of this discrepancy.

To identify the shortest wavelength range that provides $A_V$ values consistent with $A_V^{diff}$, computed using all accretion tracers from the UVB to the NIR, we performed several checks. We  analysed if and how much the $A_V^{diff}$ estimate changes considering only UVB tracers (emission lines up to 600\,nm), only VIS tracers (emission lines with $600\,{\rm nm} < \lambda < 900$\,nm), only NIR tracers ($\lambda > 900$\,nm), and combination of these bands. 
For this analysis we selected only the $\brg$ sample. We found that the difference method fails when using only one band. Contrary, extinction estimates using UVB+VIS bands yield results consistent with the $A_V^{\mathrm{diff}}$.
When considering UVB+NIR or VIS+NIR tracers, the $A_V^{\mathrm{diff}}$ is underestimated in 80\% of the cases. We believe this is related to the fact that, when detected, the $\brg$ line often yields $\log \lacc$ values lower than those derived from other lines (see Figs. in Appendix,\ref{app:logLacclogLambda}).

\begin{table*}
\centering
\caption{\label{tab:Lacc-Av-varie1} Stellar extinction and accretion luminosities.} \resizebox{\textwidth}{!}{%
\begin{tabular}{llcrcrcrccr}
\hline
\hline
Region & Source Name                            & $A_V^W$ & $\log \lacc^{lines-W}$ & $A_V^{notW}$ &  $\log \lacc^{lines-notW}$ & $A_V^{diff}$ & $\log \lacc^{lines-diff}$ & $ \log \lacc^{lines,TW}$ & $\Delta$T\\
 &                             & mag & L$_\odot$ & mag &  L$_\odot$ & mag & L$_\odot$ & L$_\odot$ & day\\
\hline		
\hline																							
OB1			& CVSO58    		&	$0.7_{-0.6}^{+0.8} $ &   $-1.45 \pm 0.15$ &  $0.9_{-0.8}^{+1.1}$ &  $-1.45 \pm 0.18$ & $1.0_{-0.2}^{+0.2}$ & $-1.30\pm0.18$ & $-0.65\pm0.23$ & 0.26\\							
OB1			& CVSO90    		&	$2.4_{-2.2}^{+2.6} $ &   $ 0.08 \pm 0.37$ &  $1.1_{-0.9}^{+1.3}$ &  $ 0.08 \pm 0.17$ & $1.1_{-0.2}^{+0.2}$ & $-0.55\pm0.17$ & $-0.27\pm0.27$ & 1.57\\     				
OB1			& CVSO104   		&	$1.4_{-1.3}^{+1.6} $ &   $-0.89 \pm 0.36$ &  $0.2_{-0.0}^{+0.4}$ &  $-0.89 \pm 0.21$ & $0.2_{-0.2}^{+0.2}$ & $-1.38\pm0.21$ & $-0.75\pm0.26$ & 0.10\\     				
...			& ...   		&	... & ... & ... & ... & ... & ... & ...  & ... \\     				
...			& ...   		&	... & ... & ... & ... & ... & ... & ...  & ...\\     				

\hline 
\hline
\end{tabular}
  }
\begin{quotation}
  \textbf{Notes.} $A_V$ and $\lacc$ are computed using the three minimizing methodologies (see Sect.\,\ref{sect:Av}) and the HST modelling (see Sect.\,\ref{sect:revised-rel}). The last column show the difference in time between XS and HST observations. Only the first entries are shown; a complete version of this table is available at the CDS. 
  \end{quotation} 
 \end{table*}

\subsection{Brackett series} 

Similarly to Balmer and Paschen series, the Brackett, Pfund, and Humphreys series lines are believed to trace the accretion process. These lines link the  NIR with the MIR and FIR regions and may, in principle, serve as calibrators for the JWST data \citep{Salyk2013, Rigliaco2015, Rogers2024, Tofflemire2025}. 

The PENELLOPE sample contains objects where several lines of the Brackett series were detected, differently to the \href{#label}{A17} Lupus sample where only two objects were found to display Brackett series  lines.
In particular, we investigated the correlation between $\lacc^{XS}$ and $\lline$ among Brackett lines higher than $\brg$ (Br7), namely, lines from Br8 to Br21.

We measured the line fluxes of these lines as described in App.\,\ref{sect:Lacc} and computed line luminosities in the same way as for the other permitted lines. 
We then plot the result of the lines luminosity with the corresponding accretion luminosity, $\lacc^{XS}$ or $\lacc^{HST}$. 
Figs.\,\ref{fig:lacc-lline-br} and \ref{fig:lacc-lline-br_MIADM} in Appendix\,\ref{app:Br} show the correlation between $\log \lacc$ and $\log L_{\rm Br\,line}$ for a few of these lines. 
We can appreciate a qualitative correlation, which suggests that these lines trace, indeed the accretion process, but the number of non-detections (upper limits plotted as triangles) is generally high and, in most cases, higher than the detections, making difficult to draw definitive conclusions.

We decided to fit such correlation only for diagnostics detected in at least 10 sources, i.e., Br10, Br11, and Br13. 
The fit has been done using the hierarchical bayesian method from \citet{kel07}, as described in Sect.\,\ref{sect:HSTmodel}. 
The best fit of these lines is shown in Tab.\,\ref{tab:br-series}. We note that the correlation factor is between 0.6 and 0.8, suggesting moderate correlation between these Br-series lines and the accretion luminosity. 
The slope and intercept of the Br13 is in agreement when computed using the PENELLOPE and the ODYSSEUS samples. In contrast, the Br11 and Br10 relation changes significantly when using the two samples. This suggests that, considering the sources in our sample, the Br11 and Br10 relations are not well constrained.
In general, we stress that the fit we performed suffers from uncertainties due to the low statistics (N$_{\rm points}$ = 11 or 12 for the Br13 and Br11 lines), the possible contamination of photospheric lines, and more importantly the modest range amplitude in line luminosity, hence, accretion luminosity.

\section{Discussion}
\label{sect:discussion}

\subsection{The Empirical Relations for CTTS}
\label{sect:emp-rel}

Fit results shown in Eqs.\,\ref{eq:fit-lacc-no-low-acc} and \ref{eq:fit-lacc-no-low-acc-HST} indicate that the relations between $\log \lacc$ modeled from UV-excess and the corresponding $\log \lacc^{lines}$ is linear, suggesting that empirical relations can generally be used to estimate $\lacc$. 
Remarkably, our sample is composed of CTTSs from various star-forming regions, including OB1 and $\sigma$Ori. 
These regions host both a low-mass population of CTTS and massive O- and B-type stars, which emit intense UV radiation capable of inducing external photoevaporation in CTTS disks \citep{Rigliaco2009, Mauco2023, Mauco2025}. 
We also stress that the  recent work by \citet{Halstead-Willett2025}, based on a Bayesian approach to derive $\lacc$, demonstrates that the $\log \lacc - \log \lline$ relations of objects in more distant regions like Serpens, IC5146, NGC7000, etc, are in agreement with the previous results by \href{#label}{A17}. All this shows that the $\log \lacc - \log \lline$ empirical relations can be applied in a variety of star forming environments (see Fig.\,\ref{fig:logLaccComparison}).

It is, however, important to bear in mind that not all accretion tracers correlating with $\lacc$ are equally reliable for constraining $\lacc$, due to poor statistics, blending of lines, or contributions from mechanisms like winds, as observed for H$\alpha$ or the He\,I line at 471.31\,nm. 
Likewise, H7, HeI\,FeI, Ca\,II(H), and OI at 777.30\,nm are blended with other species and should also be used with caution in computing $\lacc$. 
For the remaining accretion tracers, we examined the correlation factor and spread of our empirical relations, with a higher correlation factor and a lower spread indicating a more reliable relation to estimate $\lacc$. 
Tab.\,\ref{tab:new-coeff} shows that all tracers have correlation factors above between 0.8 and 1.0. 
However, Fig.\,\ref{fig:new-rel-check} shows that the empirical relations revisited with the HST-ULLYSES data, which provide more UV information than X-Shooter, are in agreement with previous versions for most tracers. Contrary, the He\,I at 471.31\,nm, the O\,I at 844.64\,nm, and the Pa10 are not in agreement within 2$\sigma$ with the \href{#label}{A17} relation coefficients. 
These lines, when used alone, could provide uncertain values of $\lacc$. 

We also emphasize that different lines form in distinct regions within the accretion columns \citep[see for example the review by][]{har16}. Narrow components, such as Paschen series lines, generally form in the postshock region near the stellar surface, whereas broad components, such as Balmer series lines, usually form in the pre-shock region. 
Consequently, different lines may suit specific regions of interest, but characterizing the overall accretion luminosity demands incorporating the highest possible number of lines tracing both pre-shock and post-shock regions.

\subsection{The Brackett series}
\label{sect:discussion-brackett}

We also provide Br-series relations for the Br10, Br11, and Br13 lines. 
The coefficients of the Br13 line obtained using the XS- and the HST-fit are in agreement, while the coefficients of the Br10 and Br11 lines are not. This might be due to the limited line luminosity range of the Br10 and Br11 ($-5.2 < \log L_{Br10} < -3.5$, $-5.8 < \log L_{Br11}$), compared to the wider luminosity range of the Br13 ($-7 < \log L_{Br10} < -3.8$). 
Further investigation of other sources with the same techniques will improve the statistics, providing stronger results.

\begin{figure*}
    \centering
    \includegraphics[width=0.9\textwidth]{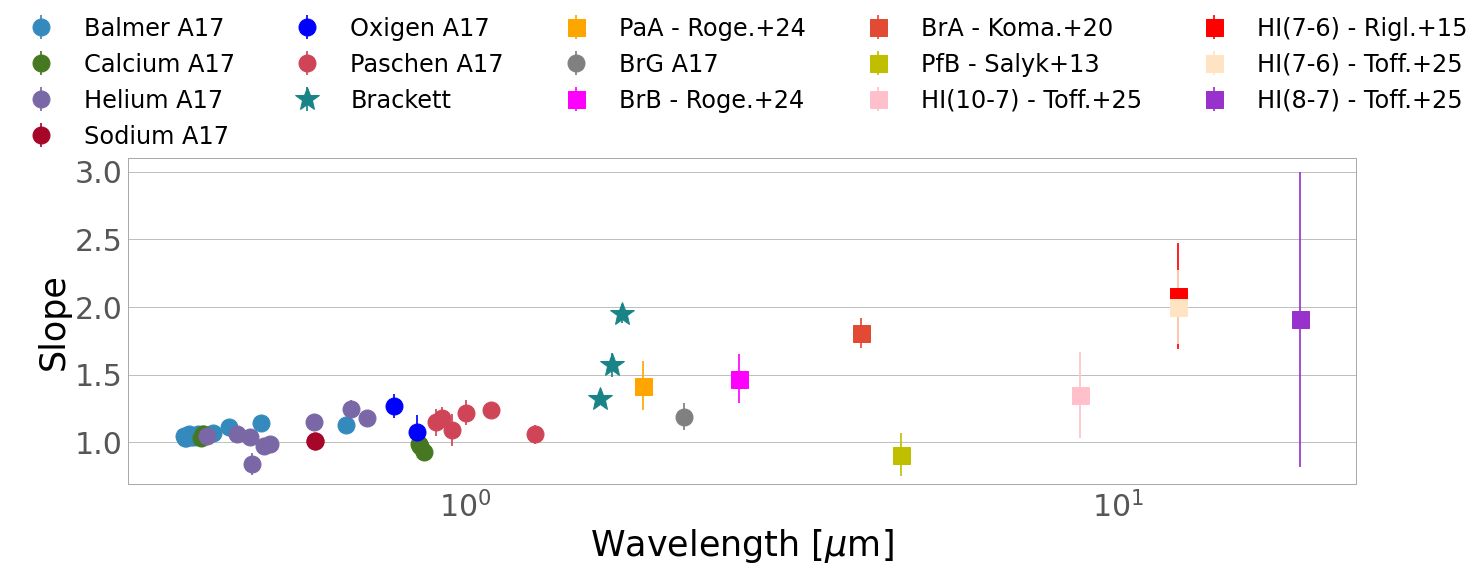}
    \caption{Slopes of empirical relations of \citet{alc17} sample (filled circles), this work (filled stars), and from the literature \citep[][filled squares]{Salyk2013, Rigliaco2015, Komarova2020, Rogers2024, Tofflemire2025}, as a function of the wavelength. Error bars are smaller than the symbols when not visible.}
    \label{fig:slope}
\end{figure*}

The tentative slopes of the Br-series that we computed with the PENELLOPE sample are between 1.3 and 1.0, steeper than the optical and NIR slopes, ranging between 0.8 and 1.3 (see Tab.\,\ref{tab:new-coeff}).
Fig.\,\ref{fig:slope} shows the correlation slopes as a function of the wavelength. Different colors trace different elements and specific HI lines as described in the legend. 
We followed up the analysis by \citet{Tofflemire2025}, where a trend in the slope as a function of the energy level of the hydrogen series has been found. They speculate that such increasing trend can be due to the fact that hydrogen at different excitation levels come from different physical conditions (such as temperature and density), thus, different lines trace the pre-shock region in different part of the accretion column or the post-shock region.

We incorporated in the analysis slopes of empirical relations of the Pf$\beta$ \citep{Salyk2013}, Hu$\beta$ \citep{Rigliaco2015}, Br$\alpha$ \citep{Komarova2020}, Pa$\alpha$ and Br$\beta$ \citep{Rogers2024}. 
The increasing trend suggested by \citet{Tofflemire2025} is less convincing when including the latter tracers, since we note a large spread in the distribution, which increases with the wavelength. However, fitting the slopes distribution as a function of their wavelength, we find a correlation factor of 0.9.  
It must be stressed that, while \citet{Tofflemire2025} compared slopes obtained by computing the accretion luminosity using the XS-fit model from X-Shooter data — which can therefore be directly compared to all the slopes up to Br10 — other empirical relations have been derived using different approaches to estimate $\lacc$, and may consequently introduce systematic biases.
More specifically, the empirical relations by \citet{Salyk2013, Rigliaco2015, Komarova2020} were fitted using $\lacc$ estimates that are not contemporaneous with the line flux measurements. Moreover, the line luminosity ranges used to fit the Pf$\beta$, H$\gamma$, H$\beta$, and H$\alpha$ relations are very narrow \citep[e.g., $-4.2 < \log L_{\rm Pa \beta} < -3.2$ for Pf$\beta$, and $-1.5 < \log L_{\rm H} < 0$ for the other H-lines;][]{Salyk2013, Tofflemire2025}. 
We also add that the low resolution of {\it Spitzer} and the delicate subtraction of H$_2$O also affect the MIR empirical relations (Baskaran et al. in prep.). 
Finally, \citet{Rogers2024} analysed a sample of CTTSs in a low-metallicity environment and derived their best-fit relations for Pa$\alpha$ and Br$\beta$ by computing $\lacc$ indirectly from the $\brg$ line using the \href{#label}{A17} relation.

All the differences — in methodology, sample selection, and luminosity range — ({\it i}) contribute to the spread observed in the slope\,$- \lambda$ distribution, ({\it ii}) affect the uncertainties in the empirical relations themselves, including the slope values. 
Only simultaneous observations of UV-excess and MIR accretion tracers of CTTSs with a suitable luminosity range will enable to provide accurate empirical relations linking the accretion luminosity to the MIR accretion tracers.

\subsection{On the $A_V$ estimate}
A critical aspect in analyzing accretion values and applying empirical relations is the estimate of $A_V$. 
Our results show that different approaches provide different results with a non-negligible spread ($0.4-0.5$) with no linear trend. 
Curiously, this is also valid for the extinction provided from two direct modeling of UV-excess, i.e., the XS-fit and the HST-fit (see Fig.\,\ref{fig:Av-miaCae}). It is possible that the spread in the $A_V$ estimate is at least partially responsible for the spread in the $\log \lacc^{HST} - \log \lacc^{XS}$ distribution (Fig.\,\ref{fig:LaccMIADMslab}). We note here that the the accretion luminosity from the HST-fit is not necessarily higher than the one provided by the XS-fit, as one would expect \citep{pittman2022} given the much extended UV spectral range of HST with respect to X-Shooter \citep{alc19}. The spread in the distribution of both the extinction and the accretion luminosity computed with the XS- and HST-fit suggests intrinsic uncertainty within which it is possible to estimate these parameters, using the PENELLOPE and ODYSSEUS methods.

Among the three different approaches to estimate $A_V$ based on empirical relations, one is more reliable than others (Sect.\,\ref{sect:Av}). 
Indeed, the $rms$ between the $A_V^{XS}$ and the extinction calculated by each method is lowest for the difference method (0.4), indicating that $A_V^{diff}$ estimates are the most similar to $A_V^{XS}$. This is also the spread present when comparing $A_V^{diff}$ and $A_V^{XS}$ with $A_V^{HST}$, suggesting 0.4 as a possible estimate of the intrinsic extinction uncertainty.

Eq.\,\ref{eq:Lacc_linesAv} shows that using $A_V^{diff}$ to deredden the line fluxes results in a linear correlation between $\lacc^{XS}$ and $\lacc^{lines}$. 
This implies that when $A_V$ is computed from observed line fluxes using the difference method, the $\lacc$ derived through \href{#label}{A17} empirical relations matches that obtained from the XS-fit. 
However, we point out that $A_V$ is constrained only when simultaneously observed accretion tracers span a broad wavelength range, ideally from the UVB to the NIR. 

We also found that in 80\% of our sources (51/64), the $\lacc$ values derived using empirical relations and dereddening the line fluxes with both $A_V^{XS}$ and $A_V^{diff}$ are consistent (see Appendix\,\ref{app:logLacclogLambda}).
For the remaining 20\%, the extinction estimate differs by at least 0.5\,mag within the error, where 0.5\,mag is the typical error on $A_V^{XS}$ but, on the other hand, very similar to the spread of 0.4\,mag that we found when comparing the XS-fit, the HST-fit, and the difference method estimates. 
In these cases, the slope of the $\log \lacc$ distribution as a function of the wavelength is flatter for 5 sources (SO1153, Sz66, Sz76, Sz98, Sz129) when using $A_V^{diff}$ to deredden the fluxes, suggesting this is a best extinction estimate for these objects. On the contrary, for 6 sources (CVSO90, CVSO176, Sz10, CVSO165, Sz130) the same distribution is flatter when using $A_V^{XS}$ to deredden the fluxes, thus $A_V^{XS}$ seems to be the best extinction estimate. (see Appendix\,\ref{app:logLacclogLambda}).

This analysis confirms that measuring the extinction on YSOs can be challenging, even with the established methods like the XS- or HST-fit. Alternative approaches, using Bayesian methods \citep{Halstead-Willett2025}, may yield more reliable results. Instead of providing a single best-fit value with an associated error, \citet{Halstead-Willett2025} method treats all stellar and accretion properties as probability distributions. It combines prior information with the observed spectral data to generate a posterior probability distribution for the parameters, using Monte Carlo Markov Chain  techniques.
This approach's main advantage is its ability to simultaneously explore the entire parameter space of the accretion model, allowing for a more robust characterization of all uncertainties and a direct measurement of the correlations between different parameters, although degeneracies among fit parameters cannot always be removed.
Our analysis also demonstrates that it is possible to estimate $A_V$ yielding accretion results in agreement with those of the XS-fit, but using empirical relations alone. 

Notably, comparing $\log\lacc^{XS}$ and $\log \lacc^{HST}$ we also find a non-negligible spread (see Fig.\,\ref{fig:LaccMIADMslab}). Indeed, $\lacc$ is used to estimate the mass accretion rate ($\macc$), which plays a key role in constraining the processes responsible for disc evolution in population synthesis models \citep[e.g.,][]{Manara2023}. Typically, published studies \citep[e.g.,][]{Lodato17, mul17,Tabone2022,somigliana2023} assume a systematic uncertainty on $\macc$ of 0.45\,dex on the values obtained fitting the UV-excess with X-Shooter. It would be important to understand whether this assumed uncertainty value is to be revised in light of the differences between $\log \lacc ^{HST}$ and $\log \lacc^{XS}$ found combining the HST and X-Shooter spectra from ULLYSES and PENELLOPE. Further investigation is needed to understand the reason of the differences observed in this work, and we will discuss this topic in a forthcoming paper.

\subsection{On the use of Empirical Relations for other types of YSOs and forming-planets}

Empirical relations for accretion have been applied not only to CTTSs but also to younger YSOs, including Class\,I and Flat Spectrum (FS) objects \citep[i.e.,][]{fio21, Fiorellino2023, Delabrosse2024, Tychoniec2024}, as well as to eruptive YSOs \citep[e.g.,][]{Hodapp2020, siwak2023, Singh2024, fiorellino2024, giannini2024}, to accreting brown dwarfs \citep[e.g.,][]{Whelan2018, Almendros-Abad2024}, and even on planets \citep[e.g.,][]{Plunkett2025, Bowler2025}. 
However, since these relations were derived for CTTSs, their application to other types of objects, which may accrete through different mechanisms, requires some examination.
According to \citet{mendigutia2015} the empirical relations between $\log \lacc$ and $\log \lline$ are a direct consequence of the $\log \lacc - \log\lstar$ correlation. Therefore, the $\log\lacc - \log\lline$ relations are not necessarily related with the physical origin of the lines. 
However, it is essential to verify this paradigm in YSOs beyond CTTS. 

Recent findings by \citet{flores2024} show that Class\,I/FS objects have magnetic field comparable to those of Class\,II YSOs. This suggests that magnetic fields, responsible for accretion on CTTSs, could play the same role also in Class\,I and FS objects, supporting the use of empirical relations for Class\,I/FS sources. 
However, some caveats exist. 
The first relates to the presence of veiling ($r$), representing the excess emission above the photosphere. Embedded YSOs, such as Class\,I and FS, often exhibit high veiling ($r \sim 50$), unlike CTTS ($r < 2$). 
For this reason, it is crucial to take into account the effect of the veiling before applying empirical relations in Class\,I, FS, or earlier-phase YSOs. Accurate veiling correction is challenging but vital, as it can significantly impact $\lacc$ estimates despite the empirical relations' accuracy.  
The second caveat stems from Class\,I and FS being embedded in their envelopes, thus obscuring their UV and optical emission and limiting their study to longer wavelengths. $\pab$ and $\brg$ are commonly used to estimate $\lacc$, and our results show that all the high-accretors in our sample exhibit $\pab$ and $\brg$ in emission, providing $\lacc$ luminosity in agreement with other accretion tracers. 
Thus, assuming Class\,I and FS are younger than CTTSs and should be high-accretors ($\log \lacc > -1$), the use of NIR empirical relations becomes a powerful tool to constrain accretion luminosity during the protostellar stage. 
However, using only NIR accretion tracers could underestimate the accretion luminosity \citep{Harsono2023} because most of the magnetospheric emission is completely blocked from view, and line fluxes are only catching a tiny fraction of $\lacc$ through scattered light on the (less extincted) cavity walls \citep{Delabrosse2024}. Thus, we agree with the conclusion of \citet{Delabrosse2024} that more complex methodologies should be used to constrain $\lacc$ on Class\,I to fully correct for occultation of the central source, and of the associated NIR emission. This is, for example, the case of the self-consistent approach used in \citet{ant08, fio21, Fiorellino2023}, based on empirical relations, integrated with the use of isochrones (or birthline) and  bolometric luminosity and veiling measurements.  
We also highlight that the samples of Class\,I protostars in both \citet{Fiorellino2023} and \citet{flores2024} are limited to those with high SNR, while fainter Class\,I objects ($m_K < 15$) have not been analyzed yet.

In the case of episodic accretion, FUors event fall outside magnetospheric accretion \citep[e.g.,][]{audard2014, vorobyov2020}, so empirical relations should not apply. 
For EXor-like YSOs, however, the spectral features are very similar to CTTSs ones. More in detail, empirical relations can be used to estimate $\lacc$ in the quiescent phase, when EXors accrete similarly to CTTSs, but not during the bursts. Indeed, \citet{CruzSaenzDeMiera2023} show that, during the last burst of Ex\,Lupi, $\lacc$ obtained with the XS-fit is up to one order of magnitude different than the $\lacc$ computed using \href{#label}{A17} empirical relations, higher or lower depending on the studied epoch. This result suggests that empirical relations obtained from CTTSs might not be applied to study EXors bursts. 

Other cases where empirical relations have been applied are those of brown dwarfs (BDs) and forming planets, whose faintness in the UV often prevents direct UV-excess modeling of the data. 
\citet{moh05} showed that the CTTS-like accretion paradigm can be extended down to BDs with masses as low as $\sim15$\,M$_\mathrm{J}$. This result has been confirmed by numerous other studies, and it is now widely accepted that stars and BDs share a common dominant formation mechanism \citep[see review by][]{Chabrier2014}.
Also, \citet{ReinersBasri2007} directly measured strong magnetic fields in BDs, on the order of kilogauss, demonstrating that BDs can generate fields comparable in strength to those of CTTSs. These results suggest that empirical relations provide reliable accretion luminosity estimates for BDs.

The situation is less clear for giant planets. Direct magnetic field measurements for exoplanets are still lacking, and observations of solar system planets set the maximum field strengths at only a few Gauss (4–14\,G) \citep{Connerney2018}. However, models suggest that hot Jupiters could have significantly stronger surface magnetic fields. For instance, \citet{Cauley2019} estimated that hot Jupiters could have surface magnetic fields ranging from 20\,G to 120\,G, 10 to 100 times stronger than those observed in the solar system and comparable to CTTS values for the stronger cases.
Nevertheless, unlike in CTTSs, magnetic fields are not believed to be the primary driver of accretion in giant planets. Core accretion is thought to proceed without the need for strong magnetic coupling, although magnetic fields may influence the dynamics of the surrounding disk \citep{Turner2014}.
Thus, even if forming planets may possess strong magnetic fields, it remains unclear whether a magnetospheric accretion scenario applies. Consequently, the reliability of accretion luminosity estimates based on empirical relations in the planetary regime is still uncertain.

\subsection{Total energy emitted in lines vs. in continuum}
\label{sect:energyloss}

In a recent work, \citet{Hashimoto2025} studied 76 brown dwarfs and very-low mass stars by assuming the magnetospheric accretion model, i.e. using empirical relations from \href{#label}{A17}, and the accretion-shock model, i.e. using empirical relations from \citet{Aoyama2021} initially created for accreting planets. The difference between the magnetospheric accretion and the accretion-shock model is that the first assumes that accretion tracers such as hydrogen lines primarily come from the accretion flow, while the latter assumes they come from the shock region. 
They find a bimodal trend between the energy per unit time emitted from the lines $E_{lines}= \Sigma L_{\rm lines, i}$ and the energy  per unit time emitted from the continuum $E_{cont} = \lacc$ when using  empirical relations from \citet{Aoyama2021}, while it confirms the presence of a correlation between $E_{lines}$ and $E_{cont}$, even for very-low mass objects, when using empirical relations from \href{#label}{A17}. In particular, they find $E_{lines} \sim E_{cont} $ for very-low mass objects, while $E_{cont} >> E_{lines}$ for the more massive objects \citep[see also][]{Zhou2021}. 

In our CTTSs sample, 
Eqs.\,\ref{eq:energy-lossXS}\,and\,\ref{eq:energy-lossHST} (see also right panels of Fig.\,\ref{fig:ratioLlinesLacc}) indicate the trend that objects with lower $\lacc$ tend to emit energy in lines close to that of the continuum. Given that low-$\lacc$ objects are typically less massive, this raises questions about the behavior of the $E_{lines}/E_{cont}$ ratio in progressively less massive objects, such as sub-stellar and planetary-mass objects.
The comparison between \citet{Hashimoto2025} and this work results suggests that the trend we note in our sample continues in the very-low mass regime, but according to the analysis by \citet{Hashimoto2025}, the trend is different if the shock model is at work in planetary-mass objects. 
We also note that for lower accretors, the contribution from chromospheric emission could account for a significant fraction of the measured line flux and may explain the trend observed in Fig. \ref{fig:ratioLlinesLacc}. To perform a direct comparison between young stars and planetary-mass objects \citep[e.g., from][]{Hashimoto2025}, it would be necessary to subtract this chromospheric contribution from the accretion tracer lines. However, as it is highly challenging to account for this contribution precisely, we emphasize that any such comparison must be interpreted carefully in light of this caveat.

\section{Conclusions}
\label{sect:conclusions}

In this study, we analyzed 64 X-Shooter spectra of CTTSs from the PENELLOPE sample. For 61 of these CTTSs, we also used quasi-contemporaneous HST spectra.
We showed that the accretion luminosities derived with the HST-fit and XS-fit methods are statistically consistent, though significant differences may arise for individual objects. 

By revisiting the $\log\lacc$–$\log\lline$ empirical relations using accretion luminosities computed via HST modeling and the HST-ULLYSES data, we confirmed the robustness of most of the relations derived from X-Shooter data. 
We also conclude that empirical relations are valid for regions in various environments.  
The spread in the extinction and accretion luminosity values derived with the two approaches suggest the existence of an intrinsic uncertainty in our current estimates of these paramenters.

We also presented a ``difference method" based on empirical relations as an independent way to constrain extinction in CTTSs. This method yields $A_V$ values comparable to those obtained via the XS-fit, along with a similar distribution of accretion luminosities. 

We highlight that for all the high-accretors (13 CTTS), we detect NIR tracers such as $\pab$ and $\brg$, with their fluxes yielding accretion luminosity estimates similar to the XS-fit. 
For low-accretors ($\log \lacc^{XS} < -1$) showing $\brg$ detection (37 CTTSs in our sample), NIR lines luminosity underestimates the accretion luminosity of 0.5\,dex on avarage in 51\% of the sources (19/37). 
This suggests that, under the assumption of being high-accretors, $\lacc$ can be reliably estimated using $\pab$ and $\brg$. This is particularly useful to constrain $\lacc$ on embedded sources not accessible at UVB or VIS bands, such as Class\,I YSOs.
In general, $\log \lacc - \log \lline$ empirical relations can be applied to Class\,I and FS protostars, provided that veiling is accurately measured and multiple accretion tracers are used. Accretion studies of eruptive YSOs, particularly EXor-type objects, can also be investigated with these relations during their quiescent phases, when their accretion resembles that of CTTSs. Similarly, empirical relations provide reliable accretion luminosity estimates for brown dwarfs, while their applicability to accreting planets remains unclear.

\section*{Data availability}
Table\,\ref{tab:Lacc-Av-varie1} is only available in electronic form at the CDS via anonymous ftp to cdsarc.u-strasbg.fr (130.79.128.5) or via \url{http://cdsweb.u-strasbg.fr/cgi-bin/qcat?J/A+A/}.

\begin{acknowledgements}
In Memoriam of Will Fischer, a pillar of the PENELLOPE-ODYSSEUS-ULLYSES collaboration and the star formation community. His scientific contributions and friendship will be greatly missed.
We thank Dr. G. Inchingolo for his assistance with the graphical design of Fig.\,\ref{fig:XS-spectrum}, and
Dr. A. Banzatti and Dr. J. F. Gameiro for the interesting discussion and feedback about the results of this work. We also thank the anonymous referee for their work to improve this manuscript. 
This work has been financially supported by Large Grant INAF-2022 “YSOs Outflows, Disks and Accretion: towards a global framework for the evolution of planet forming systems (YODA)”, by Large Grant INAF-2024 “Spectral Key features of Young stellar objects: Wind-Accretion LinKs Explored in the infraRed (SKYWALKER)”, and by the European Union (ERC, WANDA, 101039452). Views and opinions expressed are however those of the authors only and do not necessarily reflect those of the European Union or the European Research Council Executive Agency. Neither the European Union nor the granting authority can be held responsible for them. 
This work was supported by HST AR-16129 and benefited from discussions with the ODYSSEUS team, \url{https://sites.bu.edu/odysseus/}. 
This work received funding from the Hungarian NKFIH project No.~K-147380; by the NKFIH NKKP grant ADVANCED 149943, the NKFIH excellence grant TKP2021-NKTA-64. Project no.149943 has been implemented with the support provided by the Ministry of Culture and Innovation of Hungary from the National Research, Development and Innovation Fund, financed under the NKKP ADVANCED funding scheme.
EF has been partially supported by project AYA2018-RTI-096188-B-I00 from the Spanish Agencia Estatal de Investigaci\'on and by Grant Agreement 101004719 of the EU project ORP. 
JMA acknowledges support from PRIN-MUR 2022
20228JPA3A ``The path to star and planet formation in the JWST era (PATH)" funded by NextGeneration EU and by INAF-GoG 2022 “NIR-dark Accretion Outbursts in Massive Young stellar objects (NAOMY)”.
EF and JMA acknowledge support from the INAF Mini-Grant 2023 ``Investigating the planet formation: initial conditions through the mass accretion rate on protostars".
CP acknowledges funding from the NSF Graduate Research Fellowship Program under grant No. DGE-1840990. GL acknowledges support from PRIN-MUR 20228JPA3A and from the European Union Next Generation EU, CUP:G53D23000870006. 

\end{acknowledgements}

\bibliographystyle{aa} 
\bibliography{biblio.bib}

\appendix

\begingroup 
\footnotesize 
\setlength{\tabcolsep}{3pt}

%\longtab[1]{

\onecolumn
\begin{landscape}
\section{The Sample}
\label{app:sample}

%\begin{landscape}
%\longtab[1]{

\begin{longtable}{llcrccccccc}
\caption{The PENELLOPE sample.}
\label{tab:sample1} 
\\
\hline
Region & Source Name & R.A. & Dec. & $d$ & $A_V^{XS}$ & $\log \lacc^{XS}$ & $\log <\lacc>^{lines, A17}$ & $A_V^{HST}$ & $\log \lacc^{HST}$ & $\brg$ \\
 & & & & pc & mag & $\lsun$ & $\lsun$ & mag & $\lsun$ & \\
\hline
\hline
\endfirsthead
\caption{continued.}\\
\hline\hline
Region & Source Name & R.A. & Dec. & $d$ & $A_V^{XS}$ & $\log \lacc^{XS}$ & $\log <\lacc>^{lines, A17}$ & $A_V^{HST}$ & $\log \lacc^{HST}$ & $\brg$ \\
 & & & & pc & mag & $\lsun$ & $\lsun$ & mag & $\lsun$ & \\
\hline
\endhead
\hline
\hline
\endfoot
\multicolumn{11}{l}{\footnotesize $^\dagger$ from \citet{Claes2024}. $^{\rm WTTS}$ is a Weak-line T\,Tauri Star. $^*$ EChaJ0844M7833 is a brown dwarf. $^**$ AA\,Tau is a dipper prototype} \\ 
\multicolumn{11}{l}{\footnotesize observed in our sample during a dip, thus we do not analyze this source in this work. $\log <\lacc>^{lines}$ is the average $\lacc$ on all detected} \\
\multicolumn{11}{l}{\footnotesize lines, calculated as described in Section,\ref{sect:Llines-XS}. The last column flags the detection (1) or non-detection (0) of the $\brg$ line.} \\
\endlastfoot

\hline
\hline
OB1 & CVSO58 & 05:29:23.26 & $-$01:25:15.5 & $354.2 \pm 2.9$ & 0.8 & $-1.12 \pm 0.25$ & $-1.40 \pm 0.16 $ & $1.86^{0.05}_{0.06}$ & $-0.36_{0.04}^{0.04}$ & 1\\
OB1 & CVSO90 & 05:31:20.63 & $-$00:49:19.8 & $343.6 \pm 3.9$ & 0.1 & $-1.34 \pm 0.25$ & $-1.04 \pm 0.17 $ & $1.04^{0.06}_{0.85}$ & $-0.43_{0.04}^{0.29}$ & 1\\
OB1 & CVSO104 & 05:32:06.49 & $-$01:11:00.8 & $366.4 \pm 4.0$ & 0.2 & $-1.73 \pm 0.25$ & $-1.48 \pm 0.19 $ & $1.24^{0.03}_{0.03}$ & $-0.72_{0.02}^{0.02}$ & 1\\
OB1 & CVSO107 & 05:32:25.79 & $-$00:36:53.4 & $335.1 \pm 2.5$ & 0.3 & $-1.30 \pm 0.25$ & $-1.46 \pm 0.26 $ & $0.01^{0.01}_{0.00}$ & $-1.31_{0.01}^{0.01}$ & 1\\
OB1 & CVSO109 & 05:32:32.66 & $-$01:13:46.1 & $400.0$ & 0.1 & $-0.77 \pm 0.25$ & $-1.08 \pm 0.14 $ & $1.30^{0.02}_{0.02}$ & $-0.03^{0.02}_{0.02}$ & 1\\
OB1 & CVSO146 & 05:35:46.01 & $-$00:57:52.2 & $336.7 \pm 1.7$ & 0.6 & $-1.46 \pm 0.25$ & $-1.25 \pm 0.23 $ & $0.58^{0.02}_{0.03}$ & $-0.99_{0.03}^{0.02}$ & 1\\
OB1 & CVSO165 & 05:39:02.57 & $-$01:20:32.3 & $400.0$ & 0.2 & $-2.05 \pm 0.25$ & $-1.30 \pm 0.13 $ & $0.01^{0.01}_{0.00}$ & $-1.66^{0.01}_{0.01}$ & 0\\
OB1 & CVSO176 & 05:40:24.15 & $-$00:31:21.3 & $306.8 \pm 3.0$ & 1.0 & $-1.27 \pm 0.25$ & $-1.70 \pm 0.22 $ & $0.27^{0.02}_{0.02}$ & $-1.73_{0.01}^{0.01}$ & 0\\
$\sigma$Ori & SO518 & 05:38:27.26 & $-$02:45:09.7 & 390.3 & 0.1 & $-0.88 \pm 0.25$ & $-0.79 \pm 0.32 $ & $2.00^{0.01}_{0.02}$ & $-0.45^{0.01}_{0.01}$ & 1\\
$\sigma$Ori & SO583 & 05:38:33.69 & $-$02:44:14.1 & 392.3 & 1.0 & $-1.22 \pm 0.25$ & $-1.64 \pm 0.22 $ & $1.02^{0.04}_{0.04}$ & $-0.60^{0.07}_{0.08}$ & 0\\
$\sigma$Ori & SO1153 & 05:39:39.83 & $-$02:31:21.9 & 385.0 & 0.4 & $-0.30 \pm 0.25$ & $-0.98 \pm 0.24 $ & $0.49^{0.04}_{0.05}$ & $-0.42^{0.02}_{0.03}$ & 1\\
Cha\,I & SY Cha & 10:56:30.38 & $-$77:11:39.4 & $180.7 \pm 0.40$ & 0.8 & $-0.77 \pm 0.25$ & $-0.90 \pm 0.19 $ & $0.16^{0.02}_{0.02}$ & $-0.83_{0.01}^{0.01}$& 1\\
Cha\,I & CS Cha & 11:02:24.87 & $-$77:33:35.6 & $190.0 \pm 19.00$ & 0.9 & $-1.29 \pm 0.25$ & $-0.94 \pm 0.26 $ & $0.90^{0.01}_{0.01}$ & $-0.77_{0.02}^{0.02}$ & 1\\
Cha\,I & Sz 10 & 11:02:55.02 & $-$77:21:50.8 & $184.2 \pm 1.30$ & 1.0 & $-1.85 \pm 0.25$ & $-1.89 \pm 0.26 $ & $0.20^{0.03}_{0.03}$ & $-2.13_{0.02}^{0.02}$ & 0\\
Cha\,I & Hn 5 & 11:06:41.79 & $-$76:35:49.0 & $194.7 \pm 0.90$ & 1.1 & $-2.09 \pm 0.25$ & $-2.33 \pm 0.21 $ & $0.43^{0.03}_{0.04}$ & $-2.35_{0.02}^{0.02}$ & 1\\
Cha\,I & Sz 19 & 11:07:20.72 & $-$77:38:07.3 & $189.0 \pm 0.50$ & 1.5 & $-0.62 \pm 0.25$ & $-0.66 \pm 0.62 $ & $1.75^{0.05}_{0.05}$ & $0.27_{0.01}^{0.01}$ & 1\\
Cha\,I & VW Cha & 11:08:01.50 & $-$77:42:28.8 & $190.0$ & 2.3 & $-0.02 \pm 0.25$ & $-0.23 \pm 0.09 $ & - & - & 1\\
Cha\,I & VZ Cha & 11:09:23.77 & $-$76:23:20.8 & $191.1 \pm 0.60 $ & 1.7 & $ 0.04 \pm 0.25$ & $-0.35 \pm 0.13 $ & $0.91^{0.06}_{0.05}$ & $-0.39_{0.03}^{0.03}$ & 1\\
Cha\,I & WZ Cha & 11:10:53.32 & $-$76:34:32.0 & $193.2 \pm 0.60 $ & 1.0 & $-1.88 \pm 0.25$ & $-1.55 \pm 0.13 $ & $1.64^{0.04}_{0.04}$ & $-0.91^{0.02}_{0.02}$ & 1\\
Cha\,I & XX Cha & 11:11:39.67 & $-$76:20:15.0 & $192.1 \pm 0.80 $ & 0.3 & $-2.34 \pm 0.25$ & $-2.06 \pm 0.19 $ & $0.15^{0.04}_{0.04}$ & $-1.60_{0.03}^{0.03}$ & 1\\
Cha\,I & CHX18N & 11:11:46.33 & $-$76:20:08.9 & $191.6 \pm 0.40 $ & 0.7 & $-1.24 \pm 0.25$ & $-1.47 \pm 0.23 $ & $0.40^{0.02}_{0.01}$ & $-2.19_{0.01}^{0.01}$ & 0\\
Cha\,I & IN Cha & 11:12:09.83 & $-$76:34:36.4 & $192.6 \pm 0.80 $ & 0.2 & $-2.73 \pm 0.25$ & $-2.76 \pm 0.12 $ & $0.01^{0.00}_{0.00}$ & $-2.71_{0.02}^{0.04}$ & 0\\
Cha\,I & CV Cha & 11:12:27.70 & $-$76:44:22.3 & $191.8 \pm 0.50 $ & 1.1 & $ 0.17 \pm 0.25$ & $ 0.14 \pm 0.17 $ & $1.37^{0.02}_{0.02}$ & $ 0.23_{0.01}^{0.01}$ & 1\\
Cha\,I & Sz 45 & 11:17:36.97 & $-$77:04:38.1 & $189.0$ & 0.7 & $-1.34 \pm 0.25$ & $-1.51 \pm 0.30 $ & $0.63^{0.03}_{0.04}$ & $-1.31^{0.02}_{0.02}$ & 1\\
$\epsilon$Cha & HD 104237E & 12:00:09.09 & $-$78:11:42.3 & $100.2 \pm 0.3$ & 1.1 & $-2.11 \pm 0.25$ & $-2.20 \pm 0.44 $ & $1.00^{0.00}_{0.00}$ & $-2.53_{0.03}^{0.03}$ & 1\\
$\eta$Cha & RECX1 & 08:36:56.05 & $-$78:56:45.2 & 99.0 & 0.1 & $-1.44 \pm 0.25$ & $-1.87 \pm 0.37 $ & - & - & 0\\
$\eta$Cha & RECX5$^{\rm WTTS}$ & 08:42:26.93 & $-$78:57:47.4 & 99.0 & 0.2 & $-4.49 \pm 0.25$ & $-3.60 \pm 0.22 $ & - & - & 0\\
$\eta$Cha & RECX6 & 08:42:38.78 & $-$78:54:42.7 & $97.9 \pm 0.1^{\dagger}$ & $0.0^{\dagger}$ & $-3.24 \pm 0.25$ & $-2.92 \pm 0.37$ & - & - & 0\\
$\eta$Cha & ETCha & 08:43:18.43 & $-$79:05:17.7 & 99.0 & 0.1 & $-2.49 \pm 0.25$ & $-1.98 \pm 0.23 $ & $0.34^{0.03}_{0.03}$ & $-1.97^{0.02}_{0.02}$ & 1\\
$\eta$Cha & EChaJ0844M7833$^{*}$ & 08:44:09.00 & $-$78:33:45.3 & 99.0 & 0.1 & $-4.15 \pm 0.25$ & $-3.98 \pm 0.18 $ & $0.10^{0.00}_{0.01}$ & $-3.90^{0.43}_{0.43}$ & 0\\
$\eta$Cha & RECX9 & 08:44:16.24 & $-$78:59:07.6 & $99.0 \pm 10.0$ & 0.0 & $-2.69 \pm 0.25$ & $-3.24 \pm 0.15 $ & $0.01^{0.01}_{0.00}$ & $-2.90_{0.01}^{0.02}$ & 0\\
$\eta$Cha & RECX11 & 08:47:01.48 & $-$78:59:33.9 & $98.8 \pm 0.1$ & 0.1 & $-1.68 \pm 0.25$ & $-2.14 \pm 0.49 $ & $0.01^{0.00}_{0.00}$ & $-3.18_{0.01}^{0.01}$ & 0\\
Cor\,Au & RXJ1842.9-3532 & 18:42:57.98 & $-$35:32:43.2 & 151.0 & 0.7 & $-1.41 \pm 0.25$ & $-1.08 \pm 0.17 $ & $0.60^{0.00}_{0.00}$ & $-0.85^{0.01}_{0.01}$ & 1\\
Cor\,Au & RXJ1852.3-3700 & 18:52:17.31 & $-$37:00:12.4 & 147.0 & 0.0 & $-1.73 \pm 0.25$ & $-2.06 \pm 0.41 $ & $0.01^{0.00}_{0.00}$ & $-1.78^{0.01}_{0.01}$ & 1\\
Taurus & LkCa4$^{\rm WTTS}$ & 04:16:28.12 & $+$28:07:35.3 & 130.0 & 0.0 & $-4.29 \pm 0.25$ & $-2.66 \pm 0.40 $ & - & - & 0\\
Taurus & DETau & 04:21:55.65 & $+$27:55:05.7 & $128.0 \pm 0.4$ & 0.3 & $-1.63 \pm 0.25$ & $-1.30 \pm 0.26 $ & $1.17^{0.04}_{0.03}$ & $-0.85_{0.01}^{0.01}$ & 1\\
Taurus & AATau$^{**}$ & 04:31:53.45 & $+$24:22:44.4 & $134.7 \pm 1.6$ & - & - & - & - & - & 0\\
Taurus & DMTau & 04:33:48.75 & $+$18:10:09.6 & 144.0 & 0.6 & $-1.21 \pm 0.25$ & $-1.40 \pm 0.19 $ & $1.21^{0.03}_{0.04}$ & $-0.63^{0.02}_{0.02}$ & 1\\
Taurus & DNTau & 04:35:27.38 & $+$24:14:58.5 & $128.6 \pm 0.4$ & 0.4 & $-1.29 \pm 0.25$ & $-1.63 \pm 0.20 $ & $0.01^{0.00}_{0.00}$ & $-1.50_{0.03}^{0.03}$ & 1\\
Taurus & RXJ0438.6$+$1546$^{\rm WTTS}$ & 04:38:39.00 & $+$15:46:11.6 & $139.1 \pm 0.3^\dagger$ & 0.0$^\dagger$ & - & $-1.91 \pm 0.66$ & - & - & 0\\
Taurus & LkCa15 & 04:39:17.80 & $+$22:21:03.1 & 157.0 & 0.5 & $-1.63 \pm 0.25$ & $-1.59 \pm 0.26 $ & - & - & 1\\
Taurus & DKTau & 04:30:44.25 & $+$26:01:24.3 & $132.0 \pm 0.9$ & 0.6 & $-1.24 \pm 0.25$ & $-1.19 \pm 0.21 $ & $0.08^{0.05}_{0.04}$ & $-1.37_{0.01}^{0.04}$ & 1\\
Lupus & Sz66 & 15:39:28.27 & $-$34:46:18.4 & $155.9 \pm 0.7$ & 0.6 & $-2.35 \pm 0.25$ & $-2.73 \pm 0.32$ & $0.52^{0.04}_{0.04}$ & $-2.25^{+0.01}_{-0.01}$ & 1 \\
Lupus & Sz68 & 15:45:12.85 & $-$34:17:30.9 & $152.7 \pm 4.3$ & 1.0 & $-0.91 \pm 0.25$ & $-1.00 \pm 0.41$ & $1.39^{0.00}_{0.00}$ & $-0.78^{+0.01}_{-0.01}$ & 1 \\
Lupus & Sz69 & 15:45:17.39 & $-$34:18:28.6 & $152.6 \pm 1.6$ & 1.0 & $-2.02 \pm 0.25$ & $-2.36 \pm 0.31$ & $0.01^{0.01}_{0.00}$ & $-2.86^{+0.01}_{-0.01}$ & 1 \\
Lupus & Sz71 & 15:46:44.71 & $-$34:30:36.0 & $155.2 \pm 0.4$ & 0.7 & $-2.22 \pm 0.25$ & $-1.85 \pm 0.24$ & $0.35^{0.02}_{0.02}$ & $-1.87^{+0.01}_{-0.01}$ & 1 \\
Lupus & Sz72 & 15:46:44.71 & $-$34:30:36.0 & $156.7 \pm 0.5$ & 1.0 & $-1.20 \pm 0.25$ & $-0.95 \pm 0.21$ & $1.06^{0.03}_{0.03}$ & $-1.05^{+0.01}_{-0.01}$ & 1 \\
Lupus & Sz75 & 15:49:12.09 & $-$35:39:05.4 & $154.1 \pm 0.7$ & 1.0 & $-0.19 \pm 0.25$ & $-0.62 \pm 0.11$ & $0.21^{0.03}_{0.03}$ & $-0.76^{+0.01}_{-0.01}$ & 1 \\
Lupus & Sz76 & 15:49:30.72 & $-$35:49:51.7 & $159.0$ & 0.3 & $-2.55 \pm 0.25$ & $-3.01 \pm 0.23$ & - & - & 1 \\
Lupus & Sz77 & 15:51:46.94 & $-$35:56:44.5 & $155.3 \pm 0.4$ & 0.3 & $-2.03 \pm 0.25$ & $-2.12 \pm 0.24$ & $0.35^{0.01}_{0.01}$ & $-1.36^{+0.01}_{-0.01}$ & 1 \\
Lupus & RXJ1556.1-3655 & 15:56:01.90 & $-$36:55:30.3 & 158.0 & 0.8 & $-0.92 \pm 0.25$ & $-1.87 \pm 0.37$ & $0.63^{0.02}_{0.02}$ & $-0.92^{0.01}_{0.01}$ & 1 \\
Lupus & Sz82 & 15:56:09.19 & $-$37:56:06.5 & $155.8 \pm 0.5$ & 1.0 & $-0.64 \pm 0.25$ & $-1.03 \pm 0.22$ & $0.47^{0.02}_{0.02}$ & $-0.98^{+0.01}_{-0.01}$ & 1 \\
Lupus & Sz84 & 15:58:02.50 & $-$37:36:03.1 & $155.6 \pm 1.1$ & 0.0 & $-3.17 \pm 0.25$ & $-2.98 \pm 0.16$ & $0.01^{0.01}_{0.01}$ & $-2.66^{+0.01}_{-0.01}$ & 0 \\
Lupus & Sz129 & 15:59:16.46 & $-$41:57:10.6 & $160.1 \pm 0.4$ & 1.0 & $-1.04 \pm 0.25$ & $-1.35 \pm 0.38$ & $0.76^{0.02}_{0.02}$ & $-1.05^{+0.01}_{-0.01}$ & 1 \\
Lupus & RYLup & 15:59:28.37 & $-$40:21:51.6 & $158.0 \pm 15.8$ & 0.0 & $-0.85 \pm 0.25$ & $-1.06 \pm 0.74$ & $0.53^{0.06}_{0.07}$ & $-1.98^{+0.01}_{-0.01}$ & 0 \\
Lupus & SSTc2dJ160000.6-422158 & 16:00:00.59 & $-$42:21:57.2 & $159.4 \pm 0.8$ & 0.0 & $-3.22 \pm 0.25$ & $-3.40 \pm 0.21$ & $0.01^{0.00}_{0.00}$ & $-2.87^{+0.01}_{-0.01}$ & 1 \\
Lupus & Sz130 & 16:00:31.02 & $-$41:43:37.3 & $159.2 \pm 0.5$ & 0.2 & $-2.57 \pm 0.25$ & $-1.88 \pm 0.26$ & $0.06^{0.01}_{0.01}$ & $-2.56^{+0.01}_{-0.01}$ & 1 \\
Lupus & MYLup & 16:00:44.50 & $-$41:55:31.3 & $157.2 \pm 0.9$ & 0.5 & $-1.60 \pm 0.25$ & $-1.42 \pm 0.90$ & $1.27^{0.01}_{0.01}$ & $-1.85^{+0.01}_{-0.01}$ & 1 \\
Lupus & Sz97 & 16:08:21.79 & $-$39:04:21.8 & $157.3 \pm 0.6$ & 0.7 & $-2.20 \pm 0.25$ & $-2.02 \pm 0.28$ & $0.24^{0.03}_{0.04}$ & $-2.13^{+0.01}_{-0.01}$ & 0 \\
Lupus & Sz98 & 16:08:22.48 & $-$39:04:46.8 & $156.3 \pm 0.6$ & 1.5 & $-1.08 \pm 0.25$ & $-1.74 \pm 0.40$ & $0.47^{0.01}_{0.01}$ & $-1.39^{+0.01}_{-0.01}$ & 1 \\
Lupus & Sz99 & 16:08:24.03 & $-$39:05:49.8 & $158.3 \pm 1.1$ & 0.0 & $-3.11 \pm 0.25$ & $-3.27 \pm 0.23$ & $0.01^{0.01}_{0.00}$ & $-2.73^{+0.01}_{-0.01}$ & 0 \\
Lupus & Sz100 & 16:08:25.75 & $-$39:06:01.6 & $158.0 \pm 15.8$ & 0.4 & $-2.54 \pm 0.25$ & $-2.64 \pm 0.18$ & $0.40^{0.03}_{0.03}$ & $-2.45^{+0.01}_{-0.01}$ & 0 \\
Lupus & Sz103 & 16:08:30.26 & $-$39:06:11.5 & $157.2 \pm 1.0$ & 0.7 & $-2.49 \pm 0.25$ & $-2.90 \pm 0.15$ & $0.53^{0.04}_{0.04}$ & $-2.41^{+0.01}_{-0.01}$ & 0 \\
Lupus & SSTc2dJ160830.7-382827 & 16:08:30.69 & $-$38:28:27.2 & $153.4 \pm 0.7$ & 0.2 & $-1.37 \pm 0.25$ & $-1.08 \pm 0.56$ & $0.19^{0.02}_{0.02}$ & $2.49^{+0.01}_{-0.01}$ & 0 \\
Lupus & Sz104 & 16:08:30.80 & $-$39:05:49.2 & $159.8 \pm 1.1$ & 0.0 & $-3.08 \pm 0.25$ & $-3.06 \pm 0.29$ & $0.56^{0.03}_{0.04}$ & $-2.33^{+0.01}_{-0.01}$ & 1 \\
Lupus & Sz110 & 16:08:51.56 & $-$39:03:18.0 & $157.5 \pm 0.6$ & 0.3 & $-2.38 \pm 0.25$ & $-2.18 \pm 0.15$ & $0.19^{0.04}_{0.05}$ & $-2.06^{+0.01}_{-0.01}$ & 1 \\
Lupus & Sz111 & 16:08:54.67 & $-$39:37:43.5 & $158.4 \pm 0.5$ & 0.5 & $-1.93 \pm 0.25$ & $-1.77 \pm 0.18$ & $0.14^{0.02}_{0.02}$ & $-2.11^{+0.01}_{-0.01}$ & 1 \\
Lupus & Sz114 & 16:09:01.84 & $-$39:05:12.8 & $157.0$ & 0.2 & $-2.34 \pm 0.25$ & $-2.71 \pm 0.21$ & - & - & 1 \\
Lupus & Sz115 & 16:09:06.21 & $-$39:08:51.9 & $156.0$ & 0.0 & $-3.05 \pm 0.25$ & $-3.80 \pm 0.37$ & - & - & 0 \\
Lupus & Sz117 & 16:09:44.36 & $-$39:13:30.2 & $157.0$ & 0.2 & $-2.08 \pm 0.25$ & $-2.54 \pm 0.36$ & $0.10^{0.02}_{0.02}$ & $-2.03^{+0.01}_{-0.01}$ & 1 \\
Lupus & SSTc2dJ161243.8-381503 & 16:12:43.75 & $-$38:15:03.1 & $160.0$ & 0.4 & $-2.70 \pm 0.25$ & $-2.52 \pm 0.29$ & $0.26^{0.02}_{0.02}$ & $-2.20^{+0.01}_{-0.01}$ & 1 \\
Lupus & SSTc2dJ161344.1-373646 & 16:13:44.10 & $-$37:36:46.3 & $159.0$ & 0.2 & $-2.47 \pm 0.25$ & $-2.70 \pm 0.37$ & $0.01^{0.00}_{0.00}$ & $-2.24^{+0.01}_{-0.01}$ & 1 \\
\hline \hline
\end{longtable}
\end{landscape}

\endgroup

\twocolumn
\section{HST and X-Shooter quasi-simultaneous observations}
\label{app:time-variability}

Despite efforts to coordinate VLT/X-Shooter and HST observations, the datasets are not always perfectly simultaneous. Looking at the time difference between the XS and HST observations (see Tab.\,\ref{tab:Lacc-Av-varie1}), we find that 38\% (23/61) of the sources were observed on the same day, 53\% (32/61) within two days, 64\% (39/61) within one week, 72\% within one month, and 80\% (49/61) within one year. Sources with a time separation on the order of a year (or larger) correspond to HST archival observations.

\begin{figure*}
    \includegraphics[width=1.01\columnwidth]{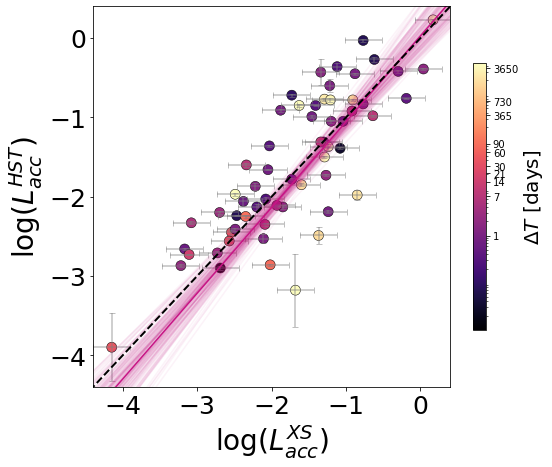}
    \includegraphics[width=0.93\columnwidth]{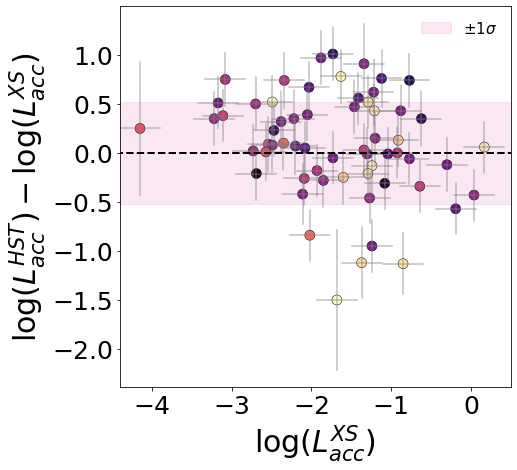}
    
    \caption{Comparison of $\lacc$ as in Fig.\,\ref{fig:LaccMIADMslab}. The colored bar shows the difference in time between XS and HST observations for each target.}
    \label{fig:Lacc_deltaT}
\end{figure*}

Since Fig.\,\ref{fig:LaccMIADMslab} shows some scatter in the distribution, we investigated whether this could be related to the time difference between the XS and HST observations. In principle, due to the intrinsic variability of YSOs and the fact that this is expected to be larger after a few days \citep[e.g.,][]{costigan2014}, targets observed closer in time are expected to yield more consistent $\log \lacc$ estimates when derived from the XS or HST fits. Conversely, outliers in the distribution may correspond to targets with larger temporal offsets between the two observations.
Nevertheless, Fig.\,\ref{fig:Lacc_deltaT} shows no clear trend between the time separation of the observations and the difference in $\log \lacc$ estimates using the XS- and HST-fit.
To quantitatively verify this, we re-examined the observed scatter by excluding sources with time lags greater than one week. The recomputed scatter ($\sigma$) is 0.46, which remains very similar to the original value of 0.50 that includes all sources. The number of sources falling outside the new $1\sigma$ range is now 7, which accounts for approximately 17\% of this new sample (39 sources). This is comparable to the 9 sources (about 15\%) that fell outside the $1\sigma$ range in the original, larger sample (61 sources). We would expect to find the agreement between the two measurements to improve. On one hand, the spread slightly decreases, while, on the other hand, the percentage of ``outliers" increases. We interpret this finding as further evidence that temporal variability is not the primary source of the spread in the $\log \lacc^{HST} vs. \log \lacc^{XS}$ distribution. We will investigate other possible sources of this scatter in a forthcoming paper.

\section{Measuring the flux of the Accretion Tracers}
\label{sect:Lacc}
The observed flux ($F_{obs}$) for the accretion tracers in every CTTS spectrum of the PENELLOPE sample were measured. 
The procedure was as follows: the local continuum was fitted with a linear function over a wavelength range of $\Delta \lambda = 2$\,nm, centered on the emission line's wavelength $\lambda_0$. 
This range was adjusted, as necessary, based on the line's intensity and to avoid overlapping emission lines or residual telluric absorption features. The line flux was then obtained by subtracting the local continuum and integrating over the emission line.

The noise of the line was calculated by multiplying the standard deviation of the local continuum ($rms$) by the wavelength element between two pixels ($\Delta \lambda$), and by the square root of the number of pixels within the wavelength range ($N_{pix}$). 
A line was considered detected if its $S/N > 3$. 
For non-detections, upper limits were computed using the following relation: 
\begin{equation}
    F_{line}^{upp} = 3 \times (\sqrt{N_{pix}} \times rms \times \Delta \lambda)
\end{equation}

As a reference, we used the same accretion tracers studied in \href{#label}{A17}. Contrary to the latter, and other previous CTTS X-Shooter samples,  PENELLOPE includes several objects in which many lines of the Brackett series have been  detected, as the PENELLOPE sample includes a larger number of high accretors with respect to the A17 sample. Hence, we also measured the flux of several of those lines.
Observed line fluxes before extinction correction are available on-line on Zenodo\footnote{https://zenodo.org/records/16419857 DOI:10.5281/zenodo.16419857}.

\section{$\log \lacc $ vs. $\log \lline$ using $\lacc$ from the HST modeling}
\label{app:MIADM}

\begin{figure*}
    \includegraphics[width=0.5\columnwidth]{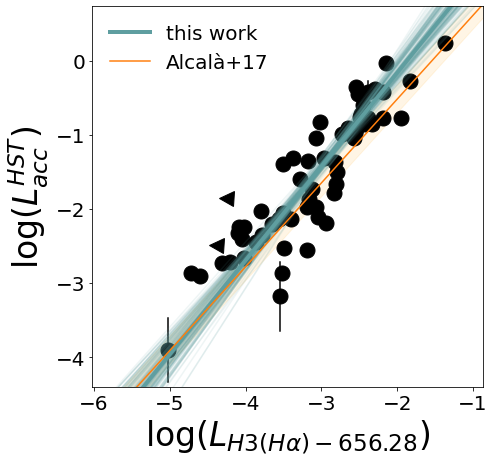}
    \includegraphics[width=0.5\columnwidth]{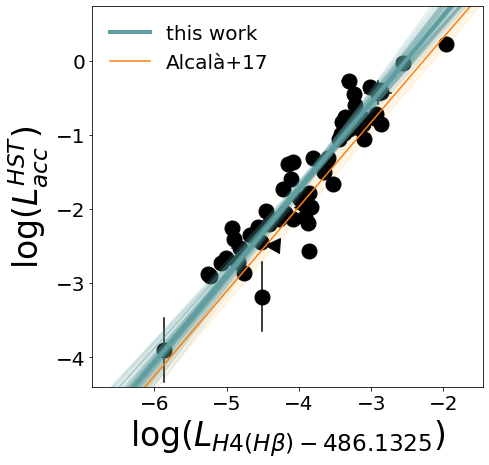}
    \includegraphics[width=0.5\columnwidth]{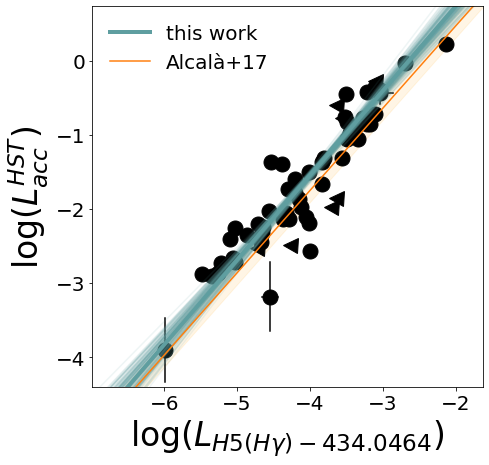}
    \includegraphics[width=0.5\columnwidth]{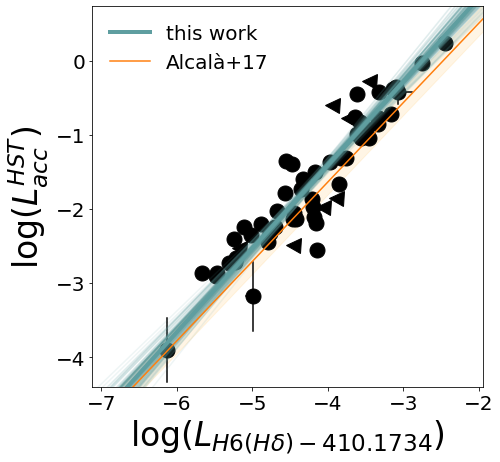}
    \includegraphics[width=0.5\columnwidth]{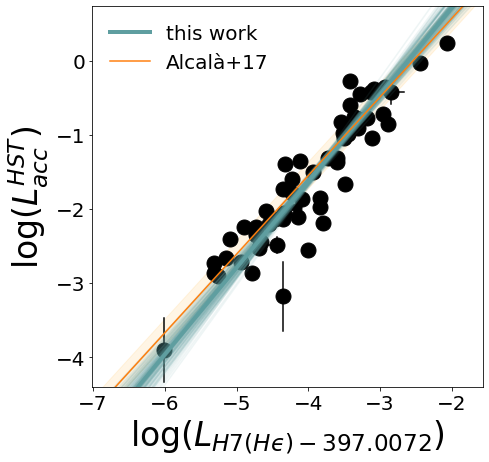}
    \includegraphics[width=0.5\columnwidth]{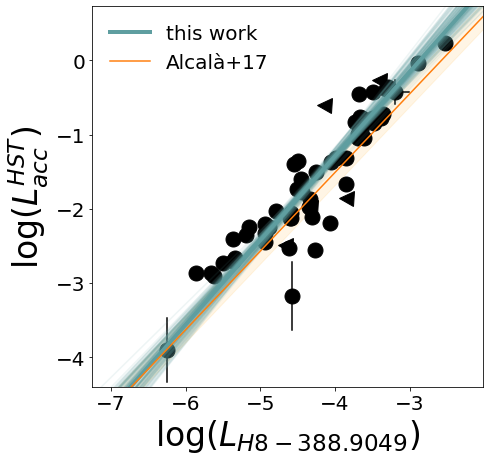}
    \includegraphics[width=0.5\columnwidth]{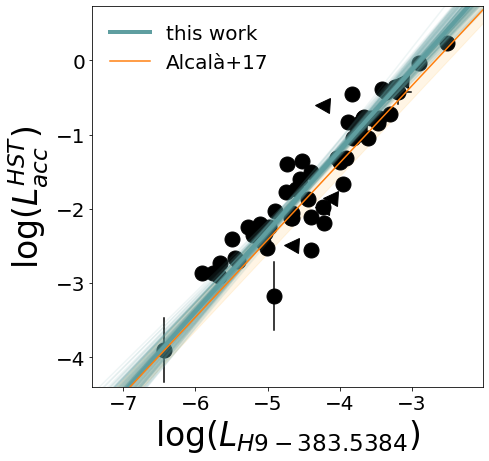}
    \includegraphics[width=0.5\columnwidth]{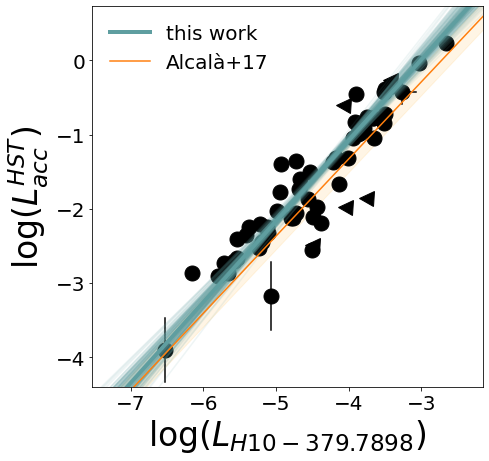}
    \includegraphics[width=0.5\columnwidth]{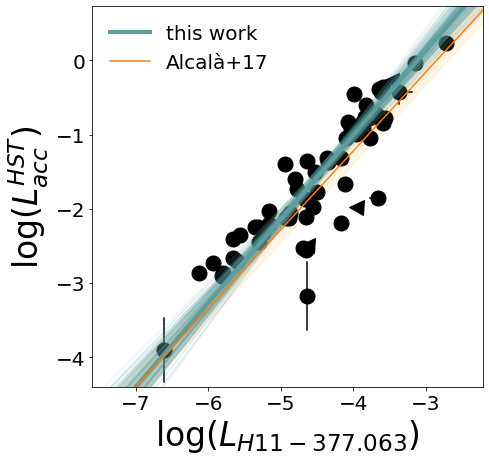}
    \includegraphics[width=0.5\columnwidth]{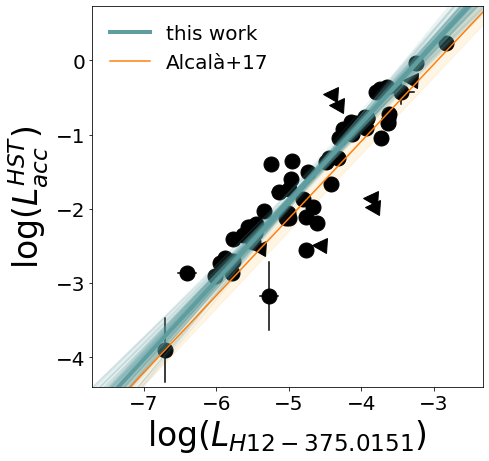}
    \includegraphics[width=0.5\columnwidth]{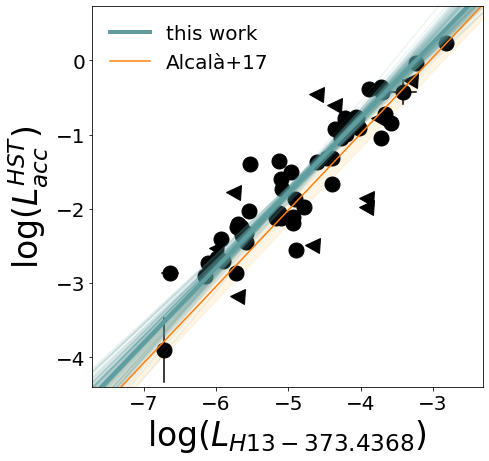}
    \includegraphics[width=0.5\columnwidth]{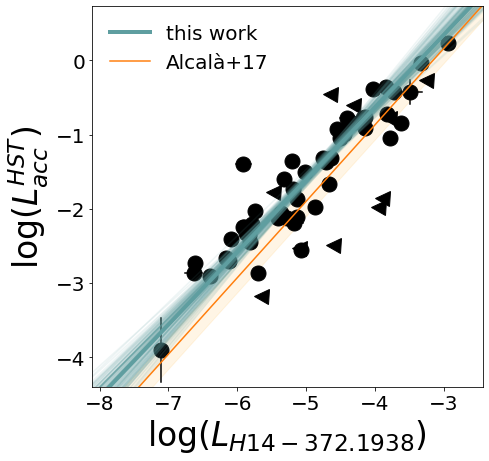}
    \includegraphics[width=0.5\columnwidth]{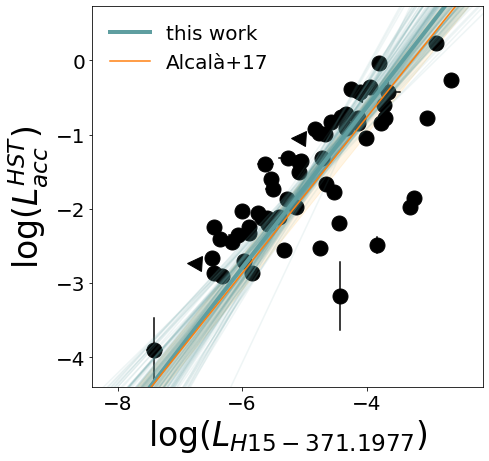}
    \includegraphics[width=0.5\columnwidth]{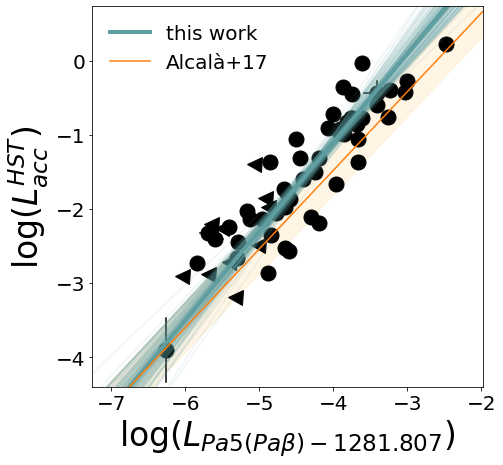}
    \includegraphics[width=0.5\columnwidth]{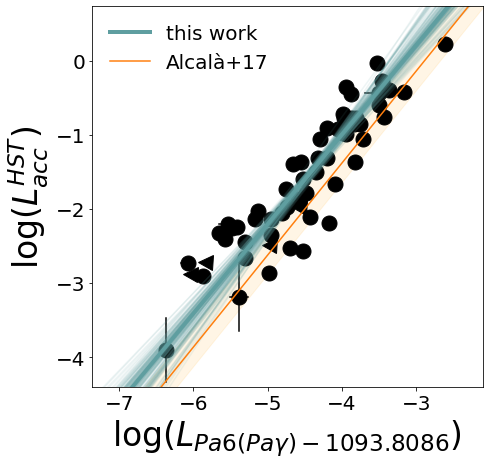}
    \includegraphics[width=0.5\columnwidth]{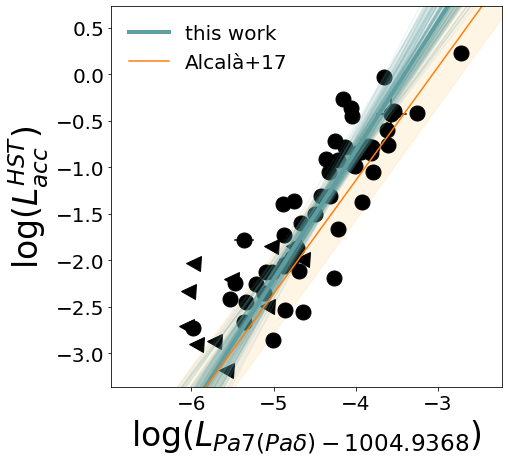}
    \includegraphics[width=0.5\columnwidth]{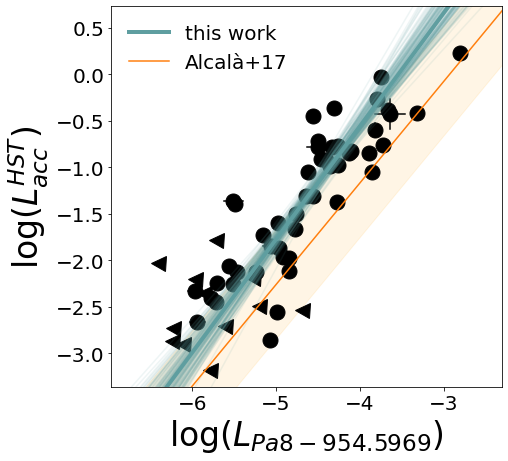}
    \includegraphics[width=0.5\columnwidth]{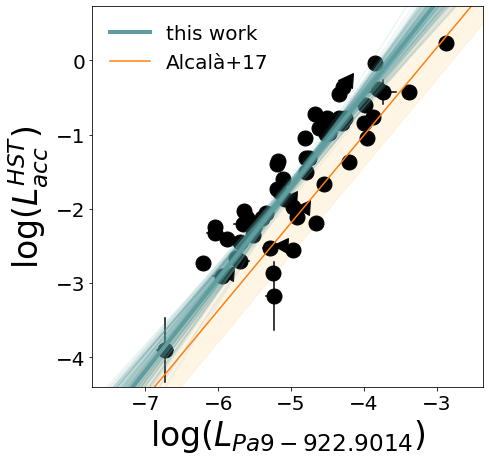}
    \includegraphics[width=0.5\columnwidth]{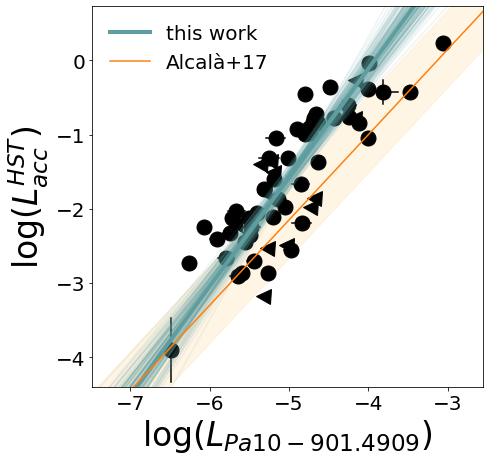}
    \includegraphics[width=0.5\columnwidth]{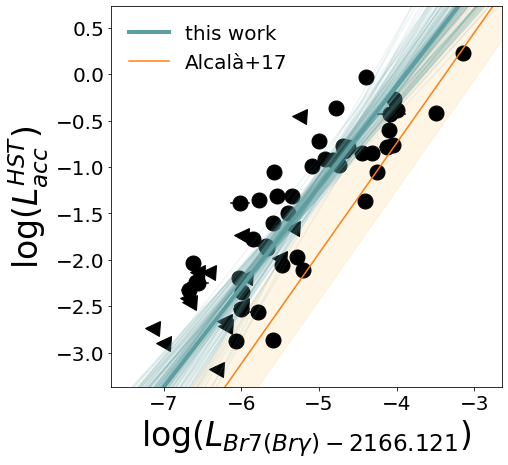}
    \caption{$\log \lacc - \log \lline$ and best fit for the Balmer series, the Paschen series and the $\brg$ line. The blue and orange lines show out best fit and the best fit from \citet{alc14}, respectively. The wavelength in nanometers of each line is shown on the x-axis.}
    \label{fig:new-rel-hydrogen}
\end{figure*}

\begin{figure*}
    \includegraphics[width=0.5\columnwidth]{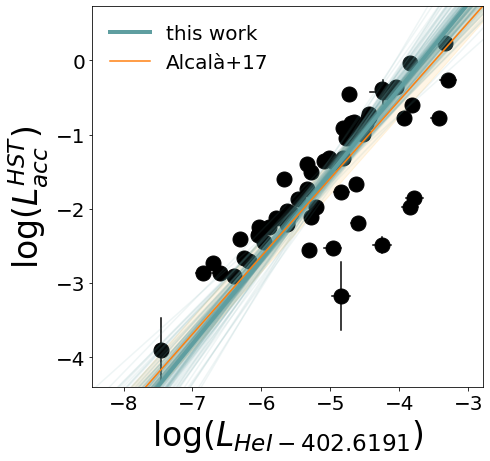}
    \includegraphics[width=0.5\columnwidth]{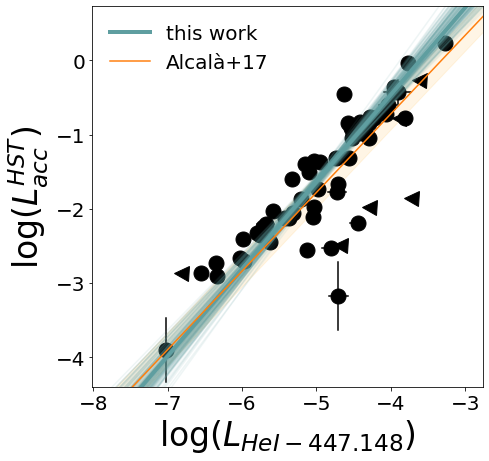}
    \includegraphics[width=0.5\columnwidth]{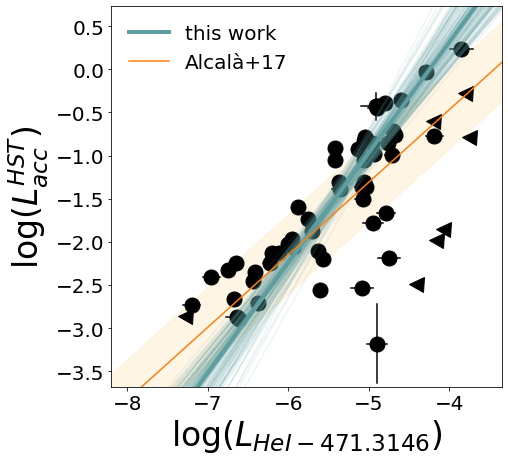}
    \includegraphics[width=0.5\columnwidth]{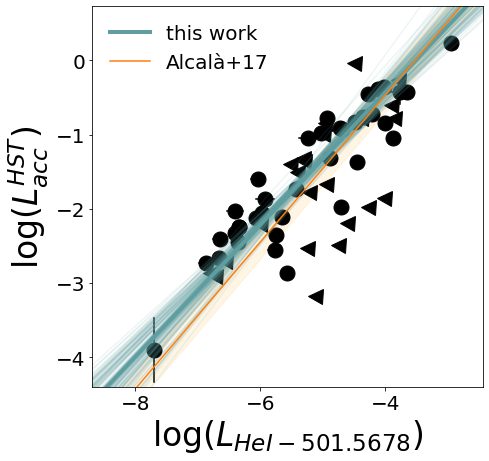}
    \includegraphics[width=0.5\columnwidth]{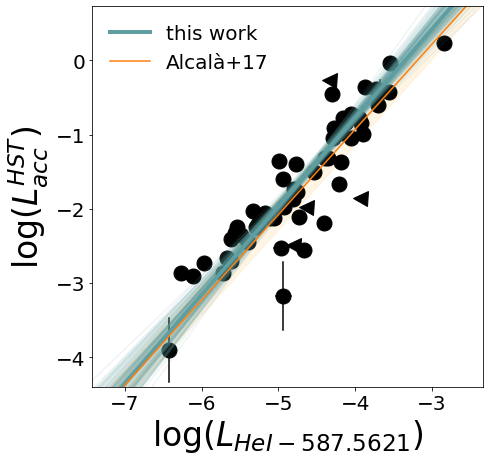}
    \includegraphics[width=0.5\columnwidth]{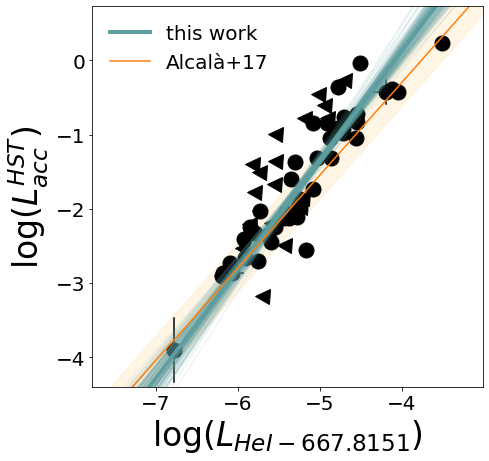}
    \includegraphics[width=0.5\columnwidth]{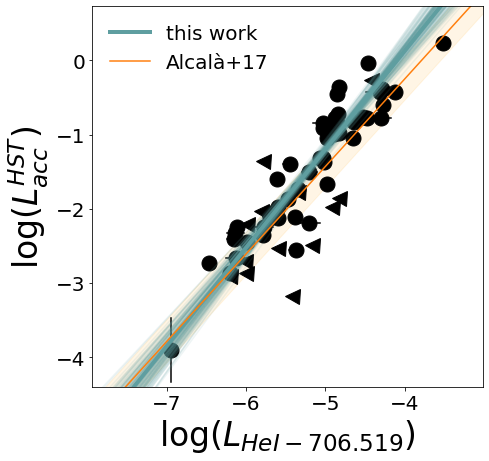}
    \includegraphics[width=0.5\columnwidth]{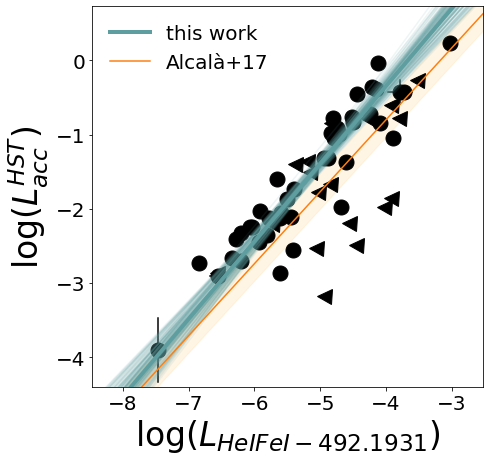}
    \includegraphics[width=0.5\columnwidth]{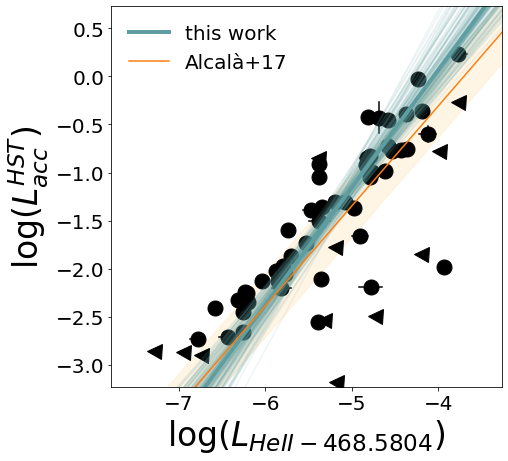}
    \includegraphics[width=0.5\columnwidth]{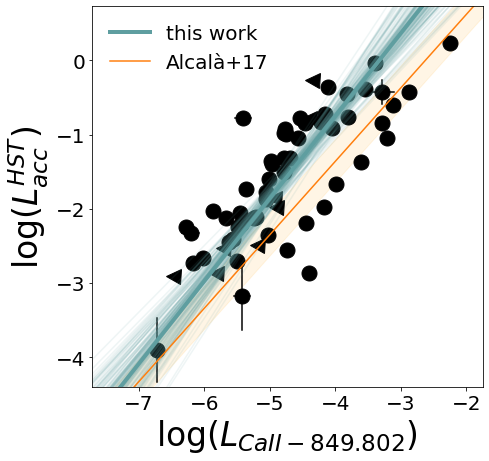}
    \includegraphics[width=0.5\columnwidth]{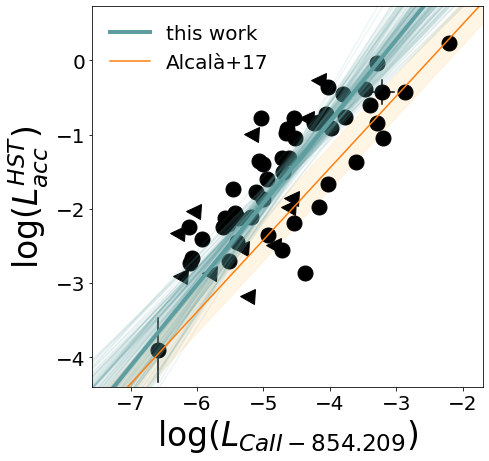}
    \includegraphics[width=0.5\columnwidth]{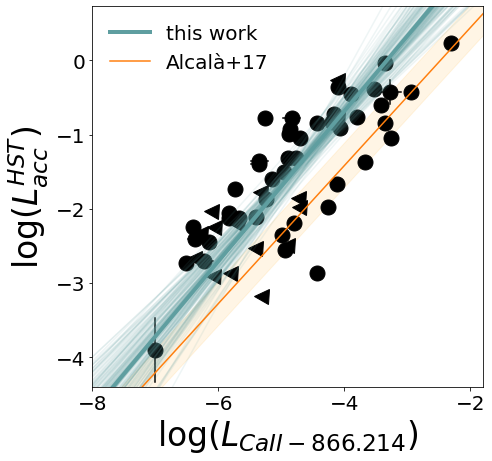}
    \includegraphics[width=0.5\columnwidth]{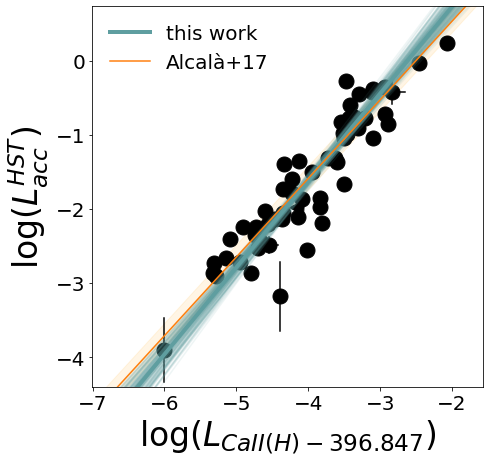}
    \includegraphics[width=0.5\columnwidth]{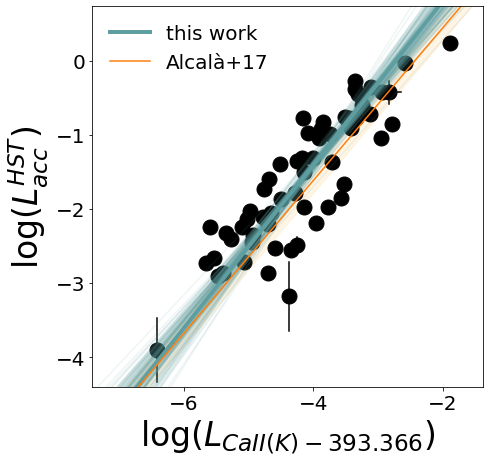}
    \includegraphics[width=0.5\columnwidth]{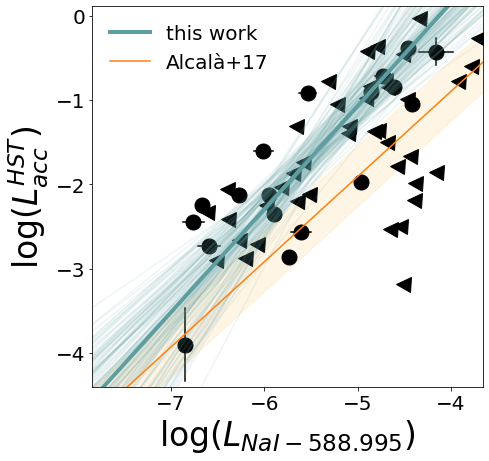}
    \includegraphics[width=0.5\columnwidth]{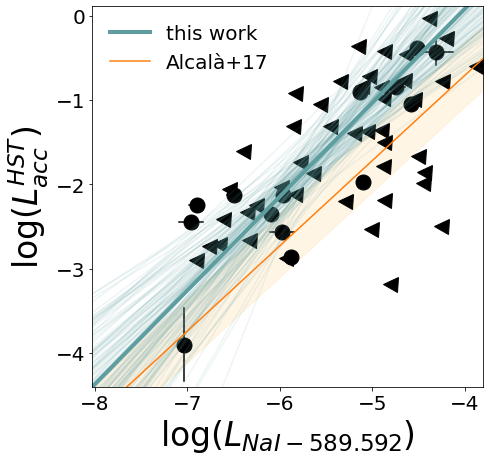}
    \includegraphics[width=0.5\columnwidth]{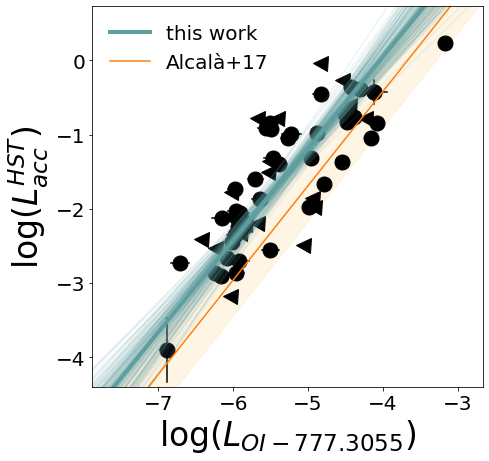}
    \includegraphics[width=0.5\columnwidth]{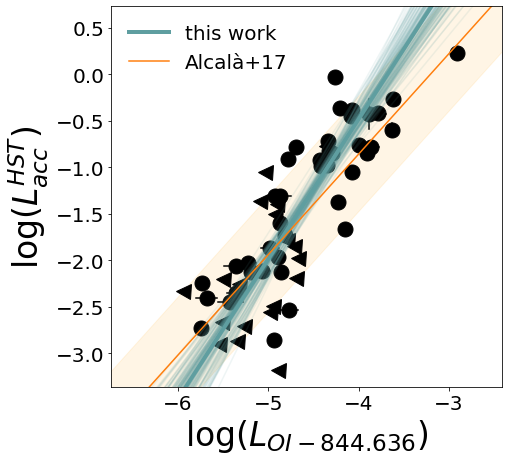}
    \caption{$\log \lacc - \log \lline$ and best fit for the Helium, Calcium, Sodium, and Oxygen accretion tracers. The blue and orange lines show out best fit and the best fit from \citet{alc17}, respectively. The wavelength in nanometers of each line is shown on the x-axis.}
    \label{fig:new-rel-other}
\end{figure*}

Figs.\ref{fig:new-rel-hydrogen}\,and\,\ref{fig:new-rel-other} show the relationships between the accretion luminosity and the line luminosity for several accretion tracers as labelled in each panel. 

For each source, we dereddened the observed fluxes of X-Shooter accretion tracers \citep[as listed in Tab.\,B.1 of][]{alc17} using extinction values derived from the HST-fit (A$_V^{HST}$, see Tab.\,\ref{tab:Lacc-Av-varie1}) and applying the \citet{car89} extinction law with $R_V = 3.1$. 
From the dereddened fluxed, the line luminosity of each accretion tracer was then computed as $L_i =4\pi d^2 F_i$, where $d$ is the distance of each source as listed in Tab.\,\ref{tab:sample1}. 

Each line luminosity was then plotted as a function of the accretion luminosity derived from HST (see Figs.\ref{fig:new-rel-hydrogen}\,and\,\ref{fig:new-rel-other}) for all the sample. 
We fit our data by employing the hierarchical Bayesian method from \citet{kel07}, which accounts for errors on both axes. To directly compare our results with the ones from \href{#label}{A17}, we do not include in the fit upper limits. 
The resulting relation is: 
\begin{equation}
\log \lacc^{i} = a_i \log \lacc^{HST} + b_i
\label{eq:fit-laccMIADM}
\end{equation}
where ``i" denotes the i-th line. Coefficients are listed in Tab.\,\ref{tab:new-coeff}. 

\begin{table*}
\centering
\caption{\label{tab:new-coeff} $\lacc - \lline$ linear fit parameters.}
\begin{tabular}{lccccccl}
\hline
\hline
Line & $\lambda_0$ [nm] & a & b & c.f. & $\sigma$  & N$_{\rm points}$ & Comments\\
\hline		
\hline						
H3(H$\alpha$)  	& 656.28   &  1.27$\pm$0.08  &   2.37$\pm$0.19  &  0.9 &   0.11$\pm$0.02 & 59 &		 \\
H4(H$\beta$)   	& 486.13   &  1.17$\pm$0.06  &   2.95$\pm$0.18  &  0.9 &   0.08$\pm$0.02 & 58 &         \\
H5(H$\gamma$)  	& 434.05   &  1.14$\pm$0.06  &   3.01$\pm$0.18  &  0.9 &   0.08$\pm$0.02 & 54 &         \\
H6(H$\delta$)  	& 410.17   &  1.13$\pm$0.06  &   3.10$\pm$0.18  &  1.0 &   0.07$\pm$0.02 & 54 &         \\
H7(H$\epsilon$) 	& 397.01   &  1.18$\pm$0.06  &   3.08$\pm$0.19  &  0.9 &   0.09$\pm$0.02 & 60 & blended \\
H8  			& 388.90   &  1.14$\pm$0.06  &   3.27$\pm$0.20  &  0.9 &   0.08$\pm$0.02 & 56 &         \\
H9  			& 383.54   &  1.09$\pm$0.05  &   3.15$\pm$0.18  &  0.9 &   0.08$\pm$0.02 & 56 &         \\
H10 			& 379.79   &  1.09$\pm$0.05  &   3.29$\pm$0.19  &  0.9 &   0.08$\pm$0.02 & 56 &         \\
H11 			& 377.06   &  1.13$\pm$0.07  &   3.52$\pm$0.24  &  0.9 &   0.11$\pm$0.02 & 58 &         \\
H12 			& 375.02   &  1.06$\pm$0.06  &   3.40$\pm$0.20  &  0.9 &   0.09$\pm$0.02 & 53 &         \\
H13 			& 373.44   &  1.01$\pm$0.06  &   3.27$\pm$0.21  &  0.9 &   0.11$\pm$0.02 & 51 &         \\
H14 			& 372.19   &  0.98$\pm$0.06  &   3.27$\pm$0.21  &  0.9 &   0.12$\pm$0.03 & 51 &         \\
H15 			& 371.20   &  1.08$\pm$0.13  &   3.68$\pm$0.49  &  0.8 &   0.49$\pm$0.10 & 58 &         \\
Pa5(Pa$\beta$) 	& 1281.81  &  1.20$\pm$0.09  &   3.68$\pm$0.31  &  0.9 &   0.14$\pm$0.03 & 51 &         \\
Pa6(Pa$\gamma$) & 1093.81  &  1.20$\pm$0.08  &   3.80$\pm$0.28  &  0.9 &   0.12$\pm$0.02 & 56 &         \\
Pa7(Pa$\delta$) & 1004.94  &  1.38$\pm$0.11  &   4.71$\pm$0.40  &  0.9 &   0.11$\pm$0.03 & 50 &         \\
Pa8  			& 954.60   &  1.22$\pm$0.11  &   4.34$\pm$0.41  &  0.9 &   0.14$\pm$0.03 & 48 &         \\
Pa9  			& 922.90   &  1.25$\pm$0.09  &   4.55$\pm$0.35  &  0.9 &   0.13$\pm$0.03 & 56 &         \\
Pa10 			& 901.49   &  1.46$\pm$0.13  &   5.70$\pm$0.53  &  0.9 &   0.13$\pm$0.03 & 49 &         \\
Br7(Br$\gamma$) & 2166.12  &  1.08$\pm$0.11  &   4.18$\pm$0.46  &  0.9 &   0.24$\pm$0.06 & 45 &         \\
HeI     		& 402.62   &  1.15$\pm$0.11  &   4.22$\pm$0.43  &  0.8 &   0.28$\pm$0.06 & 61 &         \\
HeI     		& 447.15   &  1.17$\pm$0.07  &   4.24$\pm$0.29  &  0.9 &   0.10$\pm$0.02 & 55 &         \\
HeI     		& 471.31   &  1.28$\pm$0.13  &   5.43$\pm$0.56  &  0.8 &   0.20$\pm$0.05 & 53 & blended \\
HeIFeI  		& 492.19   &  1.05$\pm$0.07  &   3.85$\pm$0.28  &  0.9 &   0.13$\pm$0.03 & 43 &         \\
HeI     		& 501.57   &  0.88$\pm$0.07  &   3.09$\pm$0.29  &  0.9 &   0.23$\pm$0.06 & 38 &         \\
HeI     		& 587.56   &  1.19$\pm$0.07  &   4.01$\pm$0.25  &  0.9 &   0.08$\pm$0.02 & 57 &         \\
HeI     		& 667.82   &  1.47$\pm$0.09  &   6.01$\pm$0.39  &  0.9 &   0.06$\pm$0.01 & 43 &         \\
HeI     		& 706.52   &  1.33$\pm$0.08  &   5.46$\pm$0.35  &  0.9 &   0.06$\pm$0.02 & 48 &         \\
HeII    		& 468.58   &  1.21$\pm$0.11  &   4.91$\pm$0.49  &  0.9 &   0.18$\pm$0.04 & 49 &         \\
CaII(K) 		& 393.37   &  1.10$\pm$0.08  &   3.01$\pm$0.25  &  0.9 &   0.19$\pm$0.04 & 61 &         \\
CaII(H) 		& 396.85   &  1.18$\pm$0.06  &   3.09$\pm$0.19  &  0.9 &   0.09$\pm$0.02 & 60 & blended \\
CaII 			& 849.80   &  1.11$\pm$0.13  &   3.71$\pm$0.47  &  0.8 &   0.38$\pm$0.08 & 53 &         \\
CaII 			& 854.21   &  1.10$\pm$0.13  &   3.57$\pm$0.45  &  0.8 &   0.34$\pm$0.08 & 49 &         \\
CaII 			& 866.21   &  1.01$\pm$0.12  &   3.37$\pm$0.45  &  0.9 &   0.46$\pm$0.11 & 46 &         \\
NaI  			& 589.00   &  1.23$\pm$0.25  &   5.07$\pm$1.09  &  0.8 &   0.32$\pm$0.14 & 18 &         \\
NaI  			& 589.59   &  1.14$\pm$0.26  &   4.65$\pm$1.14  &  0.8 &   0.38$\pm$0.26 & 14 &         \\
OI   			& 777.31   &  1.24$\pm$0.12  &   5.02$\pm$0.51  &  0.9 &   0.19$\pm$0.05 & 43 & blended \\
OI   			& 844.64   &  1.49$\pm$0.15  &   5.49$\pm$0.58  &  0.9 &   0.12$\pm$0.03 & 41 &         \\ 
\hline 
\hline
\end{tabular}
\begin{quotation}
  \textbf{Notes.} We used $\lacc$ as drawn from the HST modelling of the HST-ULLYSES data and $\lline$ obtained dereddening by $A_V^{HST}$ the observed flux of the accretion tracers detected in the PENELLOPE sample. $\lambda_0$ is the central wavelength of the line; $a$ and $b$ coefficients are the best fit of Eq.\,\ref{eq:fit-laccMIADM}; c.f. stands for correlation factor between $\lacc$ and $\lline$, $\sigma$ is the spread of the distribution, and N$_{\rm points}$ is the number of points for the fit over the total sample.
  \end{quotation}  
 \end{table*}

\section{$\lacc^{\rm lines}$ as a function of wavelength for the PENELLOPE sample}
\label{app:logLacclogLambda}
\begin{figure*}
    \includegraphics[width=0.33\textwidth]{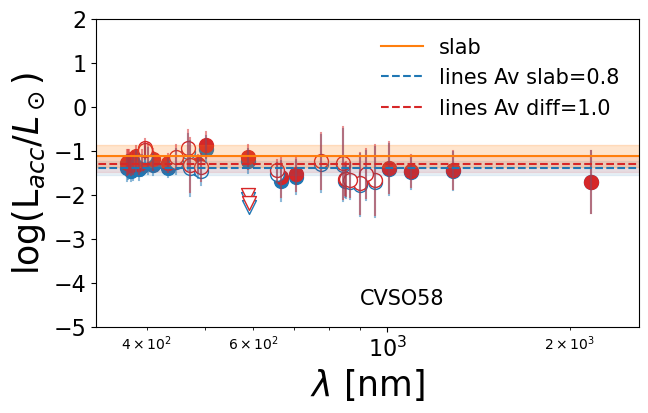}
    \includegraphics[width=0.33\textwidth]{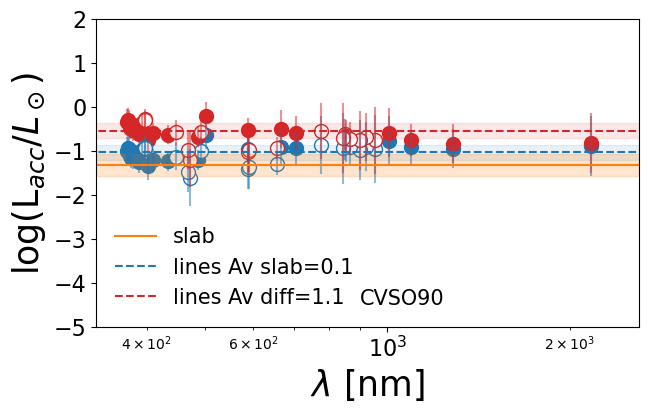}
    \includegraphics[width=0.33\textwidth]{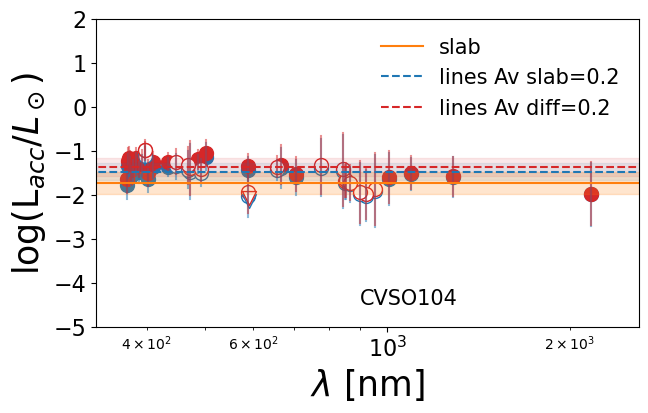}
    \includegraphics[width=0.33\textwidth]{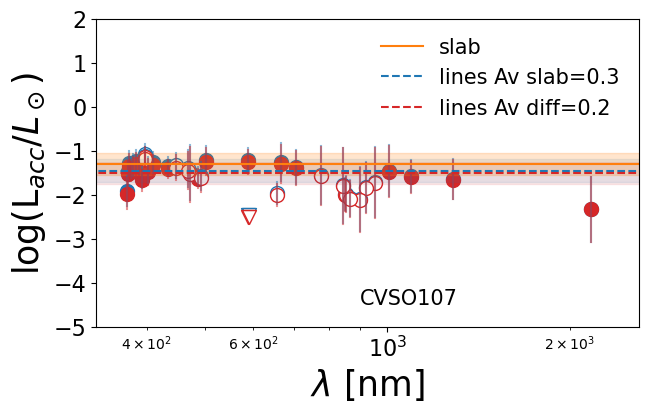}
    \includegraphics[width=0.33\textwidth]{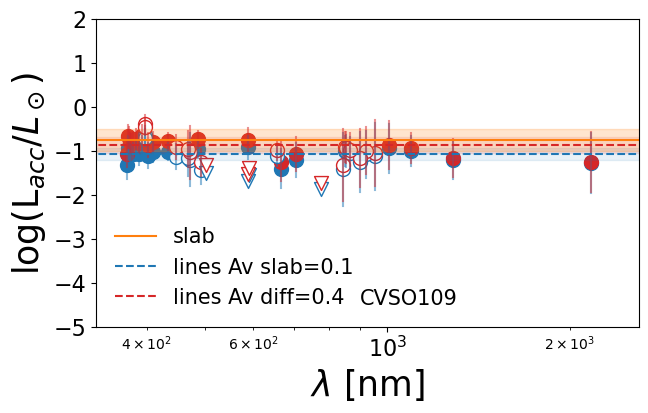}
    \includegraphics[width=0.33\textwidth]{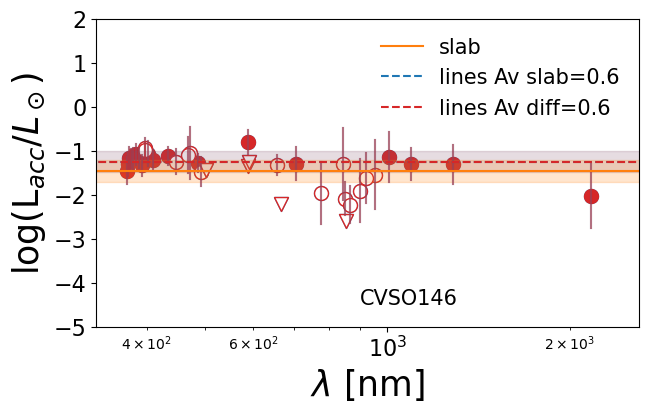}
    \includegraphics[width=0.33\textwidth]{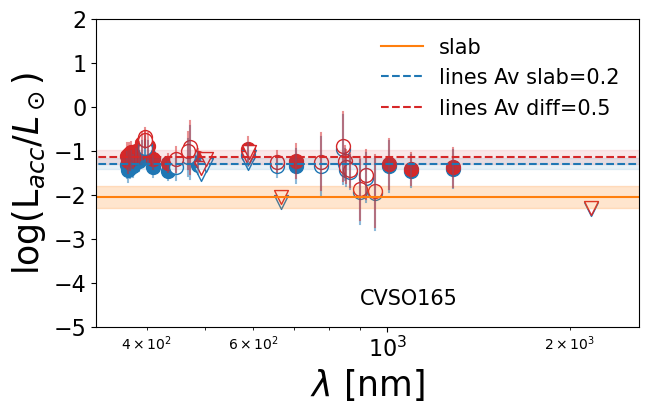}
    \includegraphics[width=0.33\textwidth]{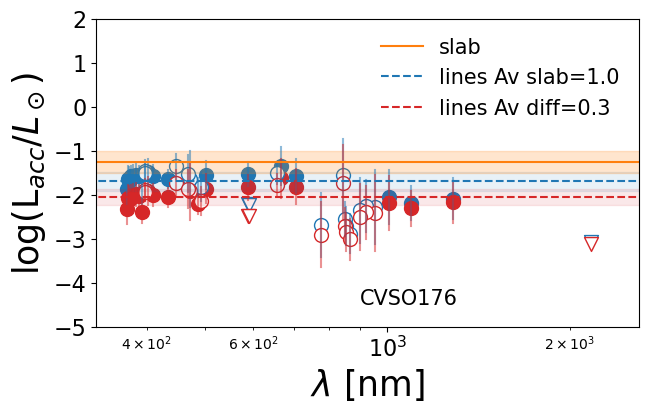}
    \includegraphics[width=0.33\textwidth]{sigmaOri_RigliacoManara_plot_diffLaccSO1153.png}
    \includegraphics[width=0.33\textwidth]{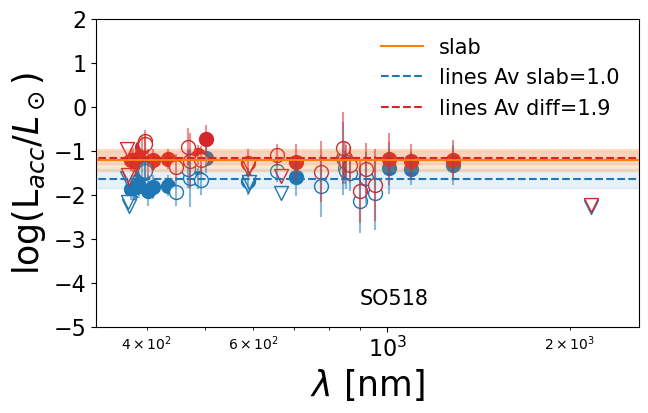}
    \includegraphics[width=0.33\textwidth]{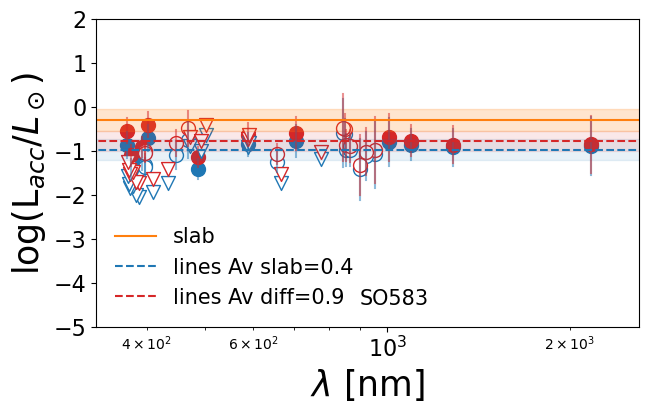}
    \includegraphics[width=0.33\textwidth]{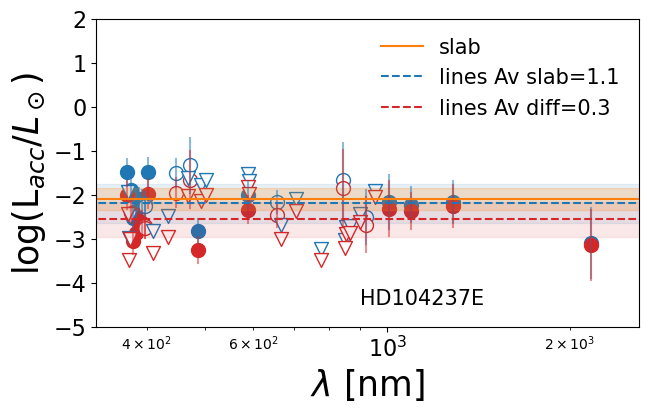}
    \includegraphics[width=0.33\textwidth]{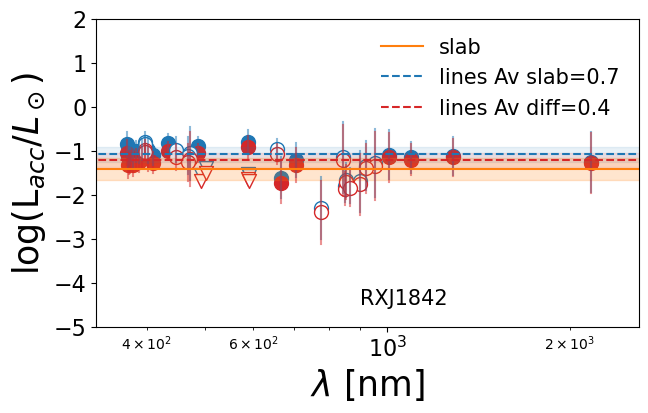}
    \includegraphics[width=0.33\textwidth]{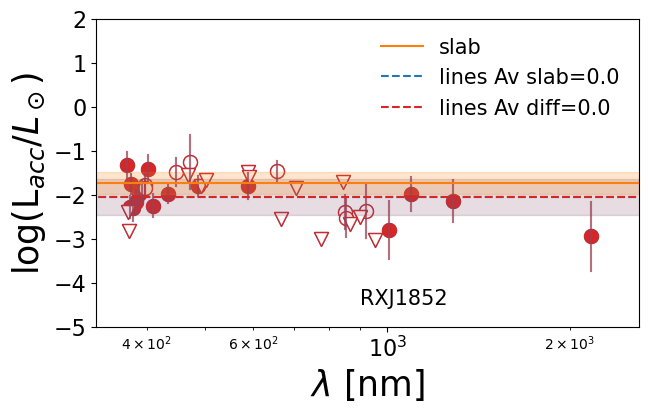}

    \caption{$\log \lacc$ as a function of the wavelength for CTTS in the OB1, $\sigma$Ori, $\epsilon$Cha and Corona\,Australis star forming regions. The orange line represents the $\log \lacc^{XS}$ computed using the XS-fit method and its associated extinction value ($A_V^{XS}$). 
The blue dots show $\log \lacc^{lines}$ (Tab.\,\ref{tab:sample1}) calculated from the accretion tracers using the empirical relations from \citet{alc14}, with fluxes dereddened using $A_V^{XS}$.  
Similarly, the red dots represent $\lacc^{lines-diff}$, computed from the same accretion tracers but with fluxes dereddened using $A_V^{diff}$ (see Sect.\,\ref{sect:Av} and Tab.\,\ref{tab:Lacc-Av-varie1}).
The blue and red dashed lines indicate the mean $\log \lacc$ values for the accretion tracers $\log L_{{\rm acc}, i}$ in their respective colors.}
    \label{fig:RMplot-ob1}
\end{figure*}

\begin{figure*}
    \includegraphics[width=0.33\textwidth]{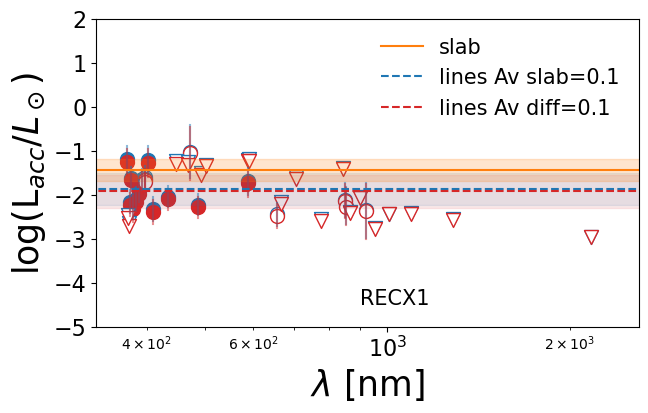}
    \includegraphics[width=0.33\textwidth]{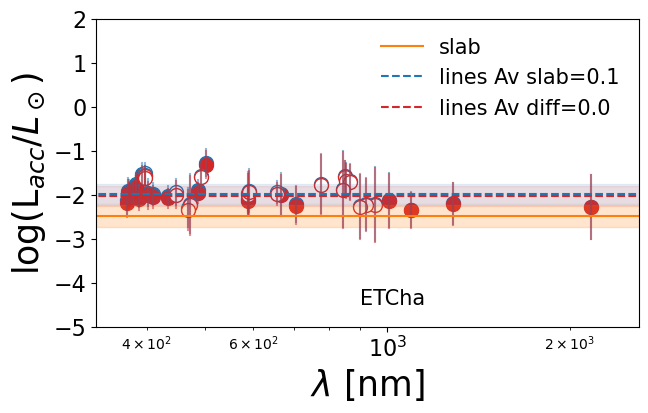}
    \includegraphics[width=0.33\textwidth]{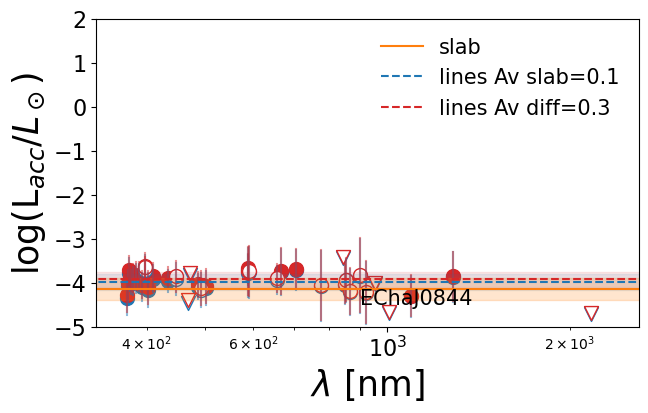}
    \includegraphics[width=0.33\textwidth]{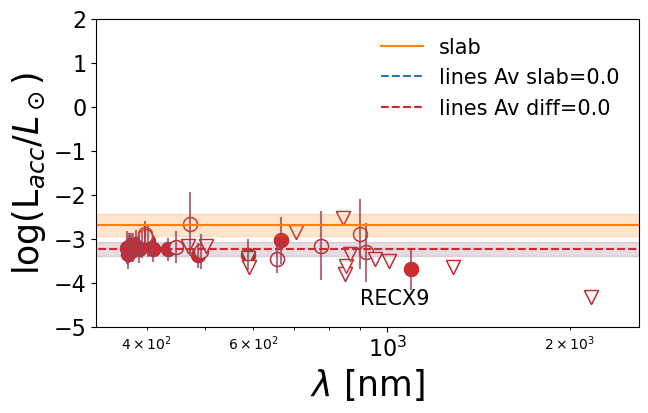}
    \includegraphics[width=0.33\textwidth]{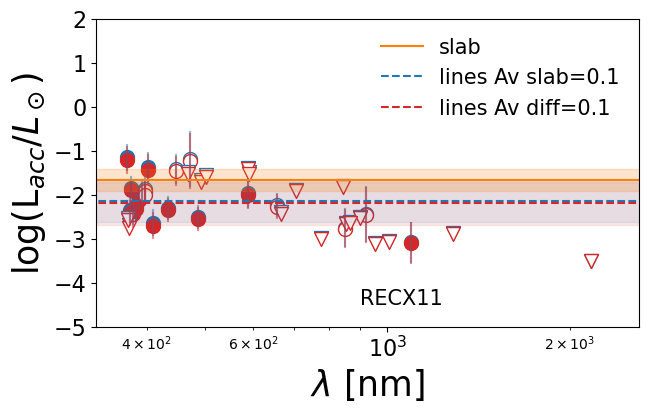}
    \includegraphics[width=0.33\textwidth]{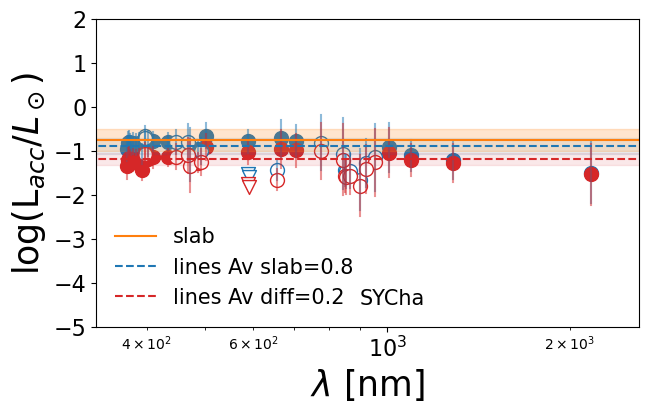}
    \includegraphics[width=0.33\textwidth]{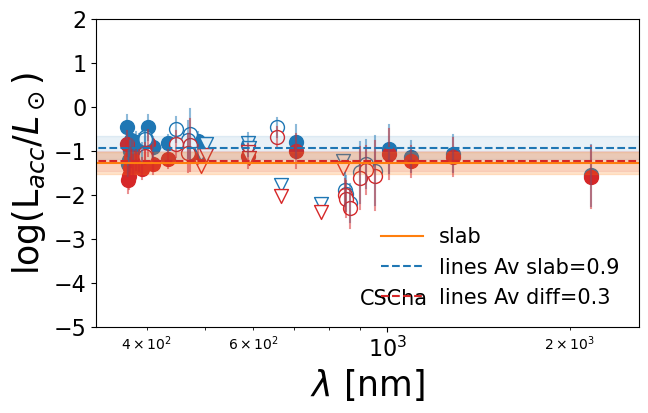}
    \includegraphics[width=0.33\textwidth]{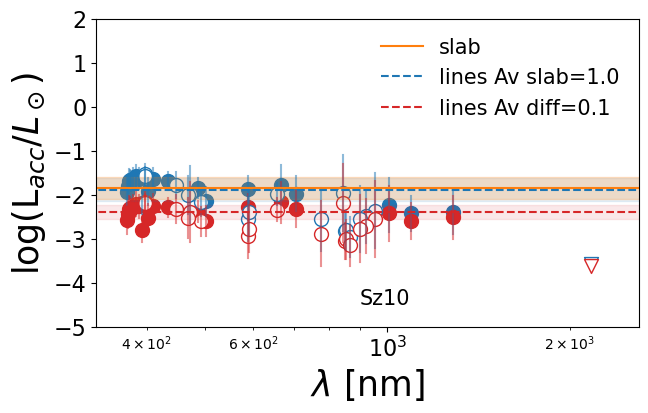}
    \includegraphics[width=0.33\textwidth]{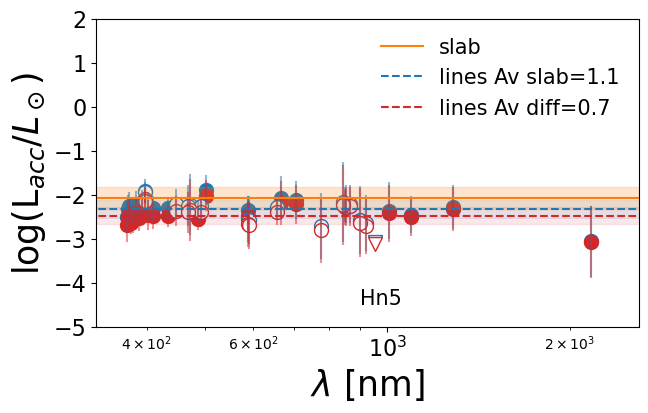}
    \includegraphics[width=0.33\textwidth]{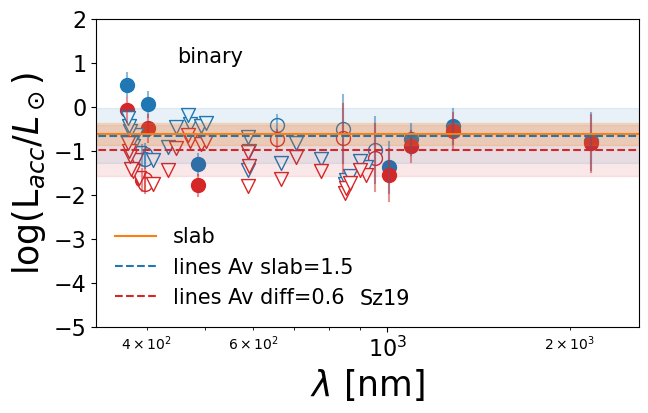}
    \includegraphics[width=0.33\textwidth]{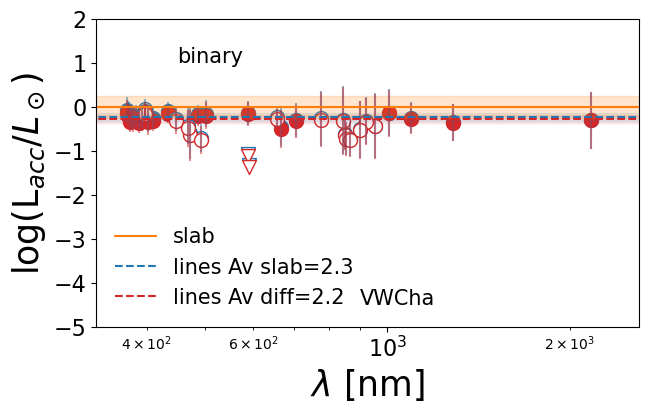}
    \includegraphics[width=0.33\textwidth]{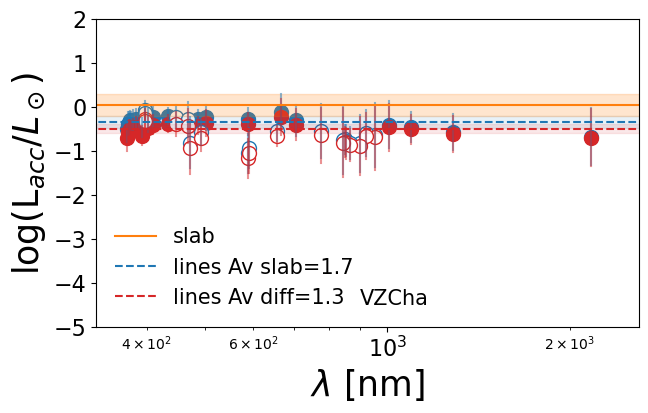}
    \includegraphics[width=0.33\textwidth]{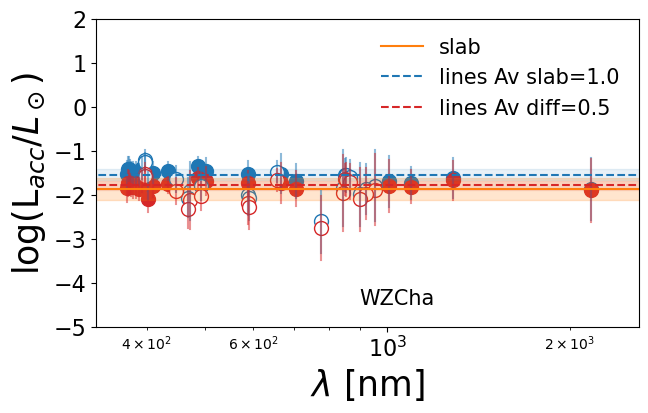}
    \includegraphics[width=0.33\textwidth]{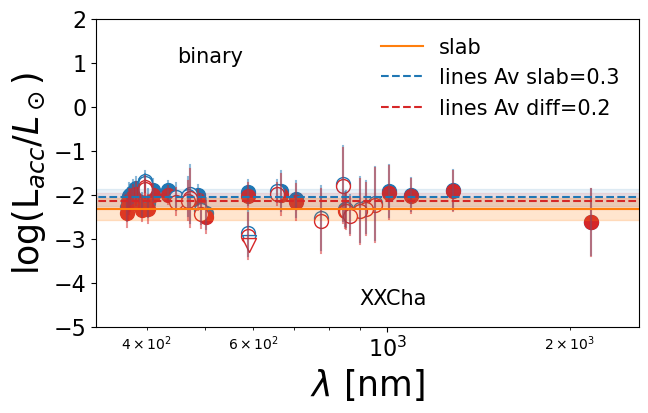}
    \includegraphics[width=0.33\textwidth]{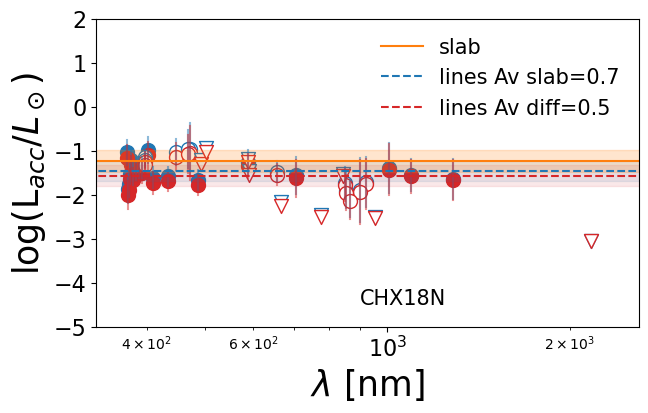}
    \includegraphics[width=0.33\textwidth]{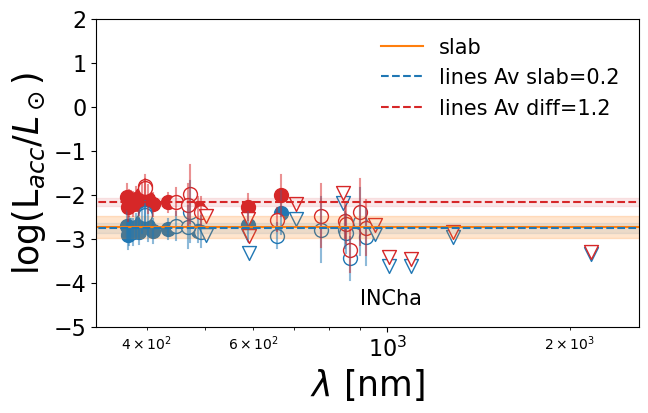}
    \includegraphics[width=0.33\textwidth]{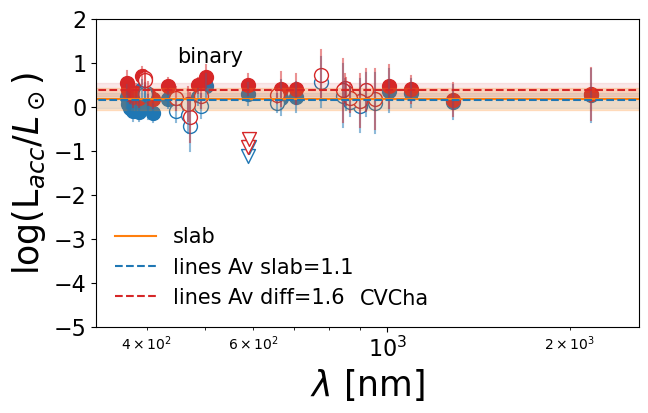}
    \includegraphics[width=0.33\textwidth]{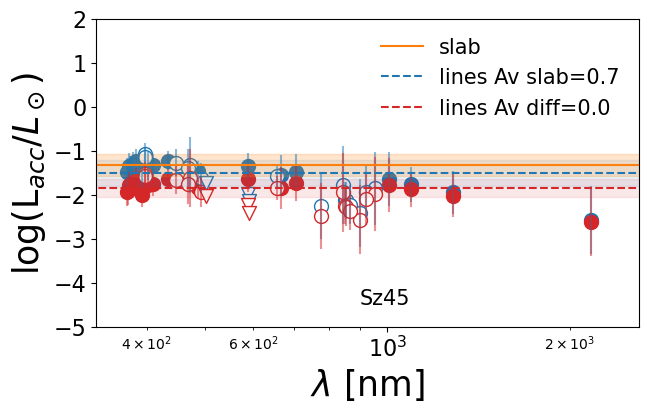}
    \caption{$\log \lacc$ as a function of the wavelength for CTTS in the $\eta$Cha and Chamaeleon\,I star forming regions. Symbols are as in Fig.\,\ref{fig:RMplot-ob1}.}
    \label{fig:RMplot-chaI2}
\end{figure*}

\begin{figure*}
    \includegraphics[width=0.33\textwidth]{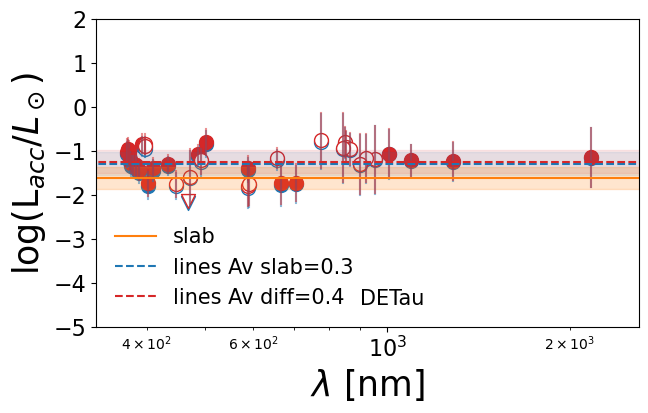}
    \includegraphics[width=0.33\textwidth]{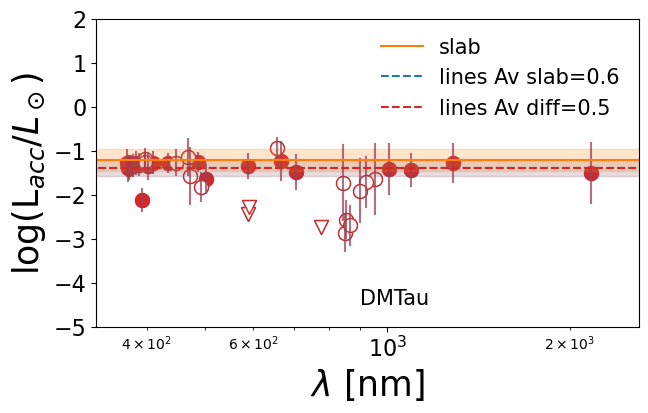}
    \includegraphics[width=0.33\textwidth]{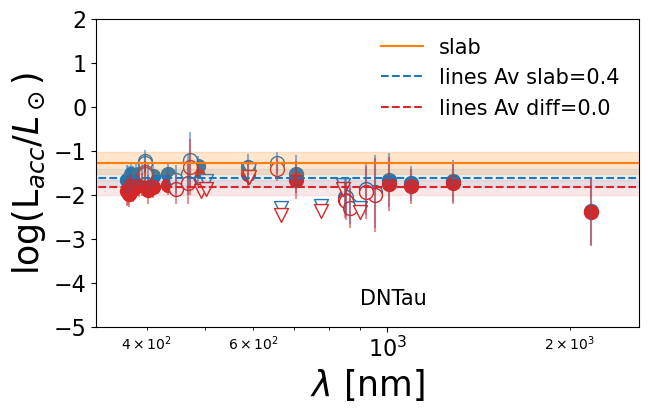}
    \includegraphics[width=0.33\textwidth]{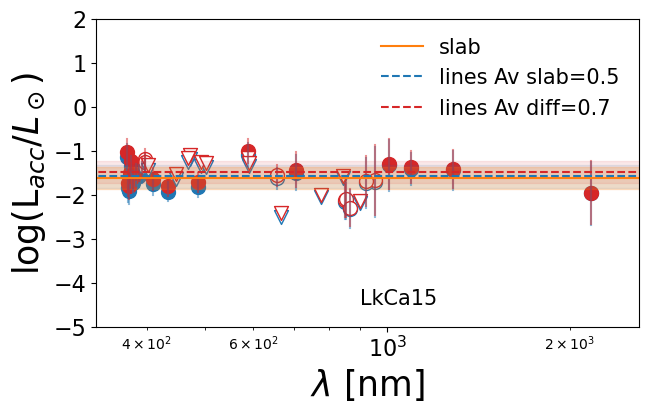}
    \includegraphics[width=0.33\textwidth]{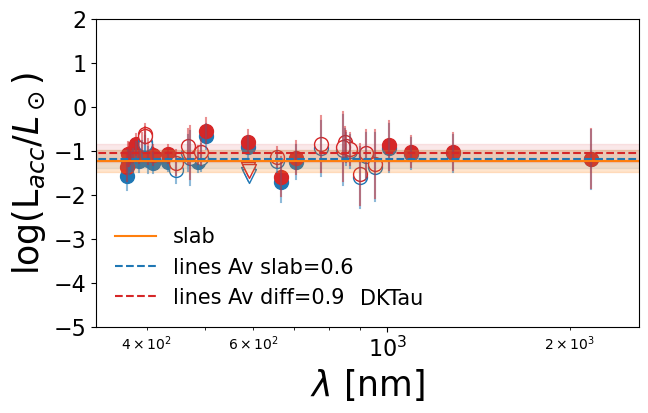}
    \includegraphics[width=0.33\textwidth]{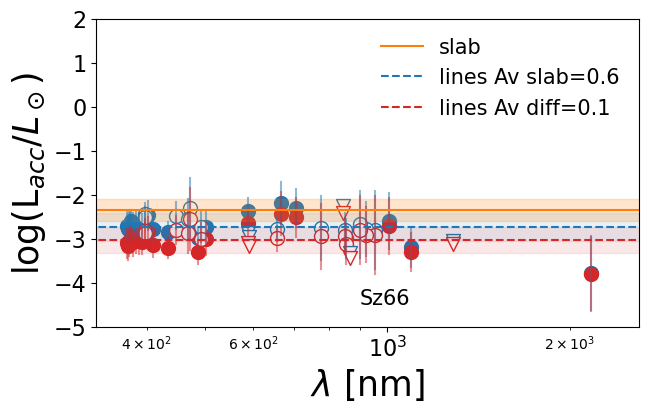}
    \includegraphics[width=0.33\textwidth]{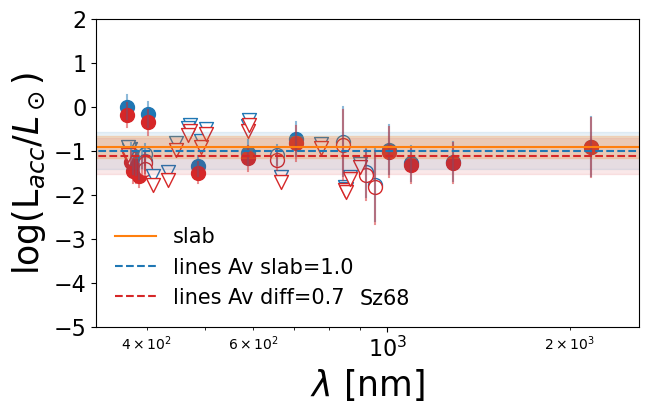}
    \includegraphics[width=0.33\textwidth]{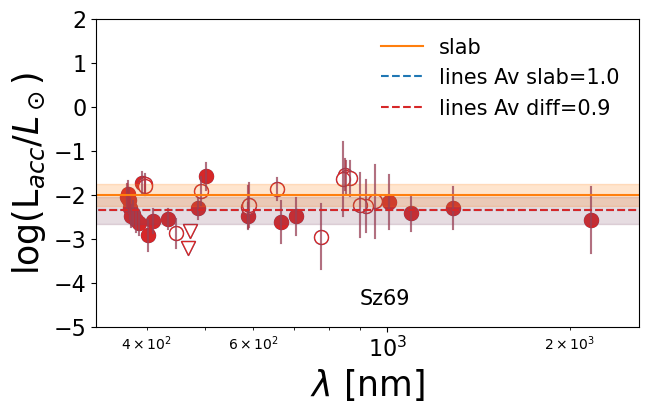}
    \includegraphics[width=0.33\textwidth]{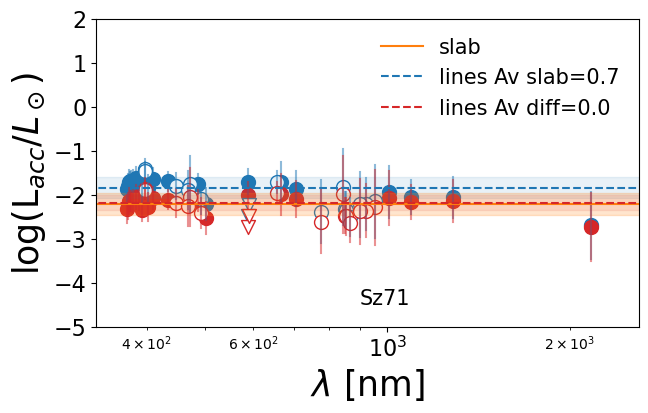}
    \includegraphics[width=0.33\textwidth]{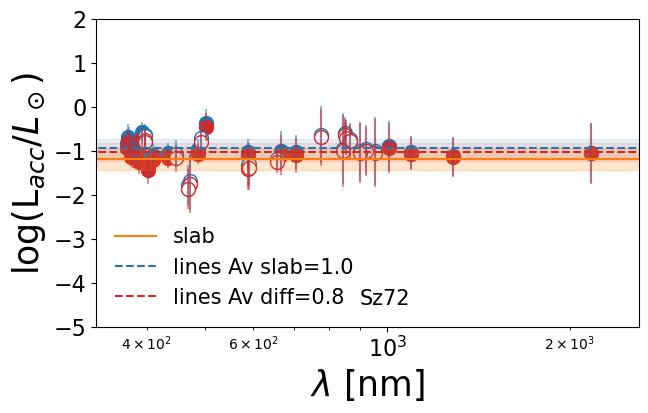}
    \includegraphics[width=0.33\textwidth]{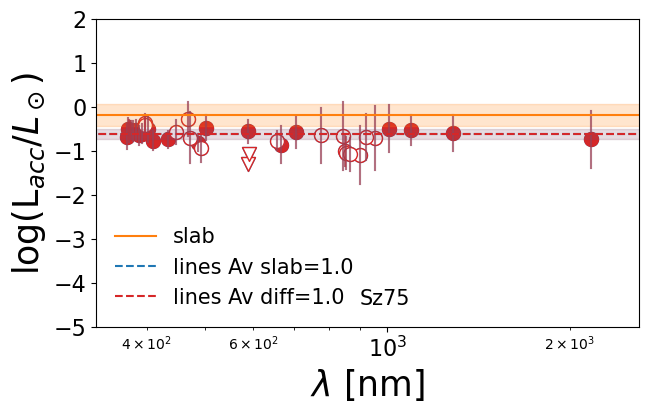}
    \includegraphics[width=0.33\textwidth]{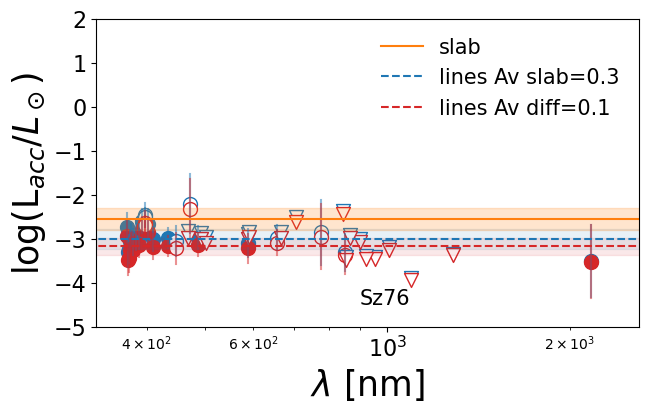}
    \includegraphics[width=0.33\textwidth]{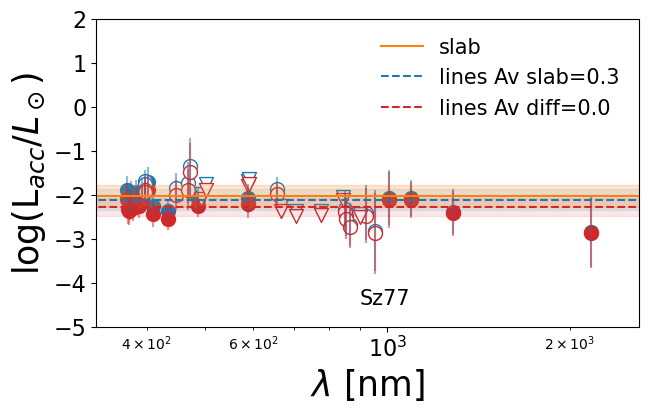}
    \includegraphics[width=0.33\textwidth]{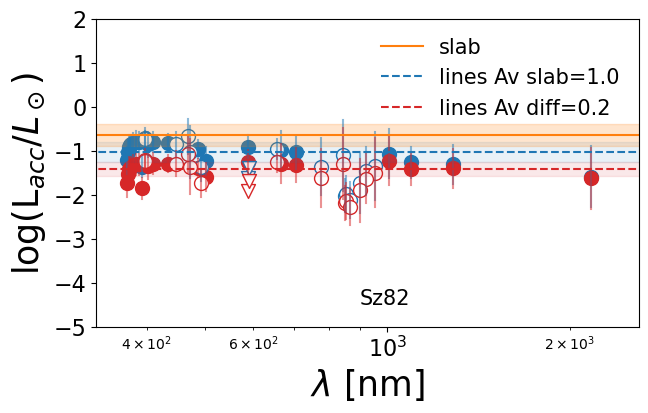}
    \includegraphics[width=0.33\textwidth]{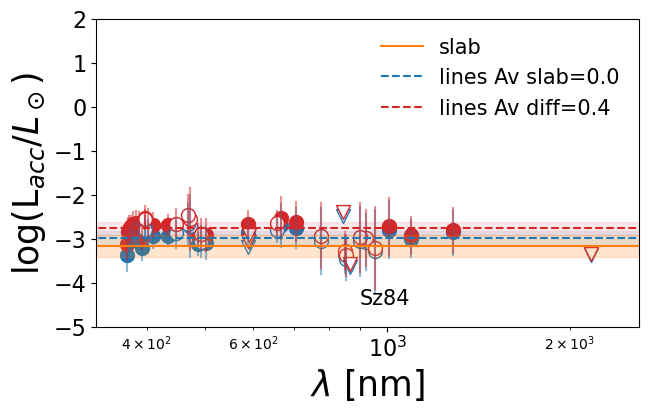}
    \includegraphics[width=0.33\textwidth]{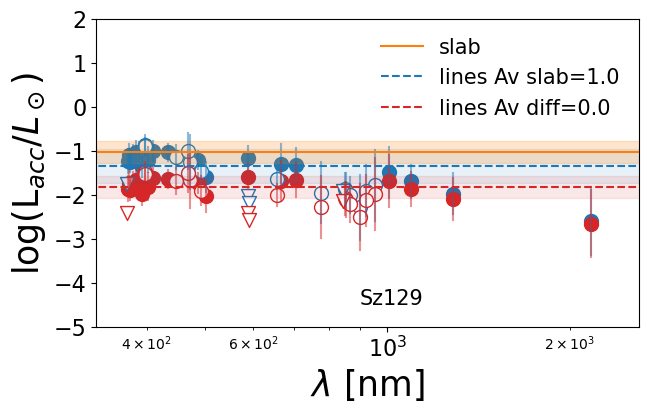}
    \includegraphics[width=0.33\textwidth]{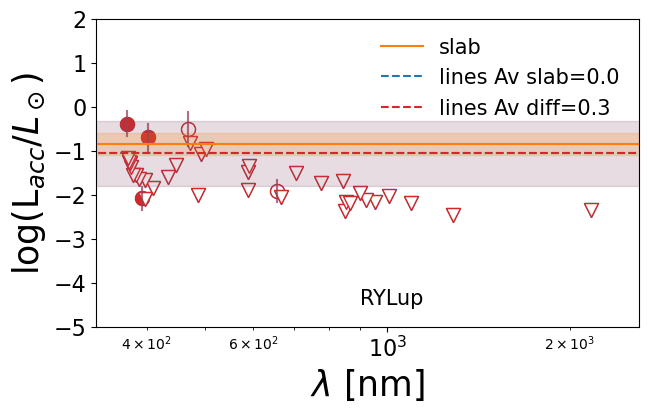}
     \includegraphics[width=0.33\textwidth]{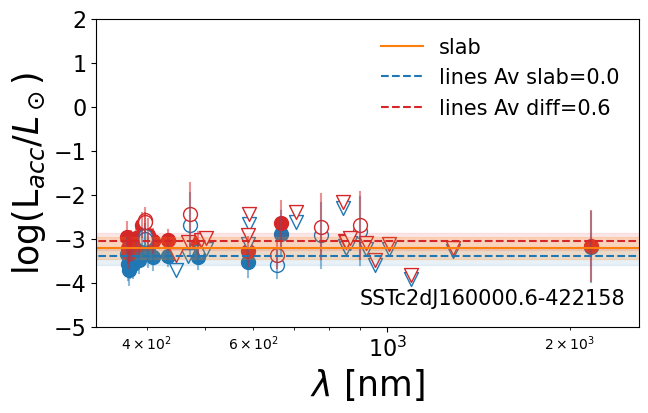}
    \caption{$\log \lacc$ as a function of the wavelength for CTTS in the Taurus\,Auriga and Lupus star forming regions.  Symbols are as in Fig.\,\ref{fig:RMplot-ob1}.}
    \label{fig:RMplot-taurus}
\end{figure*}

\begin{figure*}
       
    \includegraphics[width=0.33\textwidth]{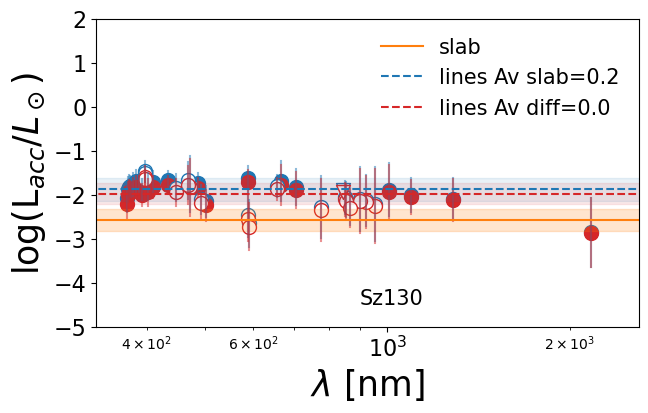} 
    \includegraphics[width=0.33\textwidth]{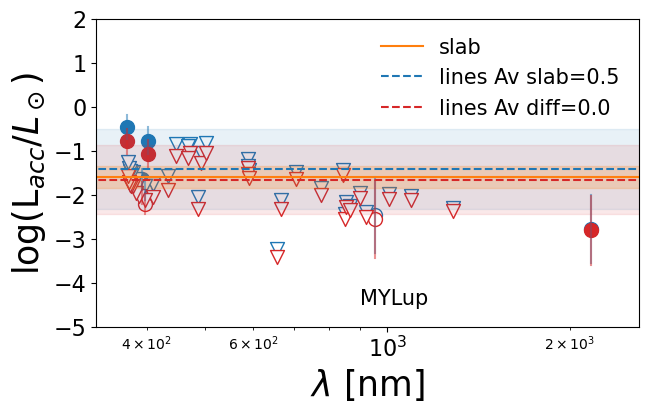}
    \includegraphics[width=0.33\textwidth]{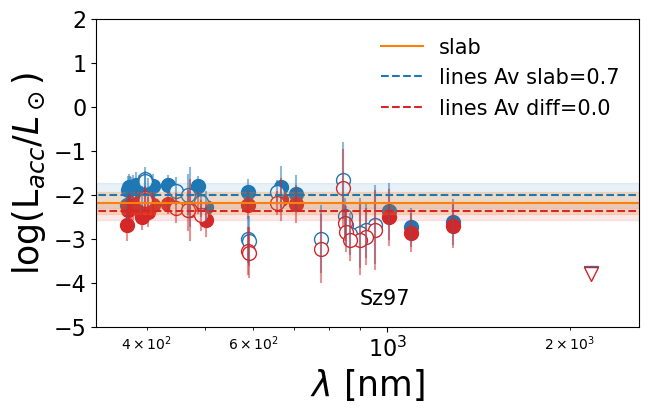}   \includegraphics[width=0.33\textwidth]{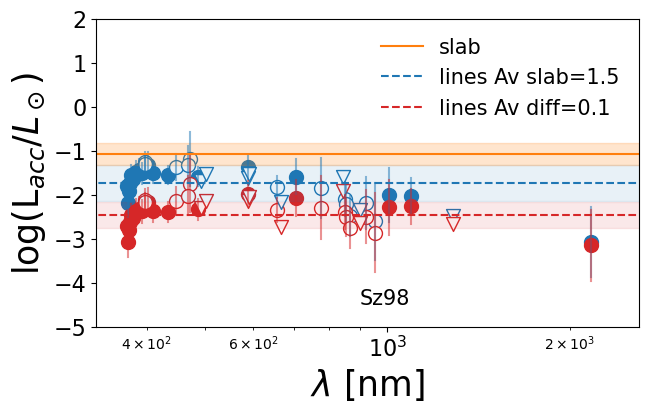}
    \includegraphics[width=0.33\textwidth]{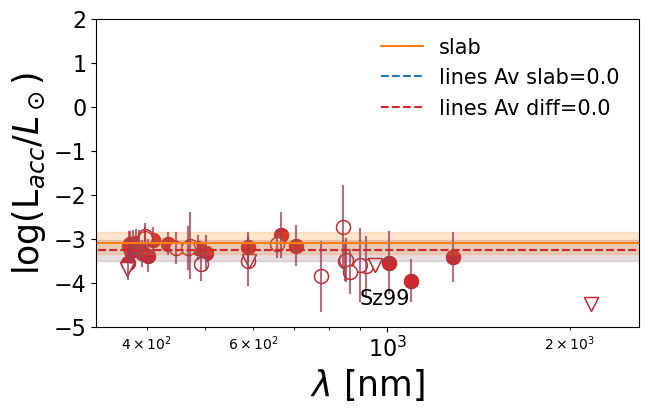}
    \includegraphics[width=0.33\textwidth]{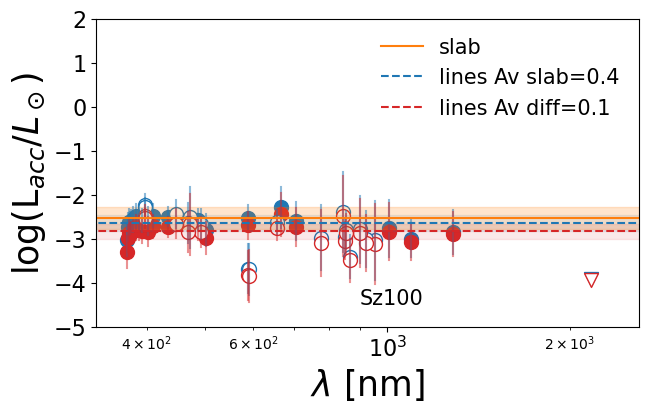}
    \includegraphics[width=0.33\textwidth]{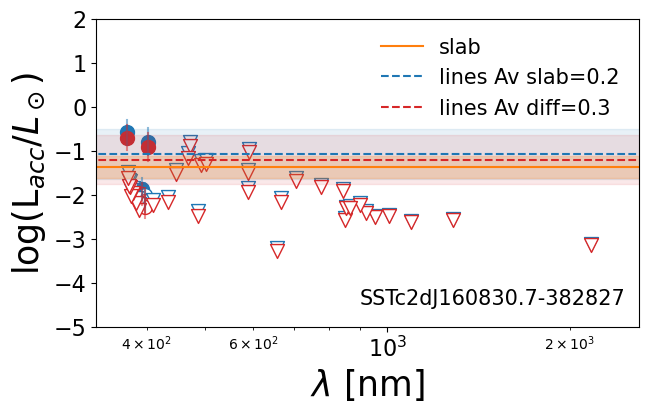}
    \includegraphics[width=0.33\textwidth]{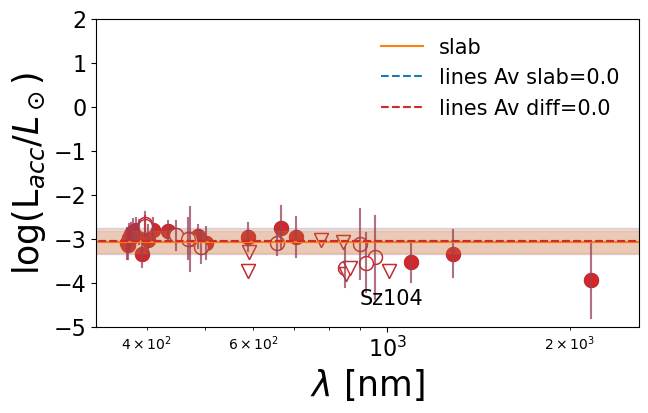}
    \includegraphics[width=0.33\textwidth]{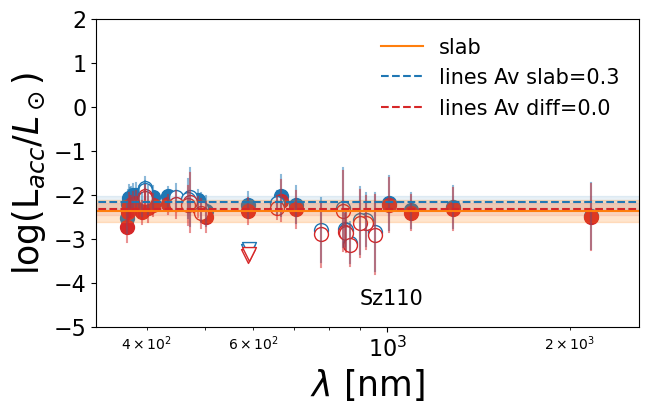}
    \includegraphics[width=0.33\textwidth]{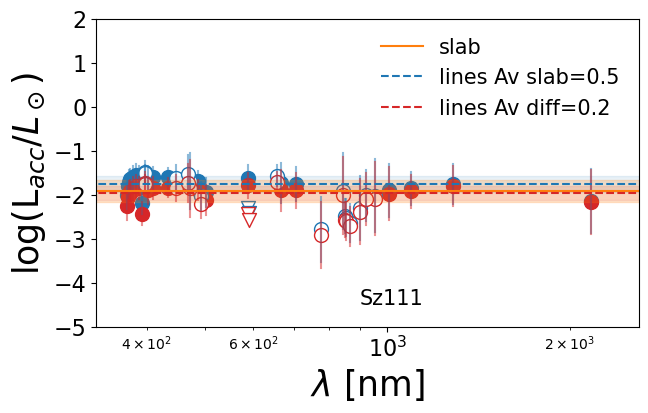}
    \includegraphics[width=0.33\textwidth]{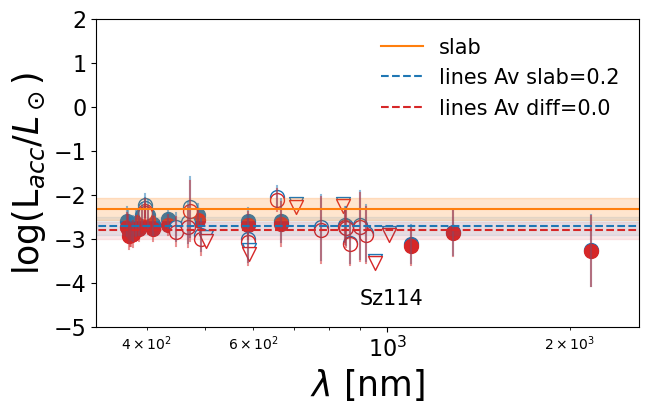}
    \includegraphics[width=0.33\textwidth]{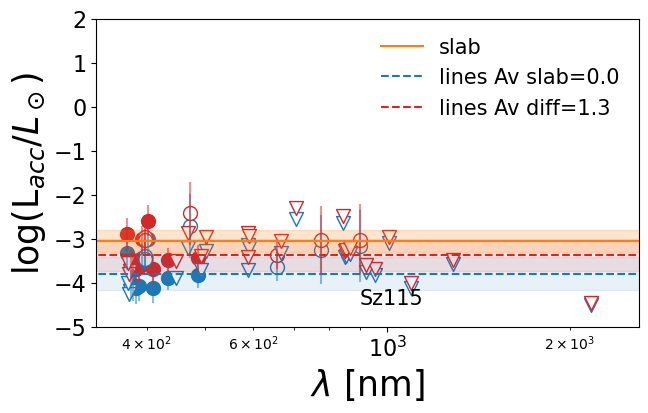}
    \includegraphics[width=0.33\textwidth]{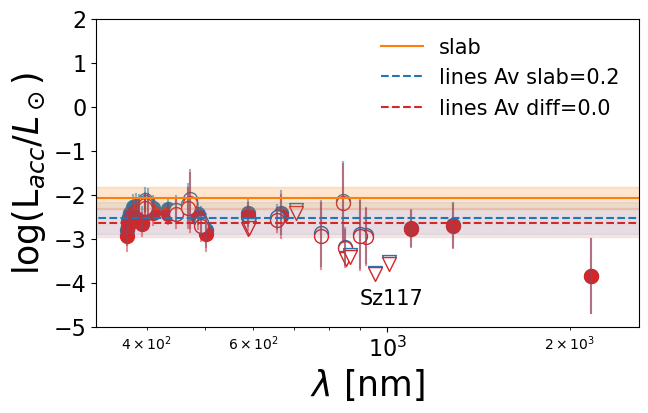}
    \includegraphics[width=0.33\textwidth]{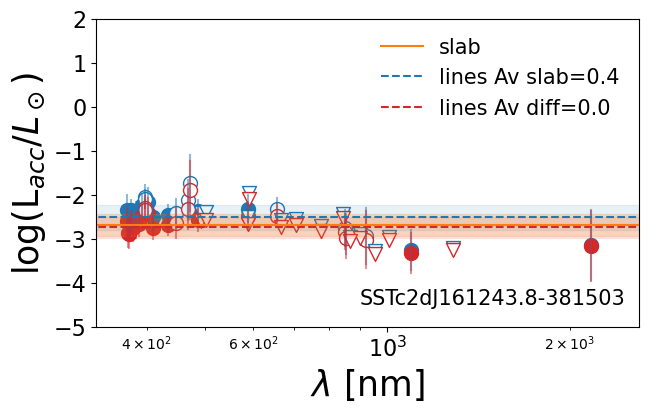}
    \includegraphics[width=0.33\textwidth]{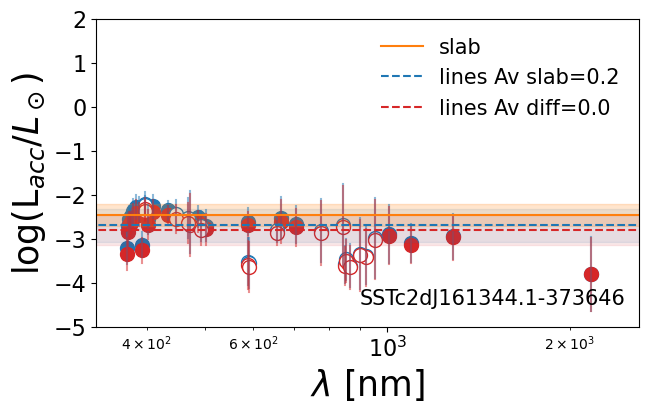}
    \caption{$\log \lacc$ as a function of the wavelength for CTTS in the Lupus star forming region.  Symbols are as in Fig.\,\ref{fig:RMplot-ob1}.}
    \label{fig:RMplot-lupus}
\end{figure*}

Figures\,\ref{fig:RMplot-ob1} to \ref{fig:RMplot-lupus} display the accretion luminosity as a function of the wavelength for every target in the PENELLOPE sample. More in detail, the horizontal orange line represent the $\log \lacc^{XS}$ value; the blue dots correspond to the $\log \lacc^{\rm line, i}$ dereddening the fluxes with $A_V^{XS}$ and using the \href{#label}{A17} empirical relations and the horizontal blue line is their average value; the red dots correspond to the $\log \lacc^{\rm line, i}$ dereddening the fluxes with $A_V^{diff}$ and using the \href{#label}{A17} empirical relations and the horizontal red line is their average value.

The sources can be classified into three groups based on plots in Figs.\,\ref{fig:RMplot-ob1} to \ref{fig:RMplot-lupus}. 
The largest group, comprising 80\% (51 out of 64) of our sources, shows agreement between the three accretion luminosities (see the caption in Fig.\,\ref{fig:RMplot-ob1}) within the errors. For these sources, the extinction value derived from the lines is very similar to the $A_V^{XS}$, and, as a consequence, the different methods for determining $\lacc$ are consistent.

The second group, representing 15\% (10 sources), shows a significant difference (up to 1\,mag) between $A_V^{diff}$ and $A_V^{XS}$. $\lacc^{XS}$ and $\lacc^{lines}$ are in agreement within eachother, but $\lacc^{lines-diff}$ is not. 
These sources include CVSO\,90, CVSO\,176, SO\,1153, IN\,Cha, Sz\,10, Sz\,66, Sz\,76, Sz\,82, Sz\,98, and Sz\,129.
For IN\,Cha, no accretion tracers are detected in the NIR ($\lambda > 1\mu$m), which suggests that the discrepancy between extinction estimates may result from the limited wavelength coverage of the tracers used. In this case, fitting the $\log \lacc$ from the accretion tracers using $A_V^{XS}$ (blue dots) and $A_V^{diff}$ (red dots) shows a flat distribution in both cases, making difficult to determine which extinction estimate is more reliable. 
For other sources, the NIR accretion tracers are detected, and the slope is flatter in one case. 
For instance, in CVSO\,90, the red dots decrease toward longer wavelengths, while using A$_V^{XS}$ results in a flat slope, suggesting A$_V^{XS}$ is a better extinction estimate for this source. 
In SO\,1153 (CVSO\,176, Sz\,10 and Sz\,82), the red dots are flat, while the blue dots show an increasing (decreasing) trend toward longer wavelengths, indicating A$_V^{diff}$ is more accurate. 
For Sz\,66, Sz\,76, Sz\,98, and Sz\,129, both red and blue dots decrease from UVB to NIR, but the slope is flatter for the red dots, suggesting A$_V^{diff}$ is the better estimate for these sources. We found no difference in stellar or accretion properties between the cases where  A$_V^{diff}$ is better and those where  A$_V^{XS}$ is better.

The third group includes three sources (5\%): CVSO\,165, RECX\,9, and Sz\,130. 
For these objects $\lacc^{lines} \sim \lacc^{lines-diff}$, but $\lacc^{XS}$ is different, even though the extinction estimated from the two methods are very similar: $|{\rm A}_V^{XS} - {\rm A}_V^{diff}| = (0 - 0.3)$\,mag. 
For CVSO\,165 and Sz\,130, the red dots show a decreasing trend from UVB to NIR, while the blue dots remain flat, suggesting that the XS method provides a better extinction estimate. However, this does not fully explain the discrepancy between $\lacc^{lines}$ and $\lacc^{XS}$.
The extinction estimate of RECX\,9 is consistently 0\,mag.

\section{Plots for the Brackett lines discussed in the text.}
\label{app:Br}

\begin{figure*}
\centering

    \includegraphics[width=0.6\columnwidth]{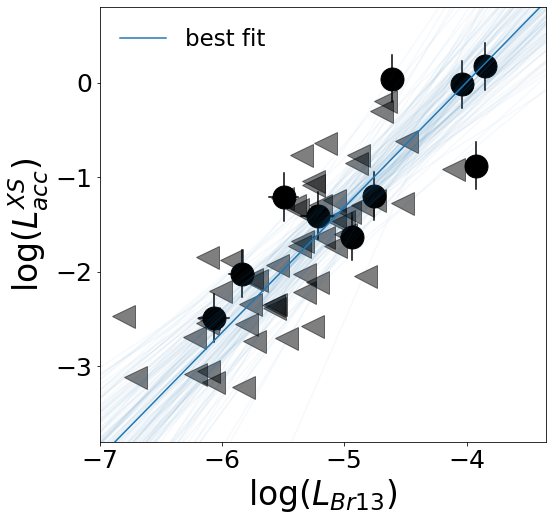}
    \includegraphics[width=0.6\columnwidth]{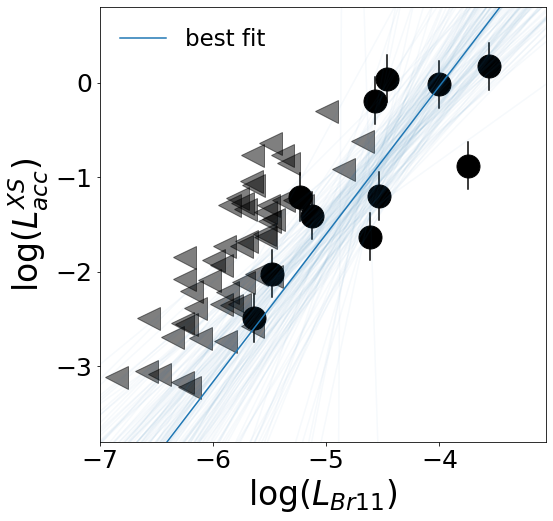}
    \includegraphics[width=0.61\columnwidth]{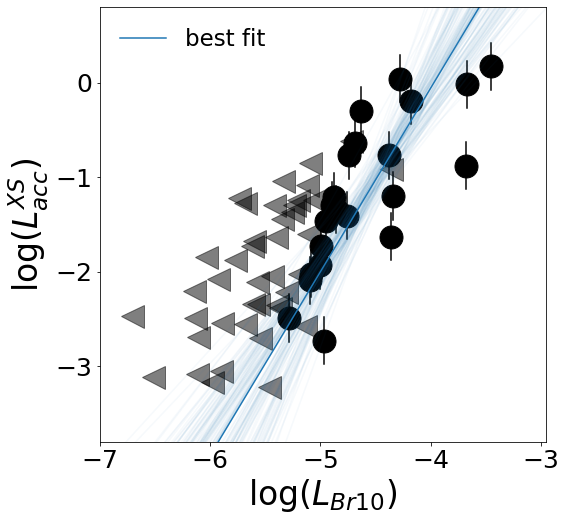}
    \caption{$\log \lacc - \log \lline$ and best fit for the Brackett series, using $A_V^{XS}$ to deredden the fluxes.}
    \label{fig:lacc-lline-br}
\end{figure*}

\begin{figure*}
    \centering
    \includegraphics[width=0.6\columnwidth]{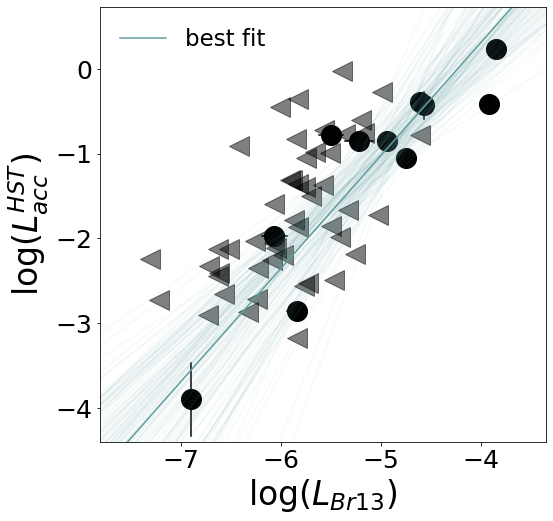}
    \includegraphics[width=0.6\columnwidth]{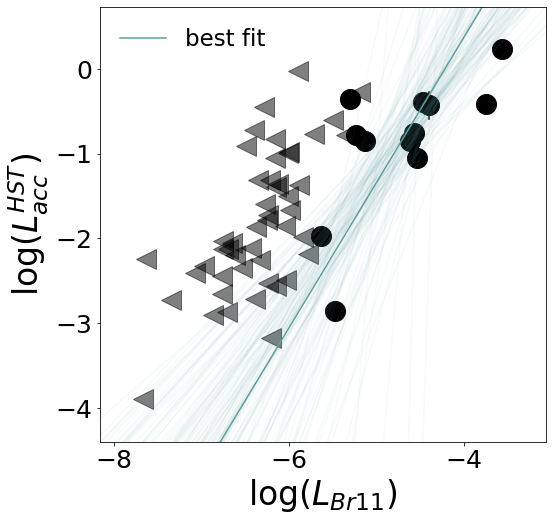}
    \includegraphics[width=0.6\columnwidth]{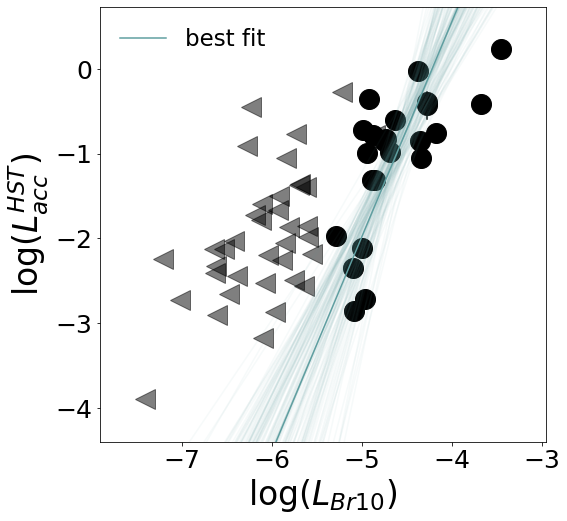}
    \caption{$\log \lacc - \log \lline$ and best fit for the Brackett series, using $A_V^{HST}$ to deredden the fluxes.}
    \label{fig:lacc-lline-br_MIADM}
\end{figure*}

Figs.\,\ref{fig:lacc-lline-br} and \ref{fig:lacc-lline-br_MIADM} shows the $\lacc - \lline$ correlation for some lines of the Brackett series derived using the results of the XS- and HST-fit, respectively. Coefficients of empirical relations are listed in Tab.\,\ref{tab:br-series} only for lines detected in more than 10 sources. 

\begin{table*}
\centering
\caption{\label{tab:br-series} $\lacc - \lline$ linear fit for the Brackett series.}
\begin{tabular}{lcccccc}
\hline
\hline
Line & $\lambda_0$ [$\mu$m] & a & b & c.f. & $\sigma$  & N$_{\rm points}$ \\
\hline		
\hline	
using $\lacc^{XS}$ and $A_V^{XS}$: \\
Br13 & 1.61 & $1.34 \pm 0.05$ & $5.38 \pm 0.44$ & 0.8 & $0.31 \pm 0.31$ & 11 \\
Br11 & 1.68 & $1.52 \pm 0.09$ & $5.98 \pm 0.58$ & 0.7 & $0.34 \pm 0.28$ & 11 \\
Br10 & 1.74 & $1.90 \pm 0.07$ & $7.52 \pm 0.42$ & 0.8 & $0.12 \pm 0.05$ & 22 \\
\hline
using $\lacc^{HST}$ $A_V^{HST}$: \\
Br13 & 1.61 & 1.33$\pm$0.05 &  5.65$\pm$0.38 & 0.8 & 0.28$\pm$0.17 & 11 \\
Br11 & 1.68 & 1.96$\pm$0.20 &  8.35$\pm$1.06 & 0.6 & 0.47$\pm$0.41 & 12 \\
Br10 & 1.74 & 2.54$\pm$0.09 & 10.71$\pm$0.50 & 0.7 & 0.14$\pm$0.05 & 23 \\
\hline 
\hline
\end{tabular}
\begin{quotation}
  \textbf{Notes.} $\lambda_0$ is the central wavelength of the line; $a$ and $b$ coefficients are the best fit of the $\log \lacc = a \, \log \lline + b$ relation; c.f. stands for correlation factor between $\lacc$ and $\lline$, and $\sigma$ is the spread of the distribution. 
  \end{quotation} 
 \end{table*}

\end{document}